%% file: ms.tex
\documentclass[iop,numberedappendix]{emulateapj}

\usepackage{graphics}
\usepackage{graphicx}
\usepackage{natbib}
\usepackage{epsfig}
\usepackage{epstopdf}

\begin{document}

\title{Low-Mass Eclipsing Binaries in the Initial \emph{Kepler} Data Release}
\slugcomment{To appear in to the Astronomical Journal - Draft version \today}

\author{J. L. Coughlin\altaffilmark{1,5}, M. L\'opez-Morales\altaffilmark{2,3}, T. E. Harrison\altaffilmark{1}, N. Ule\altaffilmark{1}, D. I. Hoffman\altaffilmark{4}}

\altaffiltext{1}{Department of Astronomy, New Mexico State University, P.O. Box 30001, MSC 4500, Las Cruces, New Mexico 88003-8001}
\altaffiltext{2}{Carnegie Institution of Washington, Department of Terrestrial Magnetism, 5241 Broad Branch Road NW, Washington, DC 20015, USA; Hubble Fellow}
\altaffiltext{3}{Institut de Ci\`encies de l'Espai (CSIC-IEEC), Campus UAB, Facultat de Ci\`encies, Torre C5, parell, 2a pl, E-08193 Bellaterra, Barcelona, Spain}
\altaffiltext{4}{California Institute of Technology, MC 249-17, 1200 East California Blvd, Pasadena, CA 91125}
\altaffiltext{5}{NSF Graduate Research Fellow; jlcough@nmsu.edu}


\begin{abstract}
We identify 231 objects in the newly released Cycle 0 dataset from the \emph{Kepler Mission} as double-eclipse, detached eclipsing binary systems with T$_{\rm eff}$ $<$ 5500 K and orbital periods shorter than $\sim$32 days. We model each light curve using the JKTEBOP code with a genetic algorithm to obtain precise values for each system. We identify 95 new systems with both components below 1.0 M$_{\sun}$ and eclipses of at least 0.1 magnitudes, suitable for ground-based follow-up. Of these, 14 have periods less than 1.0 day, 52 have periods between 1.0 and 10.0 days, and 29 have periods greater than 10.0 days. This new sample of main-sequence, low-mass, double-eclipse, detached eclipsing binary candidates more than doubles the number of previously known systems, and extends the sample into the completely heretofore unexplored P $>$ 10.0 day period regime. We find preliminary evidence from these systems that the radii of low-mass stars in binary systems decrease with period. This supports the theory that binary spin-up is the primary cause of inflated radii in low-mass binary systems, although a full analysis of each system with radial-velocity and multi-color light curves is needed to fully explore this hypothesis. As well, we present 7 new transiting planet candidates that do not appear among the recently released list of 706 candidates by the \emph{Kepler} team, nor in the \emph{Kepler} False Positive Catalog, along with several other new and interesting systems. We also present novel techniques for the identification, period analysis, and modeling of eclipsing binaries.
\end{abstract}

\keywords{stars: binaries: eclipsing --- stars: binaries: general --- stars: fundamental parameters --- stars: late-type --- stars: low-mass}

\section{Introduction}

A double-lined, detached, eclipsing binary (DDEB) is a system that contains two non-interacting, eclipsing stars, in which the spectra of both components can
be clearly seen, allowing for the radial-velocity (RV) of each component to be obtained. In these systems, the mass and radius of each star can be determined with errors usually less than 1-2\%, thus making DDEBs currently the most accurate method of obtaining masses and radii of stars. Models of main-sequence stars with masses similar to or greater than the Sun have been tested over the years using DDEBs. \citet{Popper1980} compiled available masses and radii of DDEB's with accuracies of $\le$ 15\%, up to that date, and found general agreement with stellar models, though stressed the need for more accurate observations and models. \citet{Andersen1991} provided a compilation of all available DDEB systems up to that date, with accuracies $\le$ 2\%, and showed that the masses and radii of these stars were in general agreement with the current stellar evolution models, with any discrepancies attributable to abundance variations. \citet{Torres2010} recently performed a similar review with nearly double the sample of DDEBs. They were able to show the need to include non-classical effects such as diffusion and convection in stellar models, definitively demonstrate the existence of significant structural differences in magnetically active and fast-rotating stars, test theories of rotational synchronization and orbital circularization, and validate General Relativity via apsidal motion rates. However, while observations of DDEBs have enhanced our understanding of stellar structure and evolution for stars with M $\geq$ 1.0 M$_{\sun}$, low-mass, main-sequence (LMMS) stars, (M $<$ 1.0 M$_{\sun}$ and T$_{\rm eff}$ $<$ 5800 K), have not been tested to the same extent.

Although a couple systems with late G or early K type components had been studied prior to 2000, \citep[c.f.][and references therein]{Popper1980,Andersen1991,Torres2006,Clausen2009}, only three LMMS DDEBs with late K or M type components were known \citep{Lacy1977,Leung1978,Delfosse1999}. This number had only increased to nine by the beginning of 2007 \citep[cf.][Table 1]{LopezMorales2007}. Despite the fact that the majority of main-sequence stars are low-mass, these stars are both intrinsically fainter, and physically smaller, than their more massive counterparts. Therefore, they have a lower eclipse probability and are harder to discover and study. As outlined by \citet{LopezMorales2007}, analysis of these systems showed that the observed radii for these stars are consistently $\sim$10-20\% larger than predicted by stellar models \citep{Baraffe1998} for 0.3 M$_{\sun}$ $\lesssim$ M $\lesssim$ 0.8 M$_{\sun}$. \citet{Fernandez2009} recently showed this was also likely the case for five M dwarfs in short-period eclipsing systems with an F type primary, though since the systems are only single-lined, the masses could not be determined directly. This discrepancy between the radii derived from models and from observations either reveals a flaw in the stellar models for this mass regime, or is due to differences in metallicity, magnetic activity, or interpretation of the light curve data when star spots are present \citep{Morales2008}. As to this last point, \citet{Morales2010} recently noted that improperly taking polar spots into account in the light curve modeling process may possibly cause the derivation of stellar radii a few percent larger than the true values for some of these systems. Of all of these scenarios, enhanced magnetic activity has been proposed as the principal cause of inflated radii \citep{Chabrier2007,LopezMorales2007,Morales2008}.

If enhanced magnetic activity is the principal cause of the inflated radii, shorter-period binary systems, with the stellar rotation rate enhanced by the revolution of the system, would be expected to show greater activity and thus larger radii than longer-period systems \citep{Chabrier2007}. Binary systems with component masses of 0.5 M$_{\sun}$ are expected to synchronize, and therefore be spun-up, in less than 0.1 Gyr for periods less than 4 days, and in less than 1 Gyr for periods less than 8 days \citep{Zahn1977, Zahn1994}. Thus, the discovery of LMMS DDEBs with P $\gtrsim$ 10 days, where the binary components should have natural rotation rates, is crucial to probing if enhanced rotation due to binarity is the underlying cause of this phenomenon. This theory might be supported by measurements of a couple isolated field M and K dwarf stars via very long baseline interferometry, which \citet{Demory2009} found to match stellar models. However, recently a much larger sample of nearly two dozen isolated M and K dwarf stars finds, for $\sim$80\% of the sample, larger radii than the model predictions for 0.35 $<$ M $<$ 0.65 M$_{\sun}$ \citep{Boyajian2010}, indicating that there are likely multiple causes of inflation at work, or a remaining flaw in the stellar models.


Though several more LMMS DDEB systems have been found since 2007, \citep{Coughlin2007,Shaw2007,Becker2008,Blake2008,Devor2008a,Devor2008b,Shkolnik2008,Hoffman2008,Irwin2009,Dimitrov2010,Shkolnik2010}, there are to-date only 7 well-studied systems with 1.0 $<$ P $<$ 3.0 days \citep[][and references therein]{LopezMorales2007} \citep{Becker2008,Shkolnik2008}, and only one has a larger period, at P = 8.4 days \citep{Devor2008b}. This is mostly due to the fact that ground-based photometric surveys, such as NSVS, TrES, and OGLE, are either cadence, precision, magnitude, or number limited, and thus not sensitive to long periods. The \emph{Kepler Mission}, with 3 years of constant photometric monitoring of over 150,000 stars with V $\lesssim$ 17, at 30-minute cadence and sub-millimagnitude precision, is the key to discovering a large number of long-period, LMMS DDEBs. 

In this paper we present the results of our search through all the newly available \emph{Kepler} Q0 and Q1 public data for LMMS DDEBs. Section~\ref{datasec} describes the data we use in this paper. Section~\ref{binaryidentsec} describes our binary identification technique, and Section~\ref{modelsec} describes how we model the light curves. Our selection and list of new LMMS DDEBs is presented in Section~\ref{newlmbsec}, and we present new transiting planet candidates in Section~\ref{transsec}. In Section~\ref{lmbmodelcompsec} we compare the new LMMS DDEBs with theoretical models, and conclude with a summary of our results in section~\ref{concsec}. Once accurate mass and radius values exist for a large range of both mass and period, our understanding of these objects should substantially improve, and we will be one step closer to extending to the lower-mass regime the advanced study of stellar structure and evolution that sun-like and high-mass stars have been a subject of for some time.

\section{Observational Data}
\label{datasec}

The data used in our analysis consists of the 201,631 light curves made public by the \emph{Kepler Mission}\footnote{http://kepler.nasa.gov/} as of June 15, 2010 from \emph{Kepler} Q0 and Q1 observations. All light curves can be accessed through the Multi-mission Archive at STScI (MAST)\footnote{http://archive.stsci.edu/kepler/}. The data consist of 51,366 light curves from \emph{Kepler} Q0, (observed from 2009-05-02 00:54:56 to 2009-05-11 17:51:31 UT), and 150,265 light curves from \emph{Kepler} Q1, (observed from 2009-05-13 00:15:49 to 2009-06-15 11:32:57 UT), each at 29.43 minute cadence. Individual light curves for Q0 contain $\sim$470 data points, and for Q1 contain $\sim$1,600 data points. Targets range in \emph{Kepler} magnitude from 17.0 at the faintest, to 5.0 at the brightest.

The \emph{Kepler} team has performed pixel level calibrations, (including bias, dark current, flat-field, gain, and non-linearity corrections), identified and cleaned cosmic-ray events, estimated and removed background signal, and then extracted time-series photometry using an optimum photometric aperture. They have also removed systematic trends due to spacecraft pointing, temperature fluctuations, and other sources of systematic error, and corrected for excess flux in the optimal photometric aperture due to crowding \citep{Vancleve2010}. It is this final, ``corrected'' photometry that we have downloaded for use in our analysis.

\section{Eclipsing Binary Identification} 
\label{binaryidentsec}

\citet{Prsa2010} have recently released an initial catalog of eclipsing binary stars they find in the \emph{Kepler} field from the same Q0 and Q1 data we use in this paper. They first identified EB candidates via \emph{Kepler's} Transit Planet Search (TPS) algorithm, eliminating those targets already identified as exoplanet candidates. To determine the ephemeris of each candidate, they used Lomb-Scargle, Analysis of Variance, and Box-fitting Least Squares periodogram techniques, combined with manual inspection and modification. They then culled, through manual inspection, non-EB candidates, such as pulsating and heavily spotted stars, as well as duplicates due to contamination from nearby stars, and arrive at their final list of 1,832 binaries, which are manually classified as detached, semi-detached, over-contact, ellipsoidal, or unknown. Next, they estimate the principal parameters of each system, (temperature ratio, sum of the fractional radii, e$\cdot$cos($\omega$), e$\cdot$sin($\omega$), and sin(i) for detached systems), via a neural network technique called Eclipsing Binaries via Artificial Intelligence \citep[``EBAI''][]{Prsa2008}. For our search, which focuses on the detection of LMMS DDEBs, we have devised our own DDEB identification technique, which we apply to the Q1 data. We do not use the Q0 data in this part of the analysis to avoid discrepant systematics between the two quarters, which complicate the analysis. 

\begin{figure}
\centering
\epsfig{width=\linewidth,file=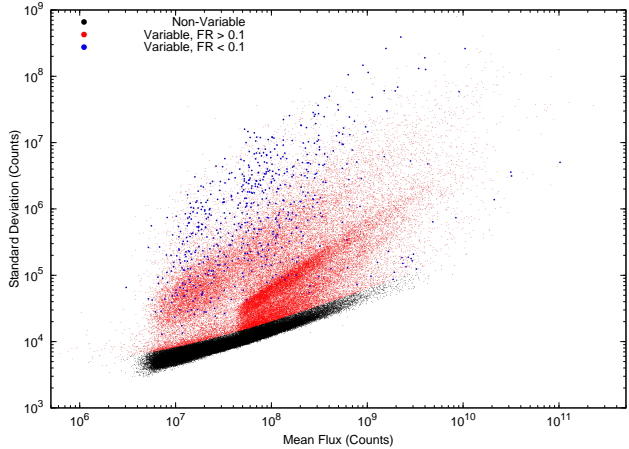}
\caption{Plot of standard deviation versus mean flux for the 150,265 stars in Q1. Black dots represent stars that vary by less than 1-sigma from a best-fit power-law to the data, and thus we classify them as non-variables. Red dots represent variables with a flux ratio greater than 0.1. Blue dots represent variables with a flux ratio less than 0.1, and thus are good candidates to be eclipsing binaries.}
\label{stdevfig}
\end{figure}

Our search consisted of two steps. The first was to identify variable stars, and to do so, we placed a light curve standard deviation limit above which the objects are classified as variables. We first subtracted an error-weighted, linear fit of flux versus time from all data, to remove any remaining linear systematic trends, and then plotted the standard deviation of each light curve versus its average flux and fit a power law. These data are shown in Figure~\ref{stdevfig}, where the black dots correspond to light curves which deviate by less than 1$\sigma$ from the standard deviation versus average flux fit, and we thus classify as non-variable. The colored dots indicate the variable candidates that deviate by more than 1$\sigma$. Next, we used the flux ratio (FR) measurement criterion, which we adapted from the magnitude ratio given in \citet{Kinemuchi2006}, and is defined as 
\begin{equation} \textrm{FR} = \frac{\textrm{maximum flux - median flux}}{\textrm{maximum flux - minimum flux}} \end{equation} 
as a measure of whether or not the variable spends most of its time above (low FR value) or below (high FR value) the median flux value. Perfectly sinusoidal variables have FR $=$ 0.50, pulsating variables, such as RR Lyrae's, have FR $>$ 0.5, and eclipsing binaries have FR $<$ 0.5. As we are principally interested in finding well detached systems with relatively deep, narrow eclipses, which thus have low FR values, we make a further cut of the systems and only examine those variables with FR $<$ 0.1, shown by blue dots in Figure~\ref{stdevfig}.

The second step of the analysis was to determine the orbital period of each candidate. This was done using two independent techniques that are both well-suited for detached eclipsing binary systems. The first is Phase Dispersion Minimization (PDM) \citep{Stellingwerf1978}, which attempts to find the period that best minimizes the variance in multiple phase bins of the folded light curve. This technique is not sensitive to the shape of the light curve, and thus is ideal for non-sinusoidal variables such as detached eclipsing binaries. The downside of this technique is that if strong periodic features exist in the light curve, which do not correspond to the period of eclipses, such as rapidly varying spots, stellar pulsations, or leftover systematics, they can weaken the signal of the eclipse period. We use the latest implementation given by \citet{Stellingwerf2006}, and determine the best three periods via this technique to ensure that the true period is found, and not just an integer multiple, or fraction, thereof. 

The second technique we use is one we invented specifically for detached eclipsing binaries, and call Eclipse Phase Dispersion Minimization (EPDM). The idea behind EPDM is that we want to automatically identify and align the primary eclipses in an eclipsing binary, thus finding the period of the system. To accomplish this, EPDM finds the period that best minimizes the dispersion of the actual phase values of the faintest N points in a light curve, i.e. the very bottom of the eclipses. Since EPDM only selects the N faintest points in a light curve, it is not affected by systematics or periodic features that do not correspond to the period of eclipses, assuming the systematics do not extend below the depth of the eclipses. The technique works for all binary systems with equal or unequal eclipse depths, and transiting planets, both with either zero or non-zero eccentricity. Computationally, EPDM is significantly faster than traditional PDM techniques. For a detailed and illustrative explanation of this new technique, please see Appendix~\ref{epdmappendsec}. We use EPDM to find the three best fit periods for each system as well, for the same reasons as we did with PDM.

\begin{figure}
\centering
\epsfig{width=\linewidth,file=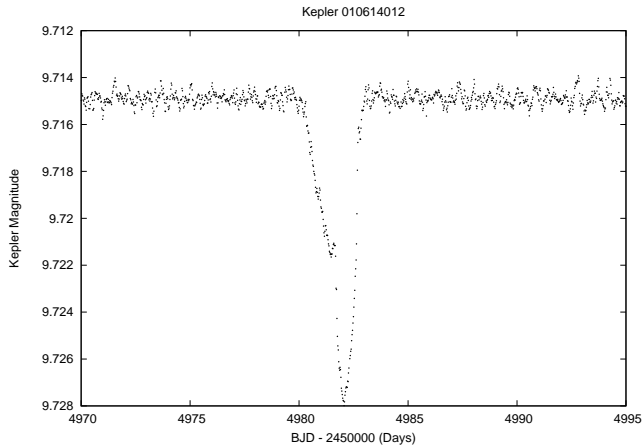}
\caption{Kepler 010614012. An apparent red giant, (T$_{\rm eff}$ = 4859K, logg = 3.086, [M/H] = -0.641, R$_{\star}$ = 5.708 R$_{\sun}$), with a very unusual, shallow, eclipse-like feature.}
\label{Kepler010614012}
\end{figure}

We identify 577 EB candidates in the Q1 data. Of these, 486 are listed by \citet{Prsa2010} as detached eclipsing binaries, and 20 are identified as semi-detached eclipsing binaries. The 71 remaining candidates were manually inspected by examining both the raw and phased light curves at the 6 best periods found via PDM and EPDM. Of these 71 remaining candidates, 48 turned out to be false positives with significantly large, sharp systematic features, and one is an apparent red giant, (Kepler 010614012, T$_{\rm eff}$ = 4859K, logg = 3.086, [M/H] = -0.641, R$_{\star}$ = 5.708 R$_{\sun}$), with an unusual, asymmetrical, eclipse-like feature that lasts for $\sim$3 days with a depth of 1.2\%, shown in Figure~\ref{Kepler010614012}. This does not appear to be a systematic feature due to its very flat out of eclipse baseline, contiguous nature, long duration, and the actual time at which the feature occurs, compared to the majority of other objects with strong systematics. The remaining 22 targets are: 2 transiting exoplanet candidates contained in the recently released list of 306 candidates by \citet{Borucki2010}, 3 already published transiting planets, (Kepler-5b, Kepler-6b, and TrES-2b), 7 shallow eclipsing systems with primary eclipse depths ranging from 1.4\% to 5.7\%, visible secondary eclipses ranging from 0.05\% to 4.6\%, and periods ranging from 4.7 to 45.3 days, the already published transiting hot compact object Kepler 008823868 \citep{Rowe2010}, a 6.4 day eclipsing binary with T$_{\rm eff}$ = 5893K and eclipse depths of 38.4\% and 12.2\% (Kepler 006182849), and 8 transiting exoplanet candidates with transit depths ranging from 0.75\% to 4.9\%, and periods ranging from 2.5 to 24.7 days. For the 7 new extremely shallow eclipsing systems, we list their \emph{Kepler} ID numbers, periods, effective temperatures, surface gravities, and primary and secondary eclipse depths in Table~\ref{shallowebtab}, and note they could be of interest for follow-up due to the potential to contain brown dwarf or extremely low-mass secondaries, or even anomalously hot exoplanet companions. Of the 8 transiting candidates, only one is listed in the \emph{Kepler} false positive catalog\footnote{http://archive.stsci.edu/kepler/false\_positives.html}, Kepler 011974540. None of them are in the list of the 306 released candidates by \citet{Borucki2010}, nor are among the 400 planetary candidates currently reserved for follow-up observations \citep{Borucki2010}. These will be further discussed in Section~\ref{transsec}.

\section{Light Curve Modeling}
\label{modelsec}

Since the system parameters determined by \citet{Prsa2010} are only estimates and do not incorporate spots, and since we seek to obtain as accurate physical parameters as possible, we modeled each system using a robust global minimization scheme with a commonly used, physically detailed eclipsing binary modeling code. We took all 314 detached eclipsing binaries with T$_{\rm eff}$ $<$ 5500K and that are publicly available, (5 systems are still proprietary), identified from both our search and the \citet{Prsa2010} catalog, combined Q0 and Q1 data if available, and via manual inspection classified systems as double-eclipse (i.e. contained two visible eclipses), single-eclipse (i.e. only contained one eclipse), or as spurious results that were not recognizable as eclipsing systems. (Given the errors in the KIC temperature determination, and to ensure the primary is below 1.0 M$_{\sun}$, we used 5500K as our cutoff, instead of 5800K. As well, the definition of a ``double-lined'' system is one in which the lines of both components are visible in an observed spectrum. Although in general if two eclipses are clearly visible in the photometric light curve, it is likely to be ``double-lined'', this cannot be determined without an actual spectrum. Thus, we use the term ``double-eclipse'' throughout the paper, with the assumption that when observed spectroscopically, the majority of these systems will be observed as ``double-lined''.) 

\begin{deluxetable}{cccccc}
\tablewidth{0pt}
\tabletypesize{\scriptsize}
\tablecaption{Period, Effective Temperature, Surface Gravity, and Eclipse Depth Estimates for the 7 New Extremely Shallow Eclipsing Systems}
\tablecolumns{6}
\tablehead{\emph{Kepler} ID & Period & T$_{\rm eff}$ & logg & Pri. & Sec.\\ & (Days) & (K) & & (\%) & (\%)}
\startdata
\input{tab1.tex}
\enddata
\tablenotetext{1}{System is listed in the \emph{Kepler} False Positive Catalog as likely to be an EB.}
\tablenotetext{2}{System has non-zero eccentricity.}
\tablenotetext{3}{Period derived assuming zero eccentricity.}
\label{shallowebtab}
\end{deluxetable}

We then used the JKTEBOP eclipsing binary modeling program \citep{Southworth2004a,Southworth2004b} to model every double-eclipse eclipsing binary system, of which there were 231, solving for the period, time of primary minimum, inclination, mass ratio, e$\cdot$cos($\omega$), e$\cdot$sin($\omega$), surface brightness ratio, sum of the fractional radii, ratio of the radii, and out of eclipse flux. In addition, we also solved for the amplitude and time of minimum of a sinusoidal term imposed on the luminosity of the primary component, with the period fixed to that of the binary, in order to account for spots. Note that in the JKTEBOP model the mass ratio is only used to determine the amount of tidal deformation of the stars from a pure sphere. Thus, it has no effect on the light curve of long-period systems, which due to their large separations are almost perfectly spherical, but must be included to properly model very short-period systems, where the tidal deformation can have a significant impact on the light curve. We used the quadratic limb darkening law, which works well for late-type stars \citep[e.g.][]{Manduca1977,Wade1985,Claret1990}, with coefficients set to those found by \citet{Sing2010} for the \emph{Kepler} bandpass via interpolation given the systems' effective temperatures, surface gravities, and metallicities as listed in the Kepler Input Catalog (KIC)\footnote{http://archive.stsci.edu/kepler/kepler\_fov/search.php}. We also fixed the gravity darkening exponent based on the effective temperature as prescribed by \citet{Claret2000}. As any contaminating flux from nearby stars in the photometric aperture has already been compensated for in the \emph{Kepler} pipeline \citep{Vancleve2010}, we set the amount of third light to 0.0. Note that third light might still exist in some systems if there is a background star or tertiary component that is unidentifiable from ground-based surveys, (i.e. less than $\sim$1$\arcsec$ separation), but since third light is usually unconstrained in a single-color light curve, we do not let it vary. If third light existed in a system and was not accounted for, the solution would result in an inclination determination lower than the true value, and therefore an over-estimation of the stellar radius. However, this should only occur in a minority of systems. For a couple binaries in our list, the light curves absolutely could not be modeled without the inclusion of third light, (i.e. very sharp eclipses with depths of less than 0.01 mag). For these cases only, we let the third light vary, and thus be a non-zero parameter. Additionally, if the effect of spots in a light curve deviates significantly from the adopted sinusoidal shape, it could affect the derived luminosity ratio to a minor extent, but it should not affect the sum of the radii.

In order to model such a large number of systems over such a large solution space, and to ensure we have found the best global solution, we adapted the JKTEBOP code to use a modified version of the asexual genetic algorithm (AGA) described by \citet{Canto2009}, coupled with its standard Levenberg-Marquardt minimization algorithm. Genetic algorithms (GA) are an extremely efficient method of fitting computationally intensive, multi-parameter models over a large and potentially discontinuous parameter space, and thus ideal for this work. For the details of how genetic algorithms work, and the specific changes we made to the \citet{Canto2009} AGA, please see Appendix~\ref{agaappendix}. 

We found that our modified AGA does an excellent job of solving well-behaved light curves, simultaneously varying all 12 aforementioned parameters over the entire range of possible solutions. For some of the systems however, strong systematics and/or variable star spots introduced a significant amount of noise, especially in systems with shallow eclipses, for which it was more difficult to arrive at a robust solution. For these systems we had to manually correct the systematics, often by either eliminating the Q0 or Q1 data, equalizing the base flux levels of Q0 and Q1 data, or subtracting out a quasi-sinusoidal variation in the base flux level due to remaining \emph{Kepler} systematics. When possible we attempted to minimize the amount of manual interference. Hopefully this will become much less of a problem with subsequent data releases. We then re-ran the AGA using a larger initial population until a good solution was found. Every light curve in the end was visually inspected to be a good fit compared to the scatter of the data points, and the obtained parameters were confirmed to be reasonable when visually inspecting the light curves.

\section{New Low-Mass Binary Candidates}
\label{newlmbsec}

In order to identify the main-sequence stars from our list of 231 candidates, and determine the best candidates for follow-up, we employ the following technique to estimate the temperature, mass, and radius of each star using the sum of the fractional radii, $r_{sum}$, and period, $P$, obtained from our JKTEBOP models, the luminosity ratio, L$_{r}$, (which is derived from the surface brightness ratio, $J$, and radii ratio, $k$, obtained from the models), and the effective temperature of the system, $T_{\rm eff}$, obtained from the KIC, with an estimated error of $\pm$200 K.

The value for $T_{\rm eff}$ given in the KIC was determined via interpolation of standard color magnitude relations as determined by ground-based, multi-wavelength photometry \citep{Vancleve2010}. Although in principle one might be able to deconvolve two separate spectral energy distributions from this photometry, in reality given the level of photometric error in the KIC and uncertainty at which binary phase the photometry was obtained, this is untenable. Instead, we assume the stars radiate as blackbodies, and that each star contributes to the determined $T_{\rm eff}$ in proportion to its luminosity. Thus, following our assumption, we obtain the following relation,

\begin{equation}
\label{teffeq}
T_{\rm eff} = \frac{L_{1}T_{1} + L_{2}T_{2}}{L_{1}+L_{2}}
\end{equation}

where $L_{1}$, $L_{2}$, $T_{1}$ and $T_{2}$  are the luminosities and effective temperatures of star 1 and 2 respectively. Still assuming the stars radiate as blackbodies, the luminosity of each star is proportional to its radius squared and temperature to the fourth power, with the temperature proportional to is surface brightness to the one-fourth power. Thus, we find that the luminosity ratio can be expressed as,

\begin{eqnarray}
\label{lumeq}
L_{r} & = & \frac{L_{1}}{L_{2}} = \frac{r_{1}^{2}T_{1}^{4}}{r_{2}^{2}T_{2}^{4}} = k^{2}T_{r}^{4} = k^{2}\left[\left(\frac{SB_{1}}{SB_{2}}\right)^{1/4}\right]^{4} \nonumber \\ & = & k^{2}\left(J^{\frac{1}{4}}\right)^{4} = k^{2}J
\end{eqnarray}

where $SB_{1}$ and $SB_{2}$ are the surface brightnesses of star 1 and star 2 respectively, and $r_{1}$ and $r_{2}$ are the fractional radii of star 1 and 2 respectively, defined as $R_{1}/a$ and $R_{2}/a$, where $R_{1}$ and $R_{2}$ are the physical radius of each star, and $a$ is the semi-major axis of, or separation between, the components. Combining equations~\ref{teffeq} and \ref{lumeq} yields the expression,

\begin{equation}
\label{finallumeq}
T_{\rm eff} = \frac{L_{r}T_{1} + T_{2}}{L_{r}+1}
\end{equation}

which has two known parameters, T$_{\rm eff}$ and L$_{r}$, and two unknown parameters, T$_{1}$ and T$_{2}$. To place a further constraint upon the values of T$_{1}$ and T$_{2}$, we make the assumption that both stars in the binary are on the main-sequence, and employ the mass, temperature, radius, and average of the V-band and R-band luminosity relations given in \citet{Baraffe1998} for 0.075 $\le$ M $\le$ 1.0 M$_{\sun}$ and in \citet{Chabrier2000} for M $<$ 0.075 M$_{\sun}$, both assuming an age of 5.0 Gyr and [M/H] = 0.0. (We average the V and R-band luminosities to obtain a very close approximation to the \emph{Kepler} bandpass.) From these models, for a given value of T$_{1}$, there is only one value of T$_{2}$ which will reproduce the observed value of L$_{r}$. Thus, there only exists one set of unique values for T$_{1}$ and T$_{2}$ that reproduces both the observed T$_{\rm eff}$ and L$_{r}$ values for the system. 

For each T$_{1}$ and T$_{2}$ then, we obtain the absolute masses and radii, (M$_{1}$, M$_{2}$, R$_{1}$, and R$_{2}$), via interpolation from the \citet{Baraffe1998} and \citet{Chabrier2000} models. Then, utilizing Kepler's 3$^{rd}$ law, given the total mass of the system, we calculate the semi-major axis, $a$, via

\begin{equation}
\label{kepeq}
a = (GM_{tot})^{\frac{1}{3}}(\frac{P}{2\pi})^{\frac{2}{3}}
\end{equation}

where $M_{\rm tot}$ is the total mass of the system, M$_{1}$ + M$_{2}$, and G is the gravitational constant. We then multiply each radius determined above by a constant so that the sum of the fractional radii derived from the JKTEBOP model, $r_{\rm sum}$, is equal to $(R_{1} + R_{2})/a$, the sum of the fractional radii when using the physical values of $M_{1}$, $M_{2}$, $R_{1}$, $R_{2}$, and $P$. This technique is robust because while individual parameters such as $i$, $J$, and $k$ can suffer from degeneracies, especially in systems with shallow eclipses, the values of $r_{sum}$ and $L_{r} = k^{2}J$, which we rely on, are firmly set by the width of the eclipses and the difference in their eclipse depths, respectively.

For clarity, we now illustrate the individual steps of this procedure using the example of an actual system, Kepler 002437452. This system was found to have T$_{\rm eff}$ = 5398 K and L$_{r}$ = 3.90 from the KIC and the JKTEBOP modeling respectively. Now, assuming the stars are main-sequence, one could choose values of T$_{1}$ = 4000 K and T$_{2}$ = 3620 K, and looking up their luminosities from the \citet{Baraffe1998} models, find that the luminosity ratio between two main-sequence stars with temperatures of 4000 K and a 3620 K is 3.90. In this case, the luminosity ratio criterion would be satisfied, but T$_{\rm eff}$ would would be $\sim$3922 K, nowhere near the measured value of 5398 K. Similarly, one could choose values of T$_{1}$ = 5400 and T$_{2}$ = 5393, and this would yield T$_{\rm eff}$ = 5398 K, but L$_{r}$ would be 1.01, nowhere near the needed value of 3.90. The unique solution that satisfies both the effective temperature and luminosity ratio constraints is that T$_{1}$ = 5591 and T$_{2}$ = 4647, which yields both T$_{\rm eff}$ = 5398 and L$_{r}$ = 3.90. Now, given these temperatures, interpolating from the \citet{Baraffe1998} models yields values of M$_{1}$ = 0.963 M$_{\sun}$, R$_{1}$ = 0.966 R$_{\sun}$, M$_{2}$ = 0.792 M$_{\sun}$, and R$_{2}$ = 0.783 R$_{\sun}$. Taking the masses, and the period of the system of 14.47184 days, and utilizing Eq.~\ref{kepeq}, we find that the semi-major axis, $a$, would be 30.1 R$_{\sun}$. Dividing the sum of the estimated physical radii by the semi-major axis just calculated, we find a value of 0.058 for the sum of fractional radii. Now, from the JKTEBOP model, this system was measured to have a sum of the fractional radii of 0.084, and so it appears that the current values for the radii are underestimated. Thus, we multiply the radii by a factor of 0.084/0.058 = 1.45, to obtain our final radii values of R$_{1}$ = 1.40 R$_{\sun}$ and R$_{2}$ = 1.13 R$_{\sun}$, with, as above, M$_{1}$ = 0.96 M$_{\sun}$, T$_{1}$ = 5591 K, M$_{2}$ = 0.79 M$_{\sun}$, and T$_{2}$ = 4647 K.

\begin{figure}
\centering
\begin{tabular}{c}
\epsfig{width=\linewidth,file=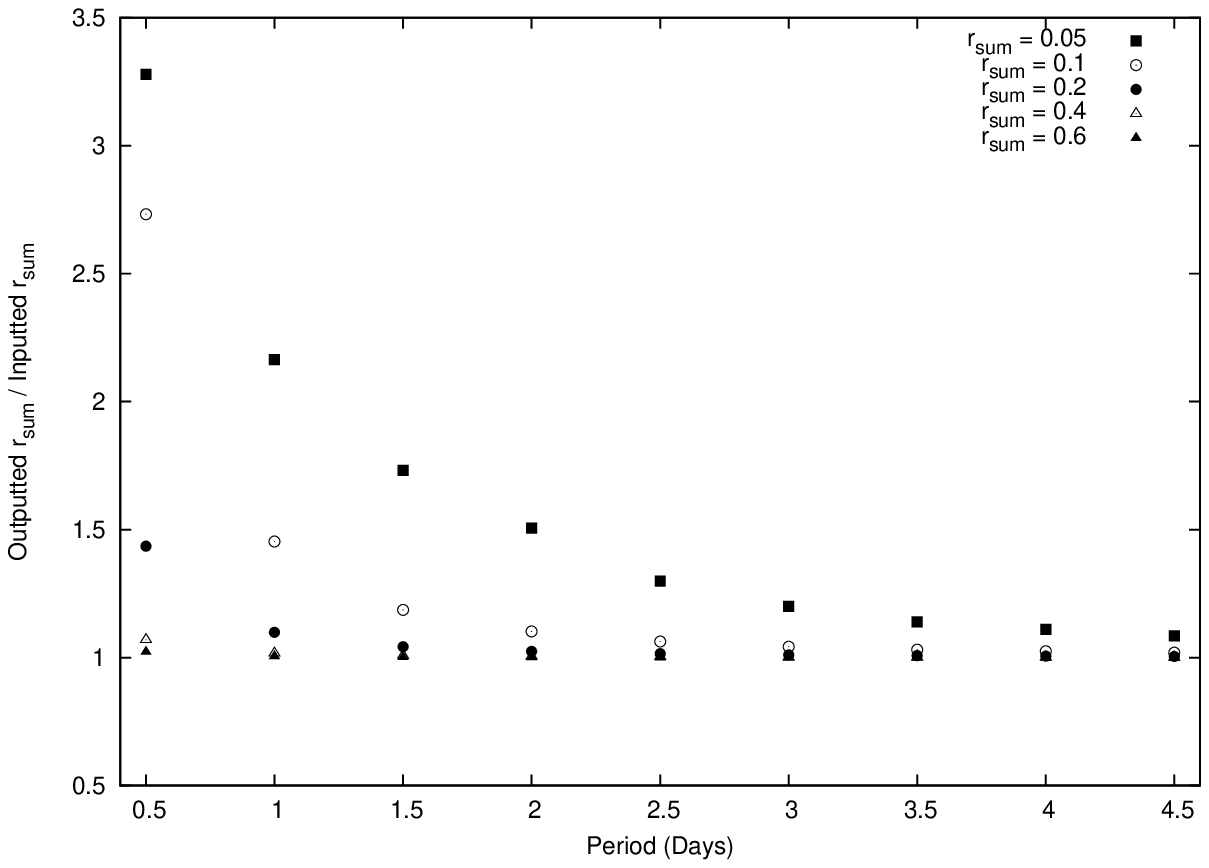} \\
\epsfig{width=\linewidth,file=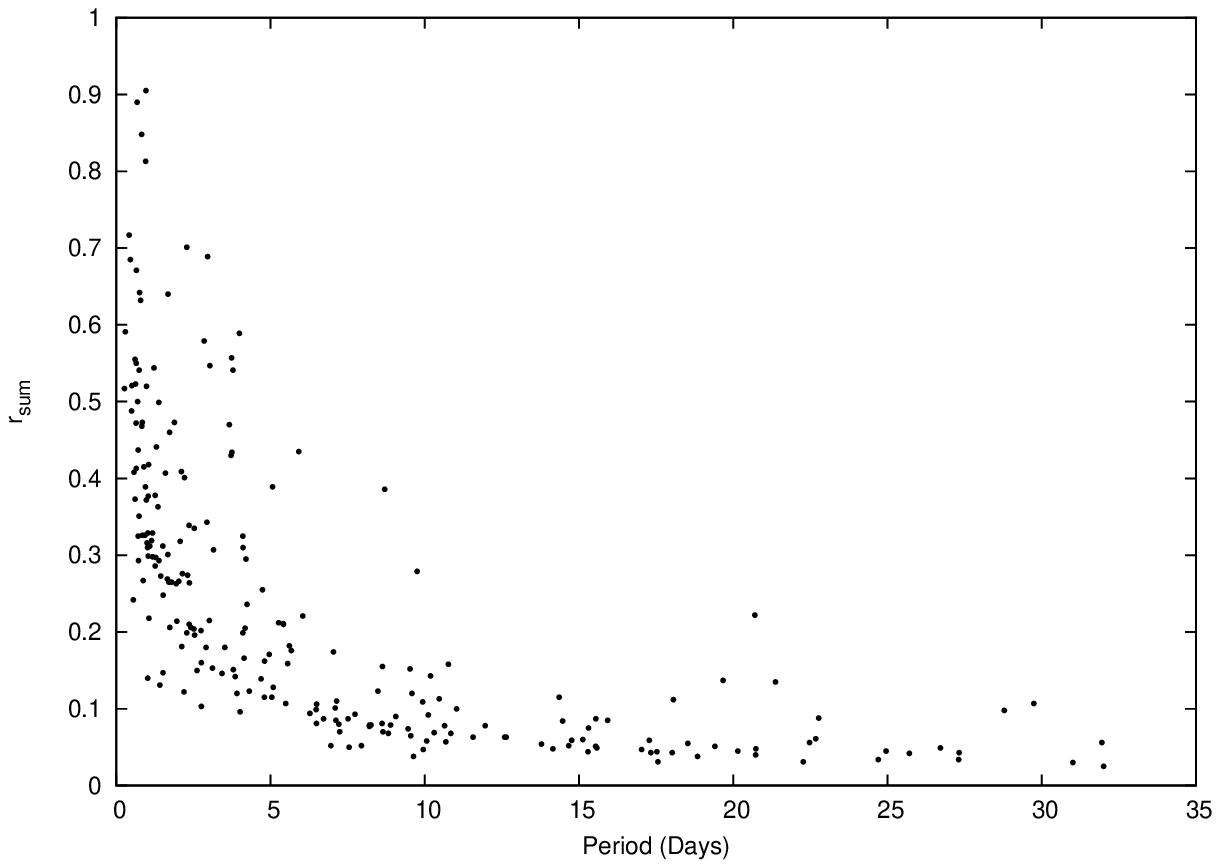} \\
\end{tabular}
\caption{Top: The effect that the 29.43 minute integration time has on the derivation on the sum of the fractional radii, r$_{sum}$, at a given period. As can be seen, only very small values of r$_{sum}$ and P yield discrepancies $\gtrsim$ 10\%, for example, combinations of P $<$ 3 days and r$_{sum}$ $<$ 0.05, P $<$ 1.5 days and r$_{sum}$ $<$ 0.1, P $<$ 0.75 days and r$_{sum}$ $<$ 0.2, etc. Bottom: The values of r$_{sum}$ versus period for the binaries we have modeled in this paper, presented in Table~\ref{ddebcandstab}. Very few of the systems, $\lesssim$ 2\%, in our sample lie in a region where they would be significantly affected by the 29.43 minute integration time.}
\label{parintfig}
\end{figure}

\citet{Kipping2010} has recently examined the effects of the long, ($\sim$30 minute), integration time of long-cadence \emph{Kepler} observations on transit light curves, and found that it can significantly alter the morphological shape of a transit curve and result in erroneous parameters if not properly taken into account in the modeling procedure. Certainly, eclipsing binaries are also affected by long integration times, namely by a ``smearing'' of the eclipses so that they appear to be shallower and have a longer duration. Qualitatively, this would result in a lower inclination and larger sum of the fractional radii, while the luminosity ratio would remain unchanged, since both eclipse depths are equally affected. To quantitatively investigate the extent to which the long integration could affect the derived parameters, we generated model light curves of a typical eclipsing binary, varying its period and the sum of the fractional radii. We then binned these light curves as if they had a 29.43 minute integration time, and the same number of data points as the Q1 \emph{Kepler} light curves. We then re-solved the light curves without accounting for the integration time, and compared the computed parameters to those used to generate the original light curve. We found that for the long-cadence \emph{Kepler} integration time of 29.43 minutes, only systems with very low values of $r_{sum}$ and P are significantly affected, as can be seen in Figure~\ref{parintfig}. These types of systems are less than 2\% of our sample. Nevertheless, we modified the JKTEBOP program to perform a numerical integration over a given exposure time, as suggested by \citet{Kipping2010}. We tested our modifications by solving the aforementioned generated light curves, now taking the integration time into account, and successfully retrieved the inputted parameters.

\begin{figure}
\centering
\epsfig{width=\linewidth,file=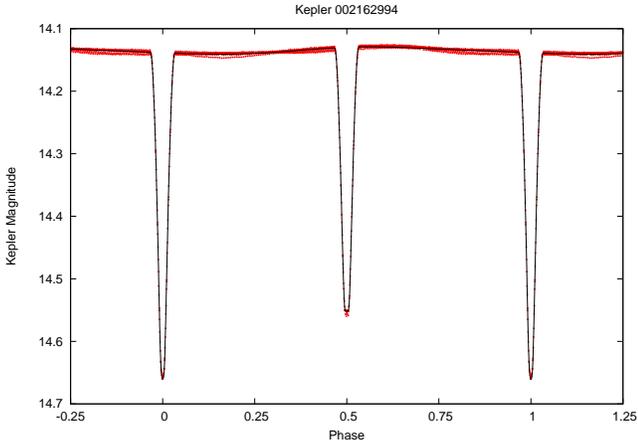}
\caption{Plots of the light curves of the 231 systems modeled with the JKTEBOP code, presented in Table~\ref{ddebcandstab}. Only the first plot, Figure~\ref{lightcurveplots}.1, is shown in the text for guidance. Figures~\ref{lightcurveplots}.1-\ref{lightcurveplots}.231 are available in the online version of the Journal.}
\label{lightcurveplots}
\end{figure}

After estimating the individual mass, radius, and temperature for each component, we re-computed the gravity and limb-darkening coefficients for each individual star, and performed a Levenberg-Marquardt minimization starting from our previously best solutions, taking into account the 29.43 minute integration time. We then repeated the processes of deriving the physical values of the components, interpolating gravity and limb-darkening coefficients, and performing a Levenberg-Marquardt minimization several more times to ensure convergence. The JKTEBOP solutions for all initial 231 candidates are shown in Table~\ref{ddebcandstab}, including the \emph{Kepler} ID number, effective temperature of the system, apparent \emph{Kepler} magnitude, magnitude range of the light curve, period, time of primary minimum, inclination, eccentricity, longitude of periastron, sum of the fractional radii, surface brightness ratio, radii ratio, luminosity ratio, amplitude of the sine curve applied to the luminosity of the primary star to account for spots, and the amount of third light. Although we list the derived surface brightness and radii ratios here, we note again that they are not always reliable on their own, and thus are combined to obtain the luminosity ratio in our analysis via Eq.~\ref{lumeq}. Plots of each of the eclipsing binaries with their model fit are given in Figure~\ref{lightcurveplots}.

As a check on the reliability of our analysis technique we took the well-studied low-mass eclipsing binary GU Boo \citep{LopezMorales2005}, and modeled only the R band light curve, (not using the radial velocity curves), via the exact same procedure as stated above in Sections~\ref{modelsec} and \ref{newlmbsec}. The only differences were that we used only the R-band luminosities from the \citet{Baraffe1998} and \citet{Chabrier2000} models, and an integration time of 2 minutes as stated in \citet{LopezMorales2005}. We used only the period, time of primary minimum, and estimated effective temperature of the system from broadband photometry provided in \citet{LopezMorales2005}, as we did for the systems in our main study. We find T$_{1}$ = 3912 K, M$_{1}$ = 0.61 M$_{\sun}$, R$_{1}$ = 0.62 R$_{\sun}$, T$_{2}$ = 3813 K, M$_{2}$ = 0.57 M$_{\sun}$, and R$_{2}$ = 0.59 via our technique. In comparison, \citet{LopezMorales2005} found with multi-color light curves and radial-velocity curves of the system, values of T$_{1}$ = 3920 K, M$_{1}$ = 0.610 M$_{\sun}$, R$_{1}$ = 0.623 R$_{\sun}$, T$_{2}$ = 3810 K, M$_{2}$ = 0.599 M$_{\sun}$, and R$_{2}$ = 0.620. The values derived from our technique using only a single color light curve are accurate to within a few percent of the very precise values derived from a study using multi-color light and radial-velocity curves, thus validating our technique.

As noted above, \citet{Prsa2010} estimated the parameters of temperature ratio, sum of the fractional radii, e$\cdot$cos($\omega$), e$\cdot$sin($\omega$), and sin(i) for detached systems, via the EBAI technique \citep{Prsa2008}. Before comparing to the parameters obtained by \citep{Prsa2010}, we note that the modeling approach between EBAI and our AGA presented in this paper have some fundamental differences. EBAI is extremely computationally efficient, but relies on a fitted polynomial to the actual data \citep{Prsa2008}, which is then compared to a neural network training set of 33,235 light curves generated by the Wilson-Devinney code \citep{Wilson1971,Wilson1993}. \citet{Prsa2008} notes that ``...the artificial neural network output is viable for statistical analysis and as input to sophisticated modeling engines for fine-tuning.'' In comparison, the use of our AGA coupled with JKTEBOP is computationally slower, but models each actual data point, obtaining an actual best-fit model while varying all physical parameters of interest over the global solution space. As well, our AGA takes into account the 29.43 minute integration time, while EBAI does not. Thus, although the EBAI technique is excellent for mining large databases, identification of light curve morphology, and obtaining estimates of parameters for statistical studies, it is not intended to model individual light curves as precisely and accurately as possible. Keeping this in mind, comparing the parameters obtained by \citet{Prsa2010} to our solutions for the same systems, we first note a moderate correlation between the sum of radii given by \citet{Prsa2010} and our results, with an average discrepancy of $\sim$20\%. However some of the \citet{Prsa2010} solutions are unphysical, (r$_{sum}$ $<$ 0.0), and visual inspection of the polyfit curves given by \citet{Prsa2010} appears to reveal a systematic underestimation of the eclipse depths. With respect to eccentricity, the parameters presented by \citet{Prsa2010} reveal an unusually large number of eccentric systems, with only 3\% of systems having e $\le$ 0.01, and 11\% of systems with e $\le$ 0.05. In contrast, our parameters show 36\% of systems with e $\le$ 0.01, and 60\% of systems with e $\le$ 0.05, which better matches the large number of systems observed that do not show any offset of secondary eclipse from phase 0.5, and no difference in the eclipse widths, indicative of a circular orbit. There is only a slight correlation between our inclination values and that of \citet{Prsa2010}, but as we previously noted, the \citet{Prsa2010} polyfit curves appear not to fit the eclipse depths well. There is practically no correlation between our values for the surface brightness ratio and EBAI's temperature ratio provided in \citet{Prsa2010}, though \citet{Prsa2010} notes that for detached systems, the ``...eclipse depth ratio is strongly affected by eccentricity and star sizes as well, rendering T$_{2}$/T$_{1}$ a poor proxy for the surface brightness ratio.''

In Table~\ref{lmbmrtab} we list the \emph{Kepler} ID number, orbital period, effective temperature of the system, and the estimated effective temperature, mass and radius of each stellar component for the 95 systems that contain two main-sequence stars, which we define as having a radius less than 1.5 times the \citet{Baraffe1998} and \citet{Chabrier2000} model relationships, and a light curve amplitude of at least 0.1 magnitudes, (suitable for ground-based follow-up and less likely to contain any third light). All of these 95 systems have both stars with masses less than 1.0 M$_{\sun}$. Note that we have ordered Table~\ref{lmbmrtab} such that Star 1 is always the more massive star, regardless if L$_{r}$ was greater or less than 1.0 in Table~\ref{ddebcandstab}. Also note that since we are using V+R-band luminosities, which best correspond to the \emph{Kepler} bandpass, one cannot always use the simple R$^{2}\cdot$T$^{4}$ relation to derive luminosity ratios from Table~\ref{lmbmrtab} to compare to Table~\ref{ddebcandstab}, since that would correspond to the bolometric luminosity. However, if one takes a system from Table~\ref{lmbmrtab}, looks up the V+R-band luminosity for each component, based on their mass and temperature, from the \citet{Baraffe1998} and \citet{Chabrier2000} models, and derives a luminosity ratio, this will exactly match the luminosity ratio in Table~\ref{ddebcandstab} from the JKTEBOP models, because the technique defines it as such. These results substantially increase the number of LMMS DDEB candidates in general, and provide 29 new LMMS DDEBs with both components below one solar mass, and at least 0.1 magnitude eclipse depths, in the heretofore unexplored period range of P $>$ 10 days. We further discuss the impact of these systems and comparison to theoretical models in Section~\ref{lmbmodelcompsec}.

\begin{figure}
\centering
\begin{tabular}{cc}
\epsfig{width=\linewidth,file=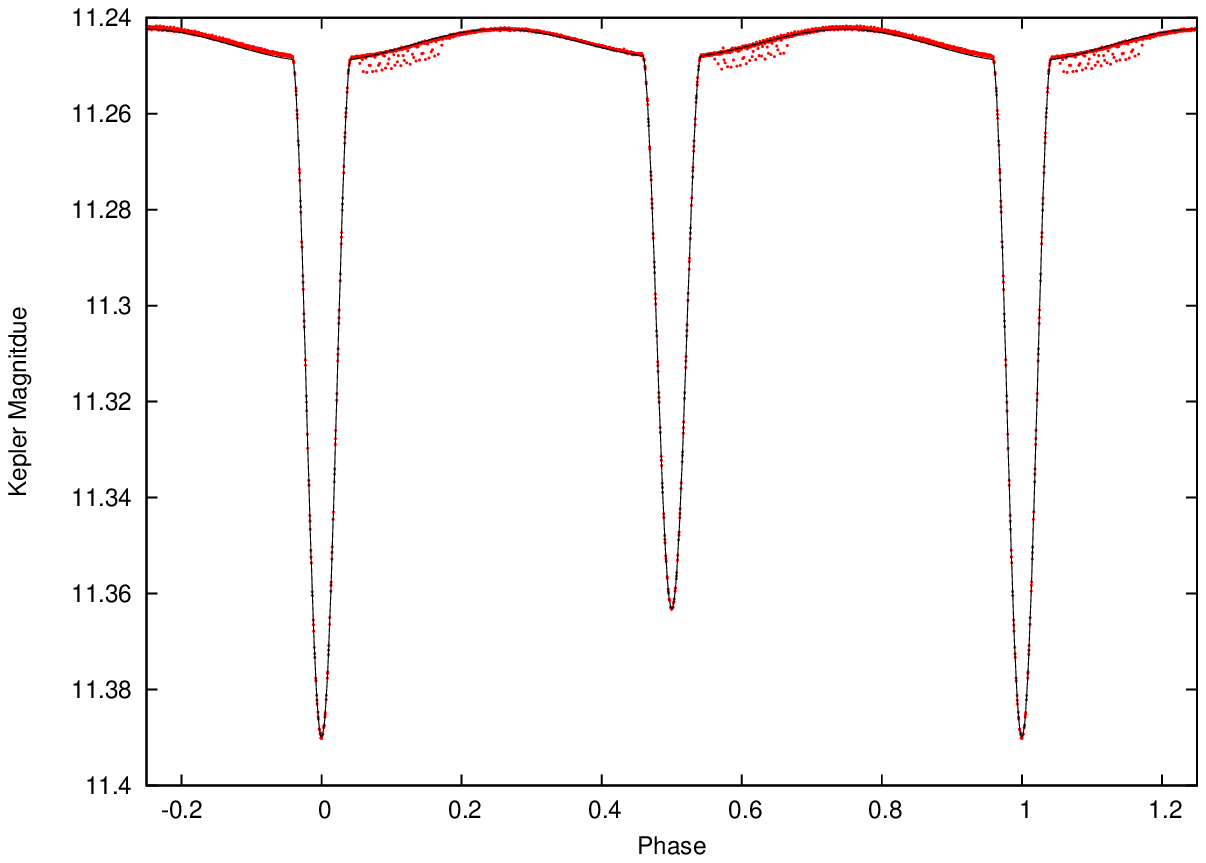} \\
\epsfig{width=\linewidth,file=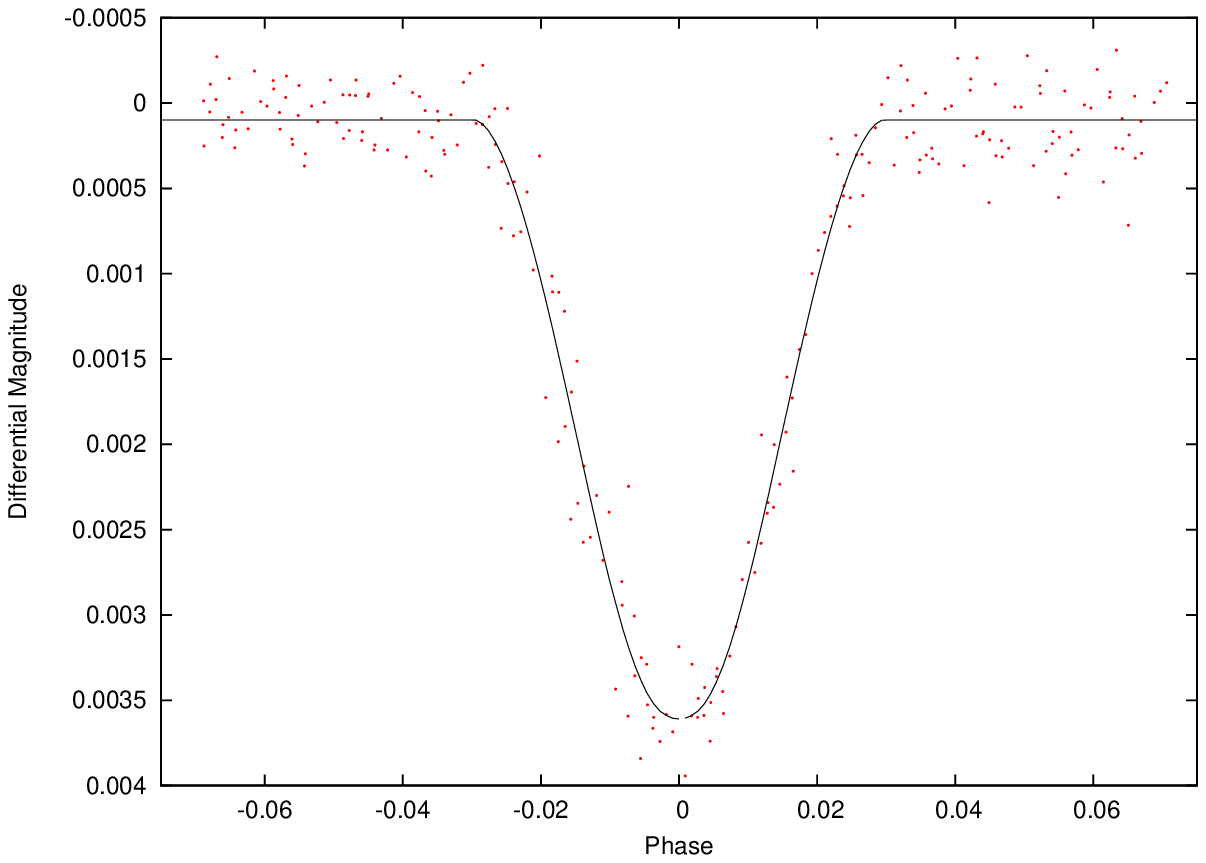} \\
\end{tabular}
\caption{Kepler 004247791. An example of a system which was determined not to be main-sequence in Section~\ref{newlmbsec}. Top: The light curve phased at its period of 4.100866 days with our best model fit. Given the shallow, wide eclipses for a $\sim$4.1 day period and T$_{\rm eff}$ = 4063K, if this system were main-sequence, it would have a combined mass of 1.28 M$_{\sun}$ and a combined radius of 3.82 R$_{\sun}$. Thus, this system contains one or more evolved stars. Bottom: The model-fit subtracted light curved phased at a period of 2.02484 days, showing a transit-like feature imposed on the light curve of the eclipsing binary. Possible explanations may include, but are certainly not limited to a background eclipsing binary with no visible secondary eclipse at 0.49376 times the orbital period of the foreground binary, a background eclipsing binary with nearly identical primary and secondary eclipses at 0.98752 times the orbital period of the foreground binary, a circumbinary transiting object, or a transiting object around one of the stars in an almost 2:1 resonant orbit with the binary.}
\label{Kepler00424fig}
\end{figure}

In Figure~\ref{Kepler00424fig} we show an example of a system which did not meet the main-sequence criterion, Kepler 004247791, which has T$_{\rm eff}$ = 4063K and a period of 4.100866 days. If this system were main-sequence, via our method, it would have a combined mass of 1.28 M$_{\sun}$ and a combined radius of 3.82 R$_{\sun}$. This can be seen by the wide, shallow eclipses for a system of this period and effective temperature. Thus, this system contains one or two evolved stars. An additional curiosity of this system is a periodic transit-like feature that is superimposed on the eclipsing binary light curve. The transit feature occurs at just slightly less than half the orbital period of the eclipsing binary, so that it is seen twice per every revolution of the eclipsing binary system, occurring at a slightly earlier phase every revolution. We subtract the model fit from the eclipsing binary, and plot the transit feature at its period of 2.02484 days in the right panel of Figure~\ref{Kepler00424fig}. Some possible explanations may include, but are certainly not limited to: 1) a background eclipsing binary with no visible secondary eclipse at 0.49376 times the orbital period of the foreground binary, 2) a background eclipsing binary with nearly identical primary and secondary eclipses at 0.98752 times the orbital period of the foreground binary, 3) a circumbinary transiting object, or 4) a transiting object around one of the stars in an almost 2:1 resonant orbit with the binary. Follow-up multi-color light curves, spectra, and radial velocities will be needed to fully characterize this interesting system.

\section{New Transiting Planet Candidates}
\label{transsec}

\begin{figure}
\centering
\epsfig{width=\linewidth,file=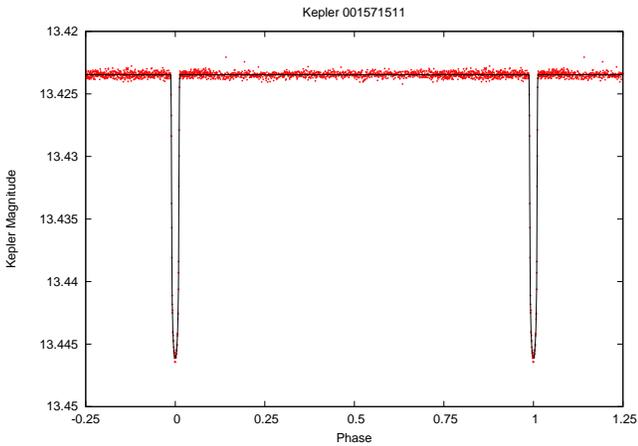}
\caption{Plots of the light curves of the 8 transiting planet candidates modeled with the JKTEBOP code, presented in Table~\ref{transmrtab}. Only the first plot, Figure~\ref{transcandslcplots}.1, is shown in the text for guidance. Figures~\ref{transcandslcplots}.1-\ref{transcandslcplots}.8 are available in the online version of the Journal.}
\label{transcandslcplots}
\end{figure}

For the 8 new transiting planet candidates mentioned in Section~\ref{binaryidentsec}, we combined Q0 and Q1 data, and modeled the transit curves using JKTEBOP, accounting for the 29.43 minute integration time, and using our modified AGA in the same manner described in Section~\ref{modelsec}. We assumed zero eccentricity and negligible flux from each planet, and interpolated the limb-darkening and gravity-darkening coefficients via the effective temperature, surface gravity, and metallicity from the relations of \citet{Sing2010} and \citet{Claret2000}. We then solved for the period, time of primary minimum, inclination, sum of the fractional radii, ratio of the radii, and the out of transit flux level. With this narrowed set of parameters, the AGA proved to be extremely quick and precise, and all fits were confirmed by eye and $\chi^{2}$ values to accurately fit the data. Plots of the transit light curves with model fits are shown in Figure~\ref{transcandslcplots}.

To estimate the physical radius of each transiting exoplanet candidate, we took the value for the radius of the host star from the KIC, and multiplied by the ratio of the radii, $k$, found from the model. In Table~\ref{transmrtab} we list the \emph{Kepler} ID number, apparent \emph{Kepler} magnitude, time of primary minimum, period, effective temperature of the star, inclination, radius of the star, and radius of the exoplanet candidate in both solar radii and Jupiter radii.

\begin{deluxetable*}{lccccccccc}
\tablenum{4}
\tablewidth{0pt}
\tabletypesize{\scriptsize}
\tablecaption{Model Parameters for the 8 Transiting Exoplanet Candidates}
\tablecolumns{10}
\tablehead{\emph{Kepler} ID & M$_{\rm kep}$ & T$_{0}$ & P & T$_{\rm eff,\star}$ & i & R$_{\star}$ & R$_{\rm p}$ & R$_{\rm p}$\\ & & (BJD-2454900) & (Days) & (K) & ($\degr$)& (R$_{\sun}$) & (R$_{\sun}$) & (R$_{\rm Jup}$) }
\startdata
\input{tab4.tex}
\enddata
\tablenotetext{1}{Listed in the \emph{Kepler} False Positive Catalog as ``velocity measurements indicate eclipsing binary''}
\label{transmrtab}
\end{deluxetable*}

As can be seen, the radii for these transiting planet candidates range from 0.56 to 2.1 R$_{\rm Jup}$, with periods between 4.1 and 24.6 days. Only one of these, Kepler 011974540, has been ruled out as a planet from follow-up RV measurements, which are needed for the rest of the candidates to confirm or refute their planetary nature. However, even if these objects turn out not to be planetary mass, they then must be either brown dwarfs or very low-mass stars, which still are valuable finds. In the case of brown dwarfs, these targets would be located within the so-called ``brown dwarf desert'' \citep{McCarthy2004}.

\section{Comparison of the New Low-Mass Binary Candidates to Models}
\label{lmbmodelcompsec}

As described in the introduction, one of the current outstanding questions in the study of low-mass stars is whether the inflated radii observed in binaries is caused by their enhanced stellar rotation, and therefore enhanced magnetic activity. We explore this problem in this section using the list of the 95 new LMMS DDEB candidates with estimated individual masses both below 1.0 M$_{\sun}$ and light curve amplitudes greater than 0.1 magnitudes, given in Table~\ref{lmbmrtab}. This sample, for the first time, provides a statistically significant number of systems with orbital periods larger than 10 days. 

\begin{figure*}
\centering
\begin{tabular}{cc}
\epsfig{width=0.4875\linewidth,file=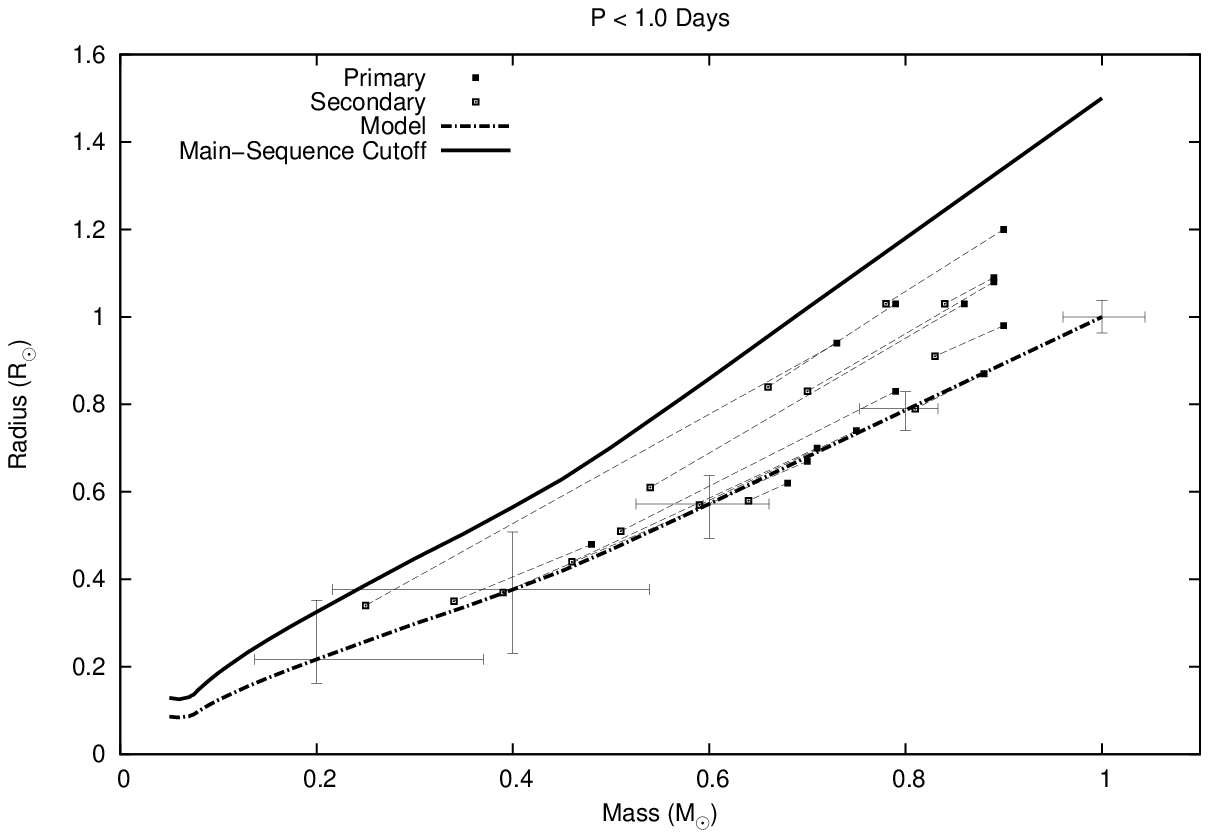} &
\epsfig{width=0.4875\linewidth,file=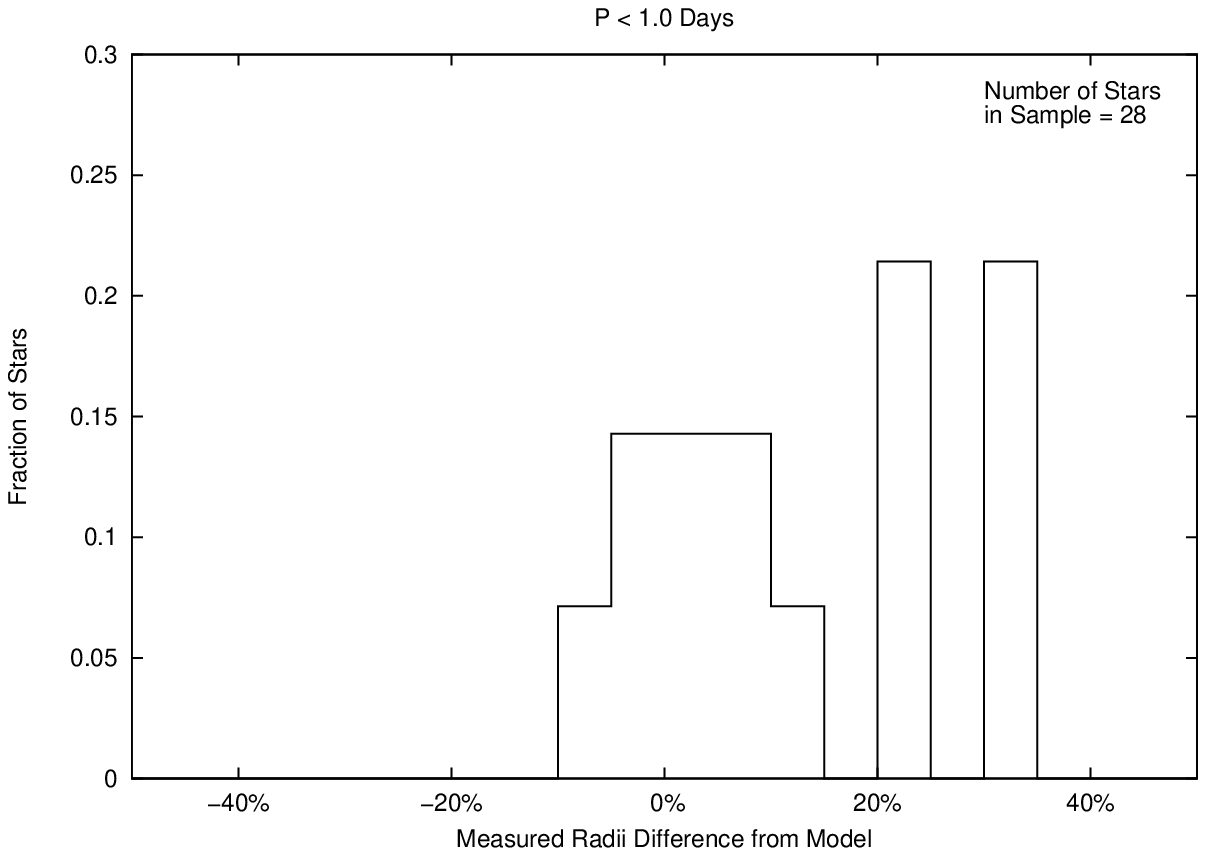} \\
\epsfig{width=0.4875\linewidth,file=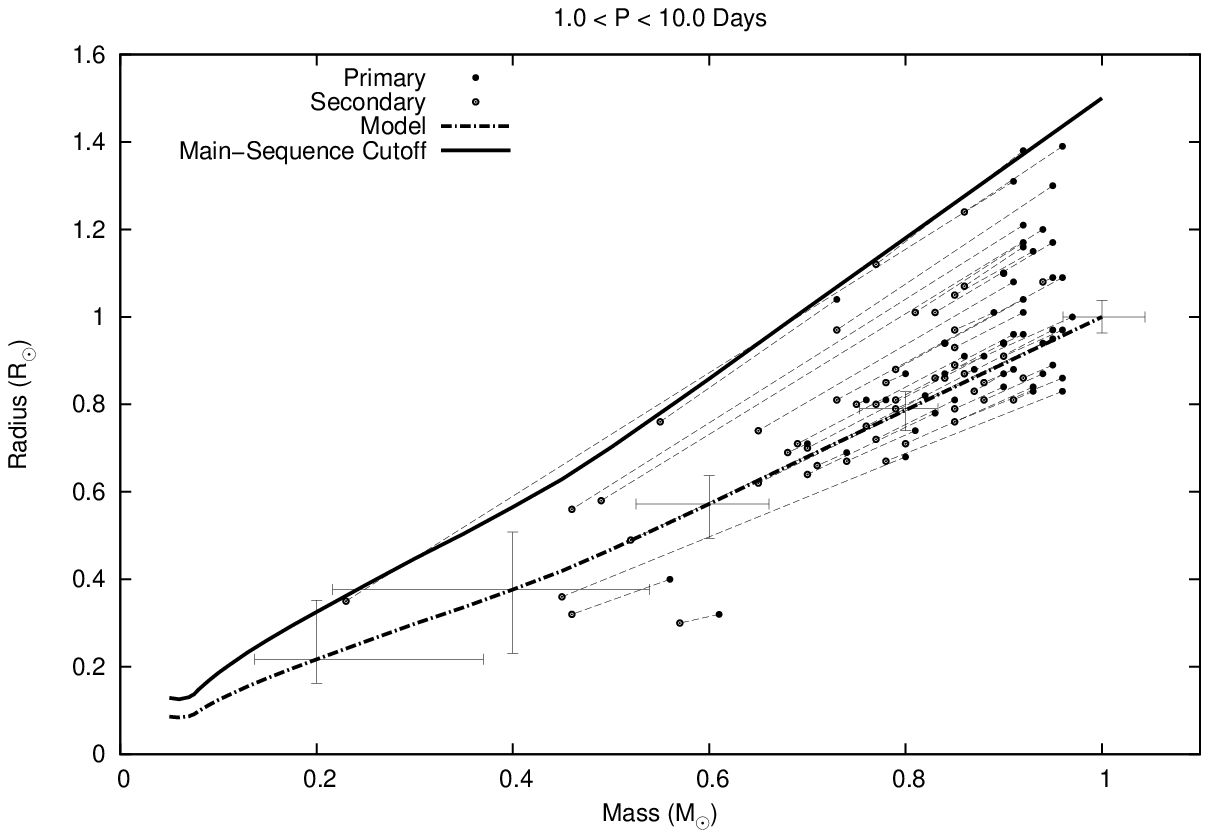} &
\epsfig{width=0.4875\linewidth,file=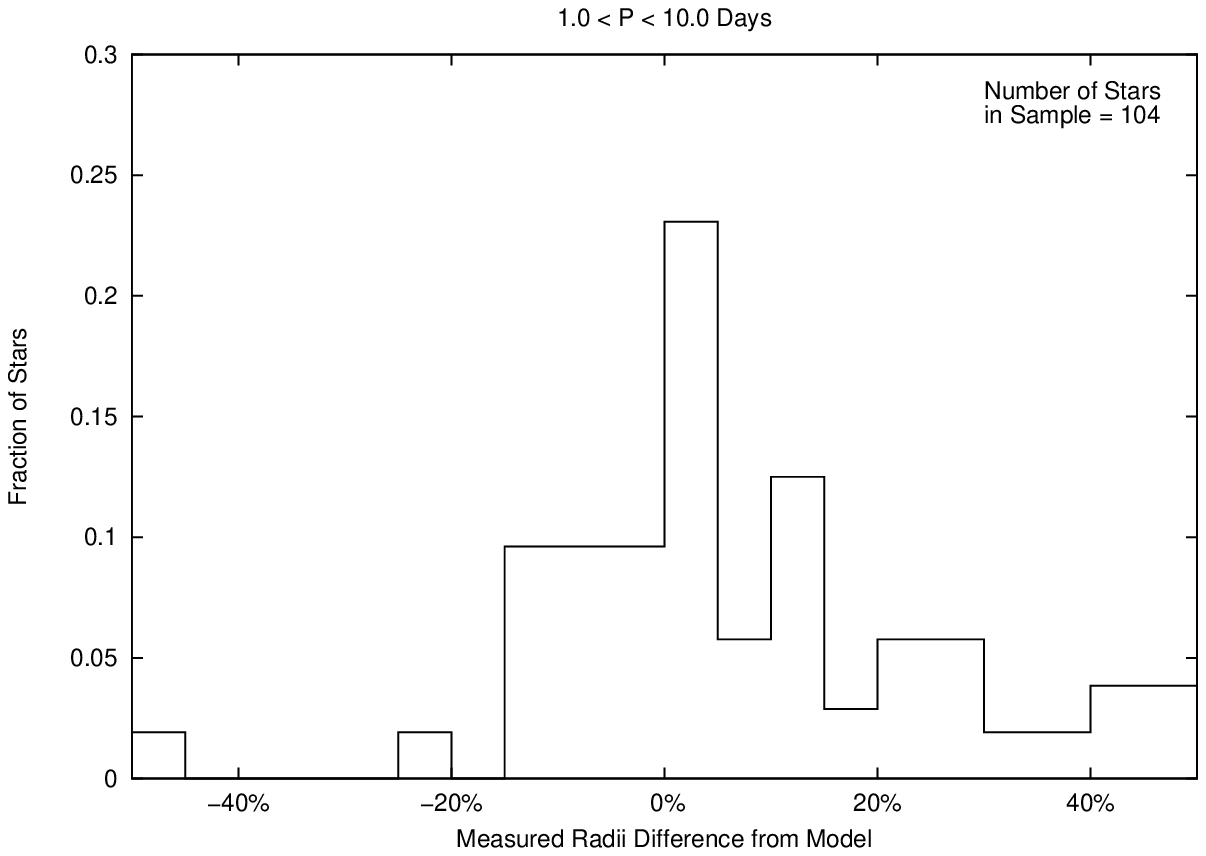} \\
\epsfig{width=0.4875\linewidth,file=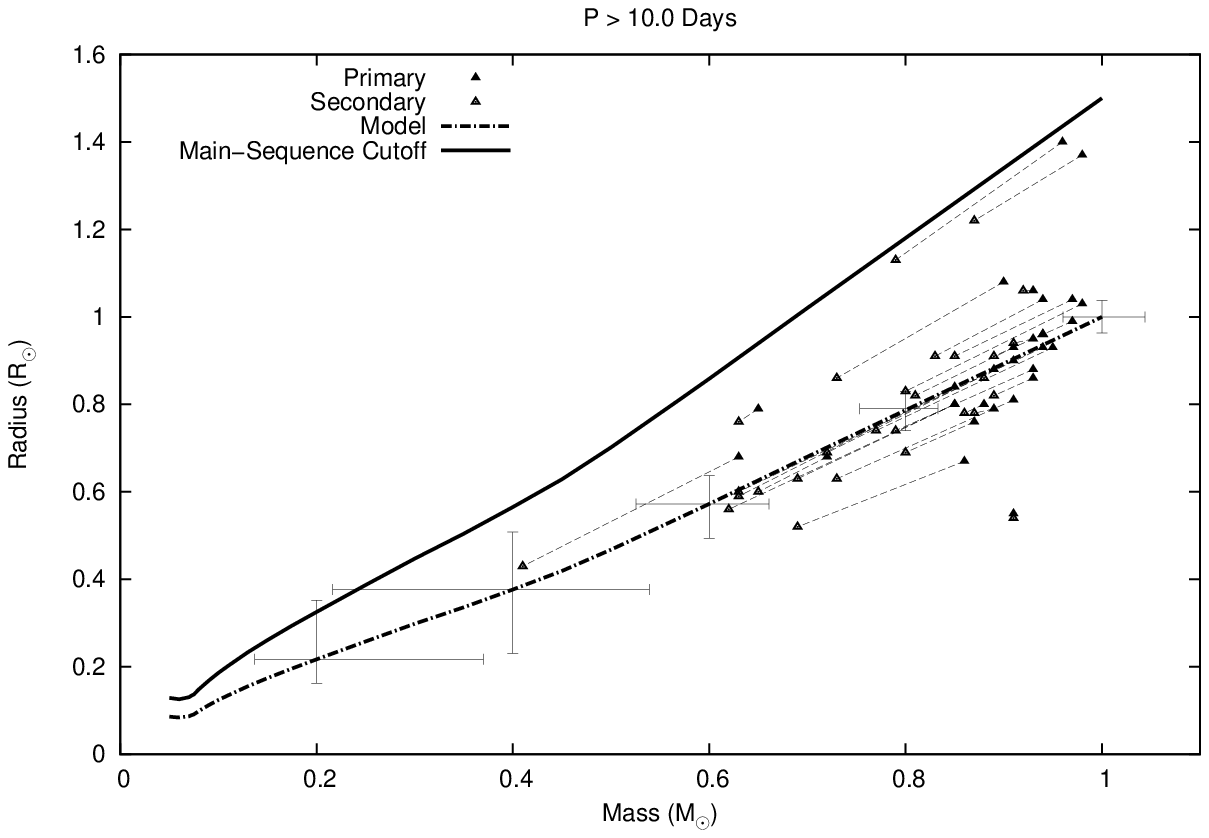} &
\epsfig{width=0.4875\linewidth,file=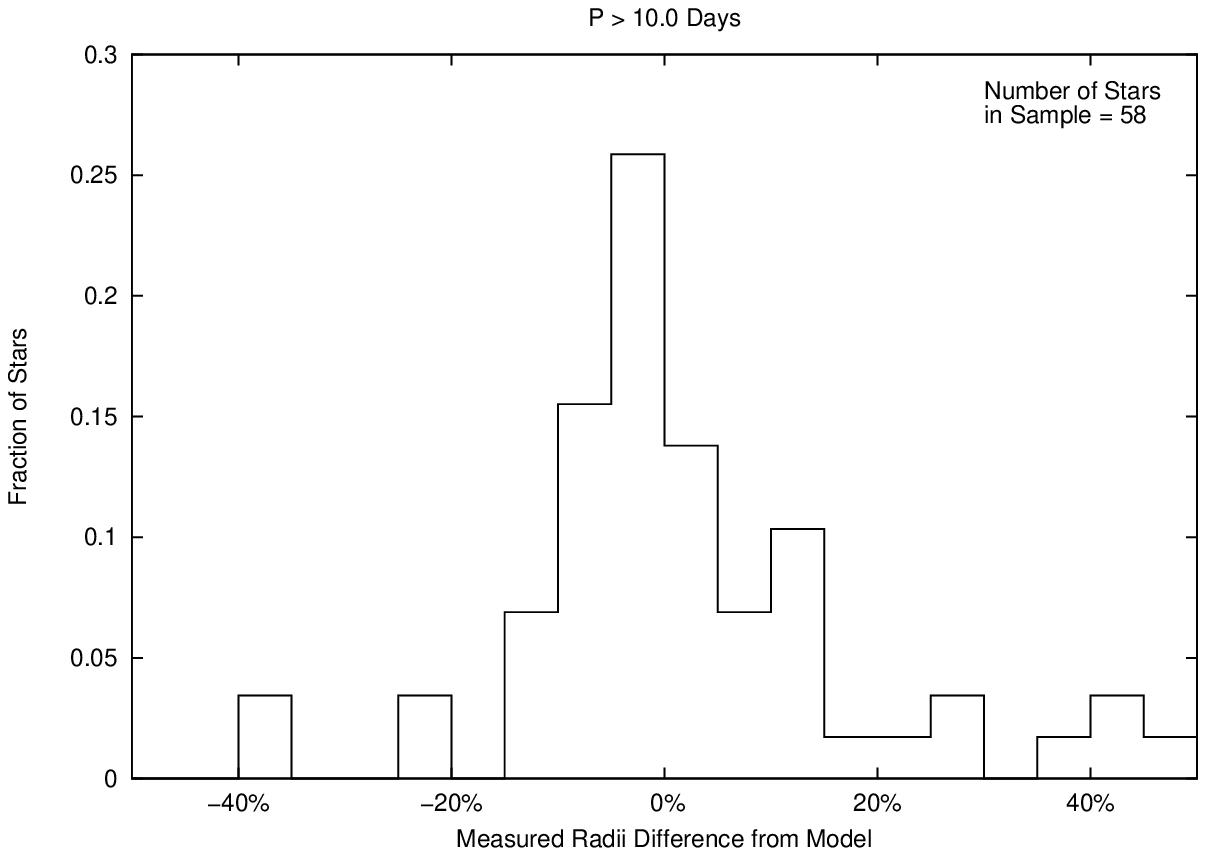} \\
\end{tabular}
\newpage
\caption{Left: Mass-radius diagrams for each binary with both components $<$ 1.0 M$_{\sun}$ and photometric amplitudes greater than 0.1 mag, as given in Table~\ref{lmbmrtab}, with systems connected by faint lines. The systems are sorted into short-period (P $<$ 1.0 days, top panel), medium-period, (1.0 $<$ P $<$ 10.0 days, middle panel), and long-period groupings, (P $>$ 10.0 days, bottom panel). The theoretical mass-radius relations of \citet{Baraffe1998} for 0.075 M$_{\sun}$ $\le$ M $\le$ 1.0 M$_{\sun}$, and of \citet{Chabrier2000} for M $<$ 0.075 M$_{\sun}$, both for [M/H] = 0.0 and an age of 5.0 Gyr, are over-plotted. The solid line shows the main-sequence cutoff criterion. The error bars indicate the error in mass and radius obtained when interpolating from the mass-temperature-radius relations with an error of 200K. Right: Histograms of the fraction of stars in the sample versus their deviance from the models for each period grouping. As can be seen by both the mass-radius relation plots and the histograms, shorter period binaries in general appear to exhibit larger radii compared to the models than longer period systems.}
\label{mrfig}
\end{figure*}

The left-side panels of Figure~\ref{mrfig} show mass-radius diagrams using the mass and radius of each binary star component estimated in Section~\ref{newlmbsec}. The LMMS DDEB candidates have been separated into three categories, with orbital periods P $<$ 1.0 day, 1.0 $<$ P $<$ 10 days, and P $>$ 10 days. Each primary and secondary in a binary pair is traced by a connecting line. We also plot in each panel of Figure~\ref{mrfig} the theoretical mass-radius relation predicted by the \citet{Baraffe1998} models for M $\ge$ 0.075 M$_{\sun}$, and the \citet{Chabrier2000} models for M $<$ 0.075 M$_{\sun}$, both for [M/H] = 0.0, and an age of 5.0 Gyrs. We have also defined a main-sequence cutoff as 1.5 times the theoretical mass-radius relation, which is illustrated by the solid line in each diagram. In the models we have used an $\alpha$ = 1.0 for M $\le$ 0.7 M$_{\sun}$ and interpolated the radius of the models for 0.7 $M_{\sun}$ $<$ M $\le$ 1.0 $M_{\sun}$ by fixing the radius of the 1.0 $M_{\sun}$ model to 1.0 $R_{\sun}$, therefore avoiding the dependence of the stellar radius with $\alpha$ between 0.7 $M_{\sun}$ and 1.0 $M_{\sun}$ \citep{Baraffe1998}. We also include in the mass-radius diagrams estimations of the error in our M and R values at several masses, computed by adding and subtracting 200 K, (the error in the T$_{\rm eff}$ determinations given by the KIC), from a given temperature and interpolating the mass and radius from the theoretical relations. Note that one of the long-period stars, Kepler 008075618, falls well below the main-sequence, with two identical components with M = 0.91 M$_{\sun}$ and R = 0.53 R$_{\sun}$. Inspection of this light curve, coupled with the light curve model, reveals that this system could in fact be a single-lined system at half the listed period.

In the figure, many of the stellar radii of binaries with P $<$ 1.0 days appear to fall above the model predictions, but as the orbital period increases, a larger fraction of the systems appear to have radii that are either consistent with or fall below the models. There certainly is a fair amount of scatter in these data introduced by the large error in the mass and radius estimations, but a histogram analysis of the radius distributions confirms these apparent trends. On the right-side panels of Figure~\ref{mrfig} we show 5\% bin-size histograms representing how many stars have a radius that deviates by a given percentage from the models. The average radius discrepancy is 13.0\%, 7.5\%, and 2.0\% for the short (P $<$ 1.0 days), medium (1.0 $<$ P $<$ 10.0 days), and long-period (P $>$ 10.0 days) systems respectively. Although a full analysis of each system with multi-color light and radial-velocity data is still needed, these preliminary estimates support the hypothesis that binary spin-up is the primary cause of inflated radii in short period LMMS DDEBs.

\section{Summary}
\label{concsec}

We present 231 new double-eclipse, detached eclipsing binary systems with T$_{\rm eff}$ $<$ 5500 K, found in the Cycle 0 data release of the \emph{Kepler Mission}, and provide their \emph{Kepler} ID, estimated effective temperature, \emph{Kepler} magnitude, magnitude range of the light curve, orbital period, time of primary minimum, inclination, eccentricity, longitude of periastron, sum of the fractional radii, and luminosity ratio. We estimate the masses and radii of the stars in these systems, and find that 95 of them contain two main-sequence stars with both components having M $<$ 1.0 M$_{\sun}$ and eclipse depths of at least 0.1 magnitude, and thus are suitable for ground-based follow-up. Of these 95 systems, 14 have periods less than 1.0 day, 52 have periods between 1.0 and 10.0 days, and 29 have periods greater than 10.0 days. This new sample of low-mass, double-eclipse, detached eclipsing binary candidates more than doubles the number of previously known systems, and extends the sample into the completely heretofore unexplored P $>$ 10.0 day period range for LMMS DDEBs. 

Comparison to the theoretical mass-radius relation models for stars below 1.0 M$_{\sun}$ by \citet{Baraffe1998} show preliminary evidence for better agreement with the models at longer periods, where the rotation rate of the stars is not expected to be spun-up by tidal locking, although, in the absence of radial-velocity measurements, the errors on the estimated mass and radius are still quite large. For systems with P $<$ 1.0 days, the average radius discrepancy is 13.0\%, whereas for 1.0 $<$ P $<$ 10.0 days and P $>$ 10.0 days, the average radius discrepancy is 7.5\% and 2.0\%, respectively. Ground-based follow-up, in the form of radial velocity and multi-wavelength light curves, is needed to derive the mass and radius of each star in each system to $\sim$1-2\%, which we have already begun to acquire. With accurate masses and radii for multiple long-period systems, we should be able to definitively test the hypothesis that inflated radii in low-mass binaries are principally due to enhanced rotation rates.

We also present 8 new transiting planet candidates. Only one of them is currently listed in the \emph{Kepler} False Positive Catalog. The remaining candidates require radial-velocity follow-up to confirm or refute their planetary nature. Even if these systems do not turn out to be planets, they then must be brown dwarf or very low-mass, late-type M dwarfs, which would still be a very valuable find. In fact, all false positive planet candidates determined by the \emph{Kepler} team will be of great interest to stellar astrophysics. We also present 7 new extremely shallow eclipsing systems, one well detached binary with deep eclipses, and one apparent red giant with an unusual eclipse-like feature. We also highlight a very unusual eclipsing binary system containing at least one evolved star and an additional transit-like feature in the light curve. Finally, the systems that we determined are not main-sequence, and we therefore did not include in the subsequent analysis, should be further studied for valuable science. Accurate mass, radius, and temperature determinations of those systems could yield valuable insights into stellar and binary evolution.

\acknowledgments
\noindent{\bf Acknowledgements}\\
We first and foremost thank the entire \emph{Kepler} team and all those who have contributed to the \emph{Kepler Mission}, without which this paper would not be possible. We also thank our second referee for a very thorough and helpful review. J.L.C acknowledges support from a NSF Graduate Research Fellowship and the New Mexico Space Grant Consortium. J.L.C., M.L.M., \& T.E.H. are grateful for funding from \emph{Kepler's} Guest Observer Program. M.L.M. acknowledges support from NASA through Hubble Fellowship grant HF-01210.01-A/HF-51233.01 awarded by the STScI, which is operated by the AURA, Inc. for NASA, under contract NAS5-26555. We also thank Robert Edmonds at New Mexico State University for inspiring us to modify the AGA to use a Gaussian distribution to create new individuals. Some/all of the data presented in this paper were obtained from the Multimission Archive at the Space Telescope Science Institute (MAST). STScI is operated by the Association of Universities for Research in Astronomy, Inc., under NASA contract NAS5-26555. Support for MAST for non-HST data is provided by the NASA Office of Space Science via grant NNX09AF08G and by other grants and contracts. This research has made use of NASA's Astrophysics Data System.

\appendix
\section{Eclipse Phase Dispersion Minimization (EPDM)}
\label{epdmappendsec}

\begin{figure}
\centering
\begin{tabular}{cc}
\epsfig{width=0.49\linewidth,file=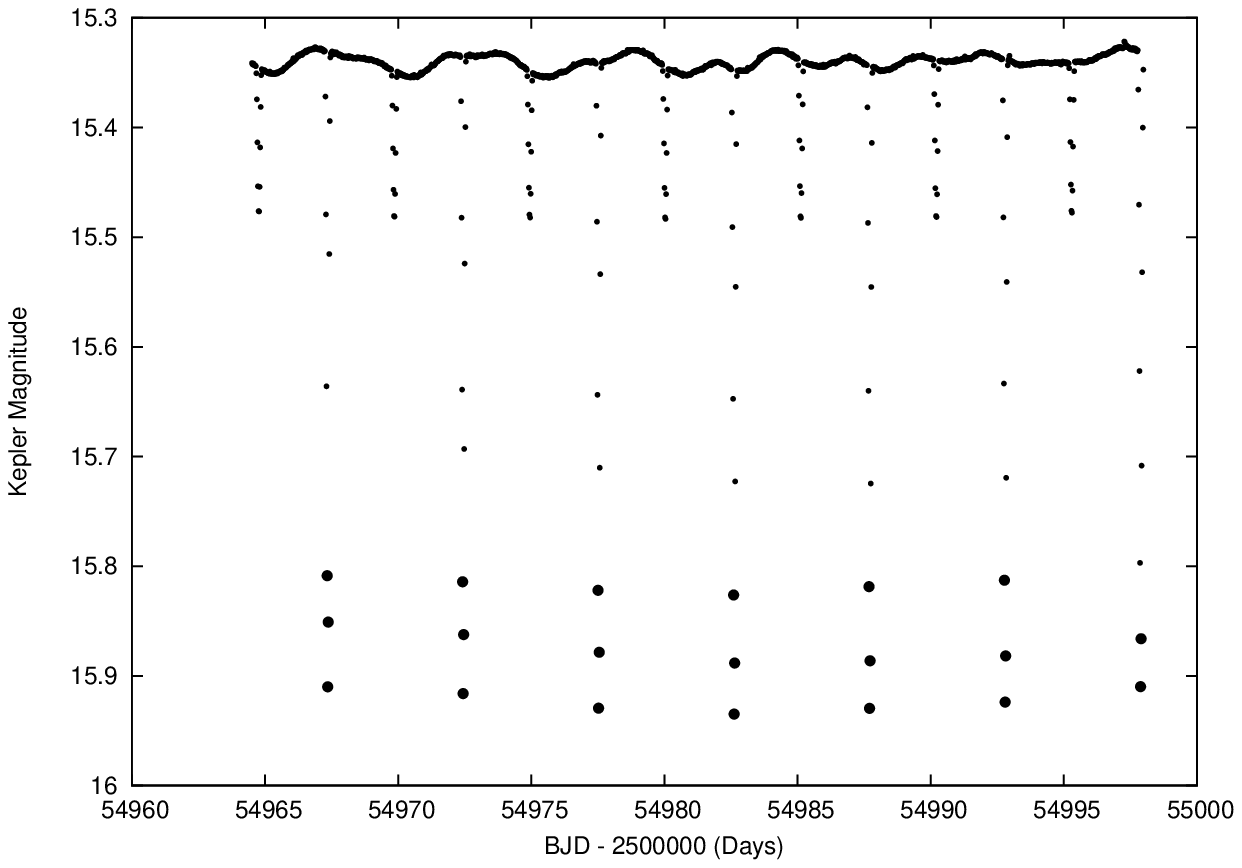} &
\epsfig{width=0.49\linewidth,file=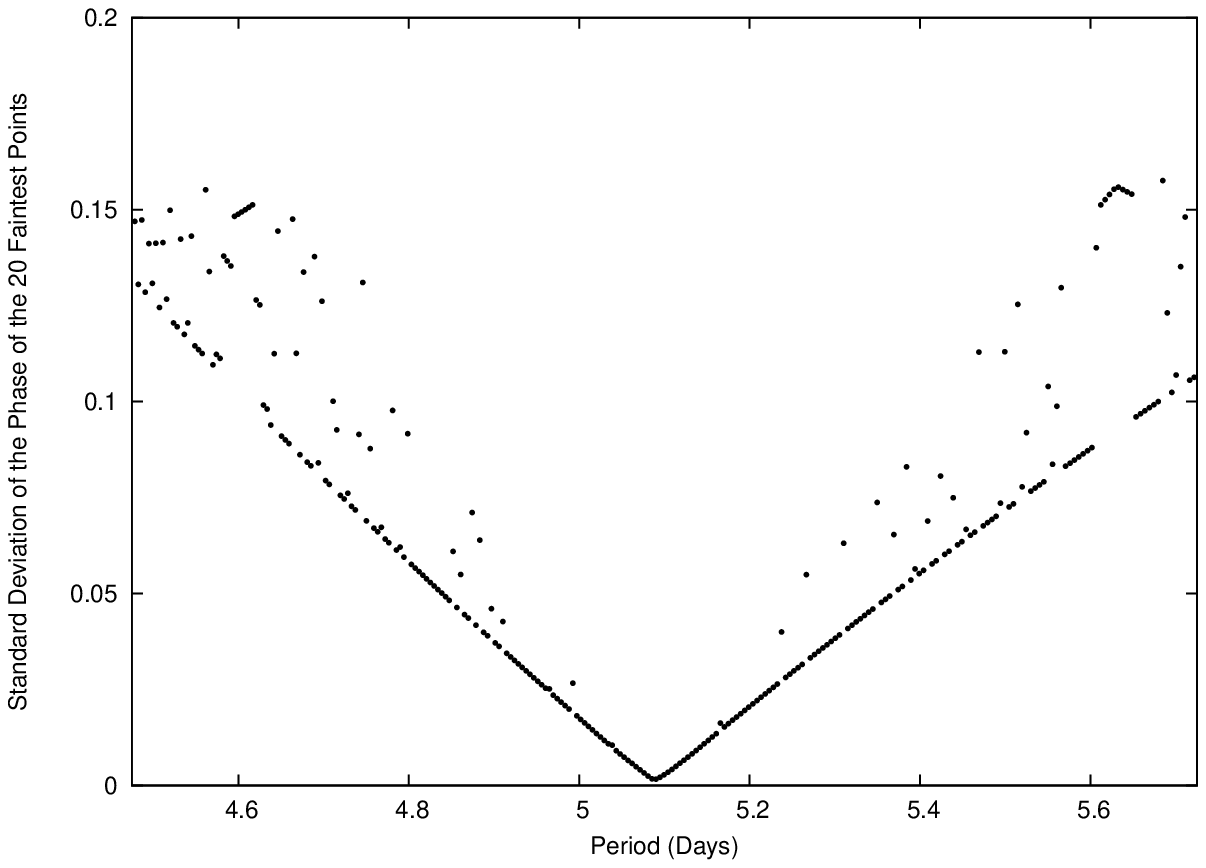} \\
\epsfig{width=0.49\linewidth,file=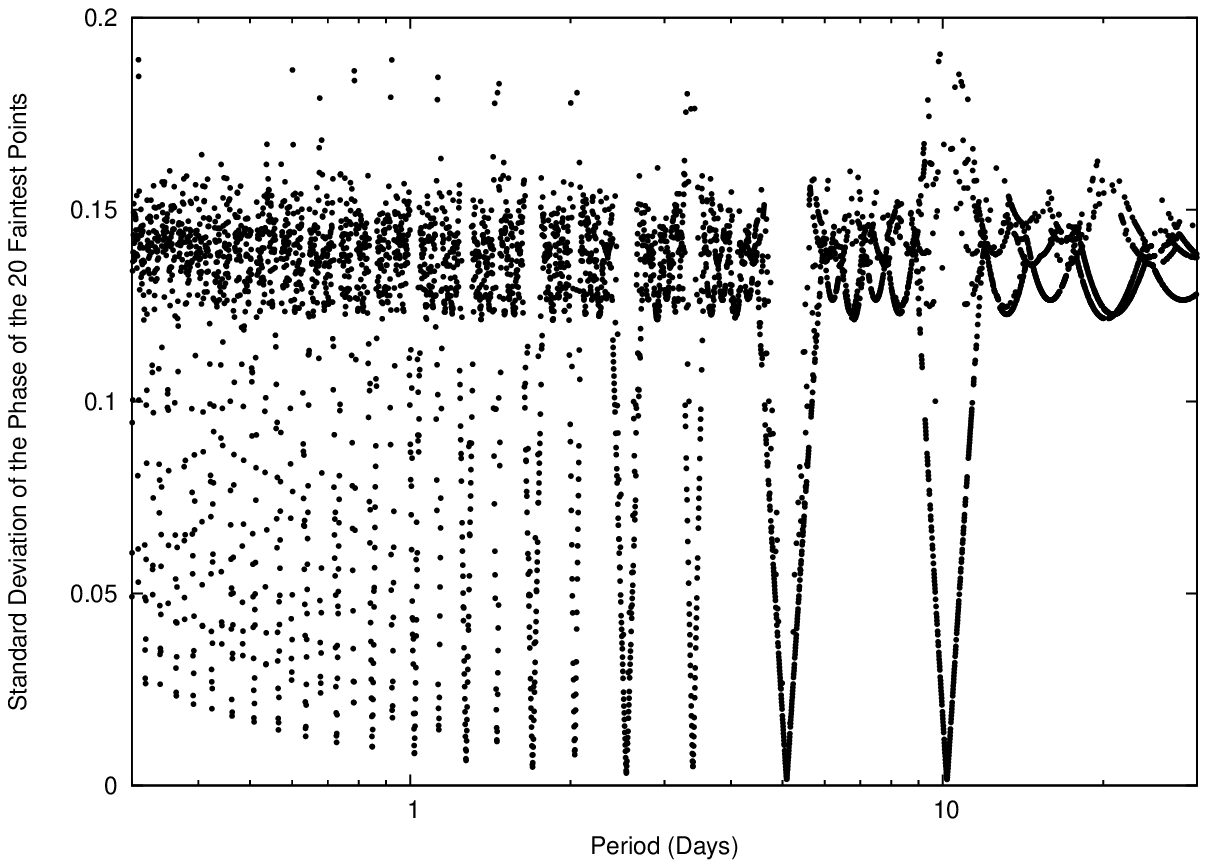} &
\epsfig{width=0.49\linewidth,file=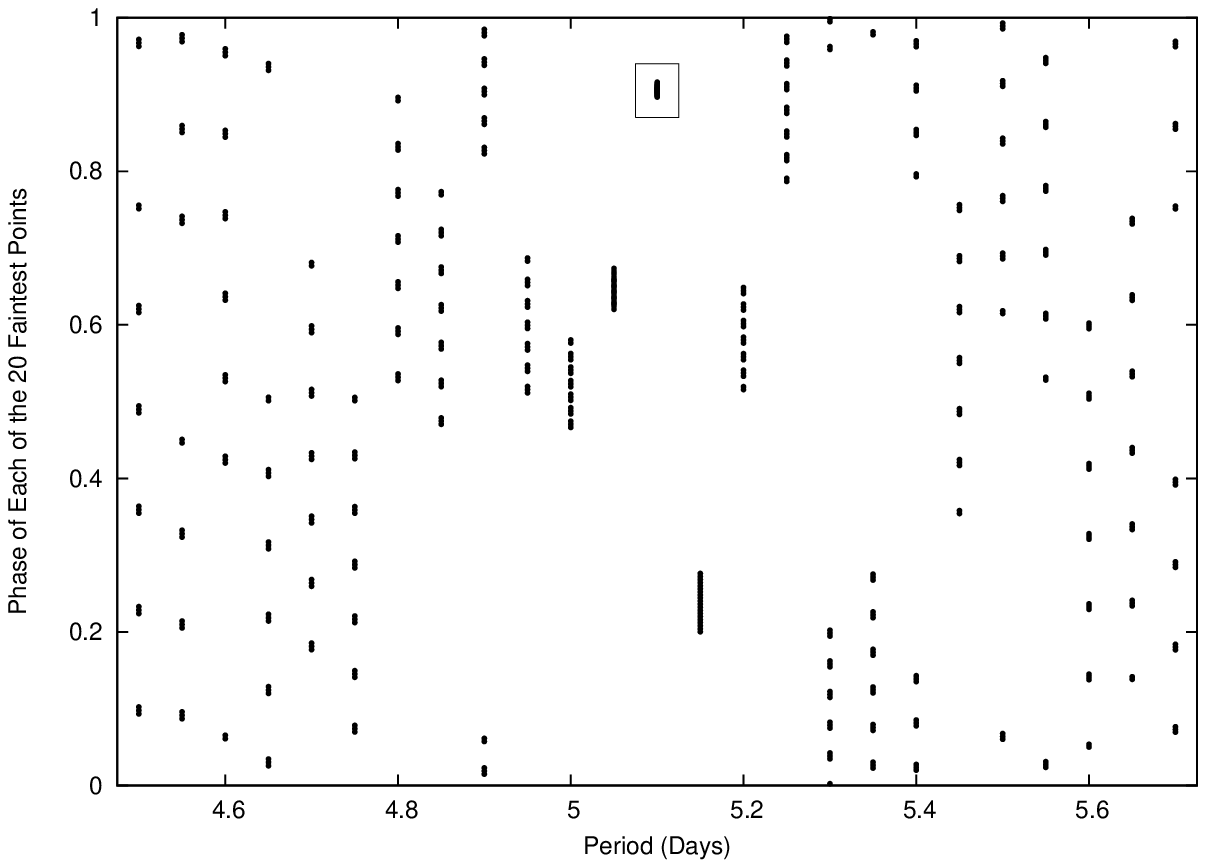} \\
\end{tabular}
\newpage
\caption{Illustration of the EPDM technique. Top-left: The unphased light curve of Kepler 006591789, with the 20 faintest points highlighted by using larger point sizes. Bottom-left: Standard deviation of the phase values of the 20 faintest points versus period for this system. As can be seen, the standard deviation approaches 0.0 at $\sim$5.1 days, and integer multiples and fractions thereof. Top-Right: The same plot as in the bottom-left panel, but with the period range restricted to show only the period with the lowest standard deviation, and true period of the system. Bottom-Right: The actual phase values for each of the 20 faintest points at multiple periods, spanning the same period range, (but with a lower period resolution, for clarity), as the plot in the top-right panel. As can be seen, as the examined period approaches the true period of the system, the phase values of the 20 faintest points strongly clump together, producing a very small standard deviation. The best period is highlighted by a box in the lower-right panel.}
\label{epdmfig}
\end{figure}

In this appendix we further explain the EPDM technique introduced in Section~\ref{binaryidentsec}. As mentioned in the text, EPDM finds the period of an eclipsing binary system by seeking the value of the period that best minimizes the dispersion in phase of the faintest N points in a light curve. To illustrate how this methods works, we show in Figure~\ref{epdmfig} the period search analysis of the LMMS DDEB candidate Kepler 006591789, which was found to have a period of 5.088435 days via the JKTEBOP model, (see Table~\ref{ddebcandstab}). The unfolded Q1 light curve is shown in the top-left panel of Figure~\ref{epdmfig}. EPDM selects the faintest 20 points of the light curve, which are highlighted by the larger points in that same panel. The number of points should be adjusted based on the quality of the data set. Too few points could result in all the points selected belonging to the same eclipse, if that one eclipse is unusually deep due to systematics or another reason, and thus EPDM will be unable to determine a period. Too many points will cause the results of EPDM to be less precise, as more points are included further away from the center of the eclipses. We have found that 20 points is a good number for \emph{Kepler} data, for which many systems do suffer from moderate systematics, as is evidently the case for Kepler 006591789, as seen by the quasi-sinusoidal variation in the baseline flux.

Having selected the faintest points from the light curve, EPDM then loops over a range of period values. In this case we choose a set of 5,000 period values that range from 0.3 to 30 days, evenly distributed in log space, so that shorter periods are as well-sampled as longer periods. At each period, the phase of each of the 20 faintest points are calculated via the following standard equation,

\begin{equation}
p = \frac{T}{P} - int(\frac{T}{P})
\end{equation}

where $p$ is the phase of a given point, with a time value, $T$, for a given period, $P$, and int() returns the argument rounded down to the nearest integer value. The standard deviation of these 20 phase values is then computed, and we are left with a standard deviation for each trial period. In the bottom-left panel of Figure~\ref{epdmfig}, we plot the standard deviation in phase of the 20 points versus each trial period. The lowest values for the standard deviation indicate the best periods, where the eclipses align in phase-space, while high values indicate bad periods. As can be seen in the bottom-left panel, the standard deviation approaches a value of 0.0 near 10.2 days, 5.1 days, 2.05 days, and decreasing fractions thereof, or period aliases. To determine the three best periods, EPDM first selects the lowest standard deviation, which in this case yields a value of 5.09004 days. It then selects the next lowest value, whose corresponding period value differs from the first by at least 10\%, and yields a value of 10.1747 days. The third period selected via the same method yields a value of 2.54402 days.

To further clarify the technique visually, in the top-right panel of Figure~\ref{epdmfig}, we show the same plot as in the bottom-left panel, but limited in period range to straddle the best period found, 5.09004 days. At the same period range, in the bottom-right panel, we plot the actual values of the phase for each of the 20 points at each period. (For ease of viewing, we use a lower trial period resolution in the bottom-right panel than the top-right panel.) As can be seen in the lower-right panel, when the trial period is far from the true period of the system, the actual phase values have a large dispersion, and range completely from 0.0 to 1.0. As the given period gets closer to the true period, the phase values begin to clump, with their dispersion decreasing as the trial period approaches the true value. Indeed, as highlighted by the box in the bottom-right panel of Figure~\ref{epdmfig}, at the best period, all the phase values are tightly grouped together at P = 5.0904 days, indicating that all the eclipses are extremely well aligned, and the period of the system has been found. 

One complication that can arise is if EPDM encounters an eccentric system with two similarly deep eclipses. In this case, when the algorithm selects the N faintest points, it will be selecting points from both eclipses. Since the system is eccentric, there is a phase offset not equal to 0.5 between primary and secondary eclipse, i.e. the two eclipses occur closer to each other in time compared to the period of the system. In this case, if we were to run EPDM as just described, in a plot like the bottom-right panel of Figure~\ref{epdmfig}, at the true period of the system there would be two groups of points, each by itself having a very small deviation, but separated from each other in phase by a large amount. Thus, the standard deviation calculation will show a much higher value than it should, and the correct period could not be found. Along similar lines, a problem arises when we consider how to calculate the standard deviation of, for example, the distribution of phase points in the bottom-right panel of Figure~\ref{epdmfig} at a period of 4.9 days, which ranges from 0.8 - 1.0, and then jumps to 0.0 - 0.05. It is clear this is a continuous group of points, which simply experiences an abrupt jump from phase 1.0 to 0.0. Although they represent a fairly good period, a calculation of their standard deviation would show a high value, and thus indicate a bad period.

To reconcile both these problems, we insert an additional step into the EPDM technique. At each trial period, EPDM searches for a reflection phase, $p_{r}$, whose value is between 0.0 and 1.0, that will allow the two distinct phase groupings to align. For each value of $p_{r}$, if the phase value of a given point is larger than $p_{r}$, a new value for the phase of the point, $p$, is calculated as
\begin{equation}
  p = p - 2.0\cdot(p-p_{r})
\end{equation}
The value of $p_{r}$ which yields the lowest standard deviation for a given trial period is the correct reflection value, and that corresponding lowest standard deviation should be assigned to that trial period. Thus, in the case of an equal depth, eccentric system, where say the N lowest points group around two phases of 0.2 and 0.4, at a value of $p_{r}$ = 0.3, the two distinct groupings would merge into a single group at phase 0.2, with a very small standard deviation at the correct period of the system. As well, in the case where a group of phase points that range from 0.9 to 1.0 and 0.0 to 0.1, $p_{r}$ allows the points to merge into a single group that only ranges from 0.0 to 0.1. In fact, we have already implemented the use of $p_{r}$ when generating the bottom-left and top-right plots of Figure~\ref{epdmfig}.

In conclusion, because EPDM only utilizes the faintest N points of a light curve, the computations are very quick, especially compared to traditional phase dispersion minimization techniques, which utilize every point in a light curve. This also allows for a more precise determination of the period, as one can apply more computing time towards finer period resolution. As well, for the same reason, EPDM is not affected by systematics or varying star spots, as long as their photometric amplitudes are not on the order of or greater than the amplitude of the eclipses. By selecting the faintest point, or the earliest of the N faintest points, one is also given a good value for the time of primary minimum. We have shown EPDM can be applied to both eccentric and non-eccentric binaries, and since a transiting planet's light curve is similar to an eclipsing binary with only one visible eclipse, the technique works equally well for transiting exoplanets. In theory, EPDM could also be applied to other variables, such as stars with rotating spots, pulsating variables, and contact binaries, although periods for these systems will be less precise than detached eclipsing binaries, due to the broader minima of those systems. In theory though, one may not have to select the faintest points of a light curve, but possibly a very narrow flux range, and achieve the same result.

\section{Genetic Algorithms for Eclipsing Binaries}
\label{agaappendix}

As mentioned in the text, in fitting our sample of eclipsing binaries, we have 12 parameters: period, time of primary minimum, inclination, mass ratio, e$\cdot$cos($\omega$), e$\cdot$sin($\omega$), surface brightness ratio, sum of the fractional radii, ratio of the radii, out of eclipse flux level, and the amplitude and phase shift of the sinusoid applied to the luminosity of the primary in order to account for spots. We aim to vary these parameters over their entire range of possible solutions, which if left to a grid search for 10$^{-3}$ precision, would require computing on the order of $\sim$10$^{36}$ light curves; a computationally prohibitive task. Standard steepest descent minimization schemes such as Levenberg-Marquardt have extreme difficulties in large, multi-parameter solution spaces, especially for eclipsing binaries as the solution space is not at all smooth and has many local minima. Thus, we need a minimization technique that is computationally efficient, not adversely affected by a non-smooth solution space, and able to find the global minimum. These criteria are superbly met by the class of optimization schemes known as Genetic Algorithms (GAs).

In a standard GA, \citep[cf.][]{Charbonneau1995}, light curve parameter sets, called individuals, for an initial population of solutions, are randomly generated within a predefined parameter space, and compared to the observational light curve. Their corresponding $\chi^{2}$ value is used as a measure of fitness for natural selection, with parameters from fit individuals bred with each other, (subjected to crossover like chromosomes), to create a second generation of new solutions, and parameters from unfit individuals eliminated. After being subject to random mutations, to maintain parameter diversity and ensure discovery of the global minimum, this second generation is compared to the observational data, and bred into a third generation of solutions. The process continues for a specified number of generations, until a satisfactorily low $\chi^{2}$ is found. \citet{Charbonneau1995} demonstrated the application of GAs to problems in Astronomy and Astrophysics, specifically fitting galactic rotation curves, finding pulsation periods in $\delta$ Scuti stars, and fitting magnetodynamical wind models with multiple critical points, showing how the GA quickly finds the global minimum, regardless of the topography of the solution space. It is this type of GA that has been already been incorporated into the ELC eclipsing binary modeling code, and used with much success \citep{Orosz2000,Orosz2002}.

\citet{Canto2009} recently proposed a new form of GA called an Asexual Genetic Algorithm (AGA). In the AGA, instead of breeding new individuals via crossover, individuals are randomly created within a small predefined parameter space, or breeding box, centered on the fittest members of the previous generation. The size of this breeding box can be shrunk over successive generations to quickly converge to the best-fit solution. As shown by \citet{Canto2009}, the AGA is computationally simpler and more precise since it does not require encoding parameters for crossover, and converges much faster than traditional GAs, without sacrificing any ability to migrate to the global solution, so long as the breeding box size does not decrease too quickly. \citet{Canto2009} first showed that it far outperformed the standard GA in both computational efficiency and final precision by solving one of the exact same problems presented by \citet{Charbonneau1995}. \citet{Canto2009} additionally demonstrated the application of the AGA to fitting the radial-velocities of extrasolar planets and the spectral energy distributions of young stellar objects.

As eclipsing binary solutions have an even larger parameter space with many local minima than most problems, we make a few modifications to the AGA described by \citet{Canto2009} to ensure discovery of the global minimum. First, while we do exactly copy the fittest 10\% of individuals of one generation to the next generation, to ensure forward progress is always made while maintaining parameter diversity, instead of picking the fittest N members of a generation, each of which breeds M offspring, to create a new generation, we randomly select individuals for breeding by weighting them by a factor of (1/$\chi^{2}$)$^{2}$. This ensures that the fittest individuals breed the most offspring, but still allows for a few less fit individuals to breed, maintaining parameter diversity and exploration of the entire parameter space. Second, instead of randomly creating new members within a breeding box of fit individuals, we randomly select a number for each parameter from a Gaussian probability distribution centered on each parameter of a fit individual. Thus, new individuals are not strictly confined to a breeding box, but merely are very likely to be created near a fit individual, and maintain a very small probability that they will be created at many standard deviations away. This mimics mutation in traditional GAs and ensures that the algorithm will not become trapped in a local minimum. Third, as suggested by \citet{Canto2009}, the standard deviation of this normal distribution is chosen for each parameter to be the standard deviation of that parameter in the entire population, times the function $0.1^{(1/\chi_{0}^{2})}$, where $\chi_{0}^{2}$ is the $\chi^{2}$ value of the fittest member of the population. This allows parameters with the greatest impact on the fit, or the smallest range of possible parameters, such as the out of eclipse flux level, to converge rapidly, while allowing parameters that are less certain to converge more slowly and thoroughly explore their parameter space. Furthermore, via this method, the standard deviation is shrunk over successive generations, so that the algorithm converges, but only very slowly initially, rapidly increasing as $\chi_{0}^{2}$ approaches 1.0, i.e. the global minimum has been found. Finally, we take the fittest 10\% of the final generation and perform a standard Levenberg-Marquardt minimization for each member, choosing the member with the resulting lowest $\chi^{2}$ value as our final solution.

\begin{figure}[h!]
\centering
\begin{tabular}{ccc}
\epsfig{width=0.315\linewidth,file=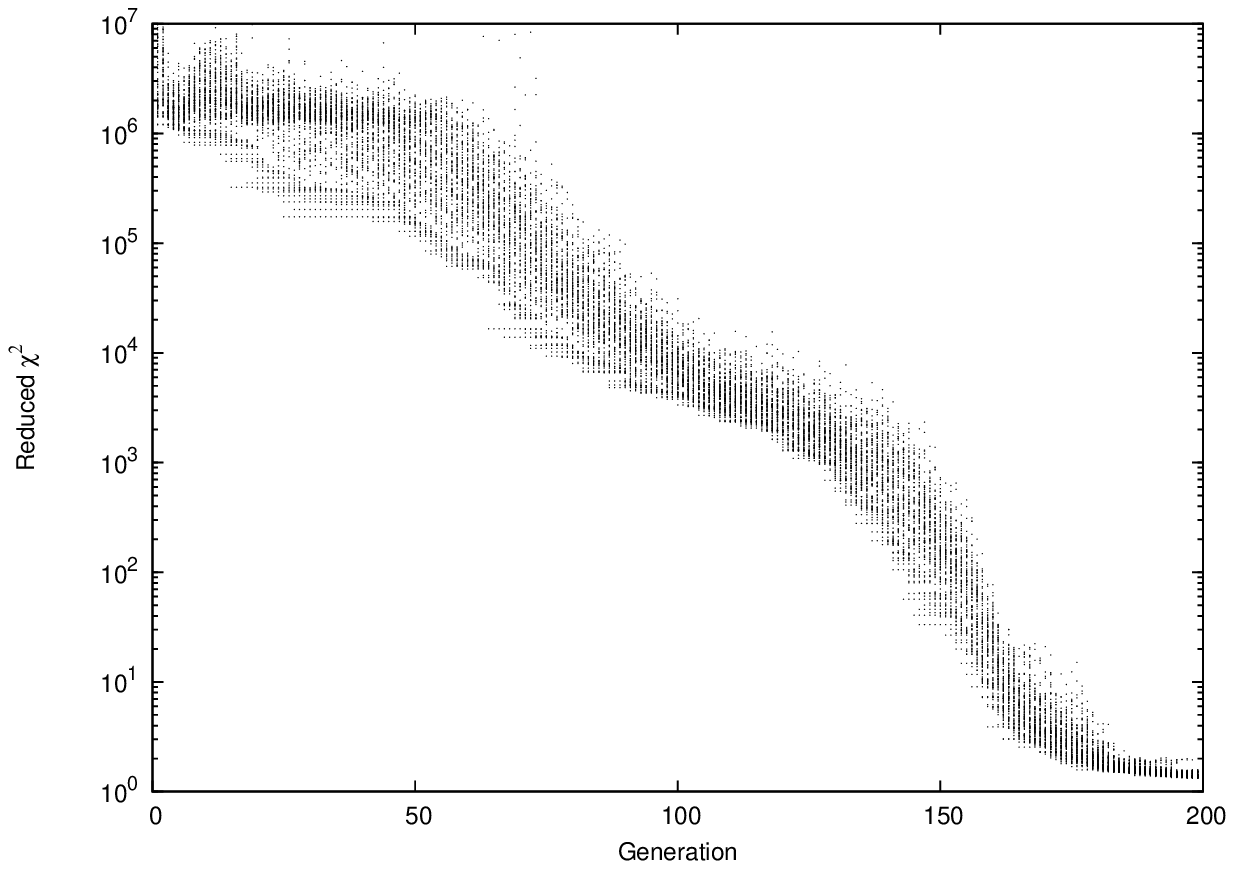} &
\epsfig{width=0.315\linewidth,file=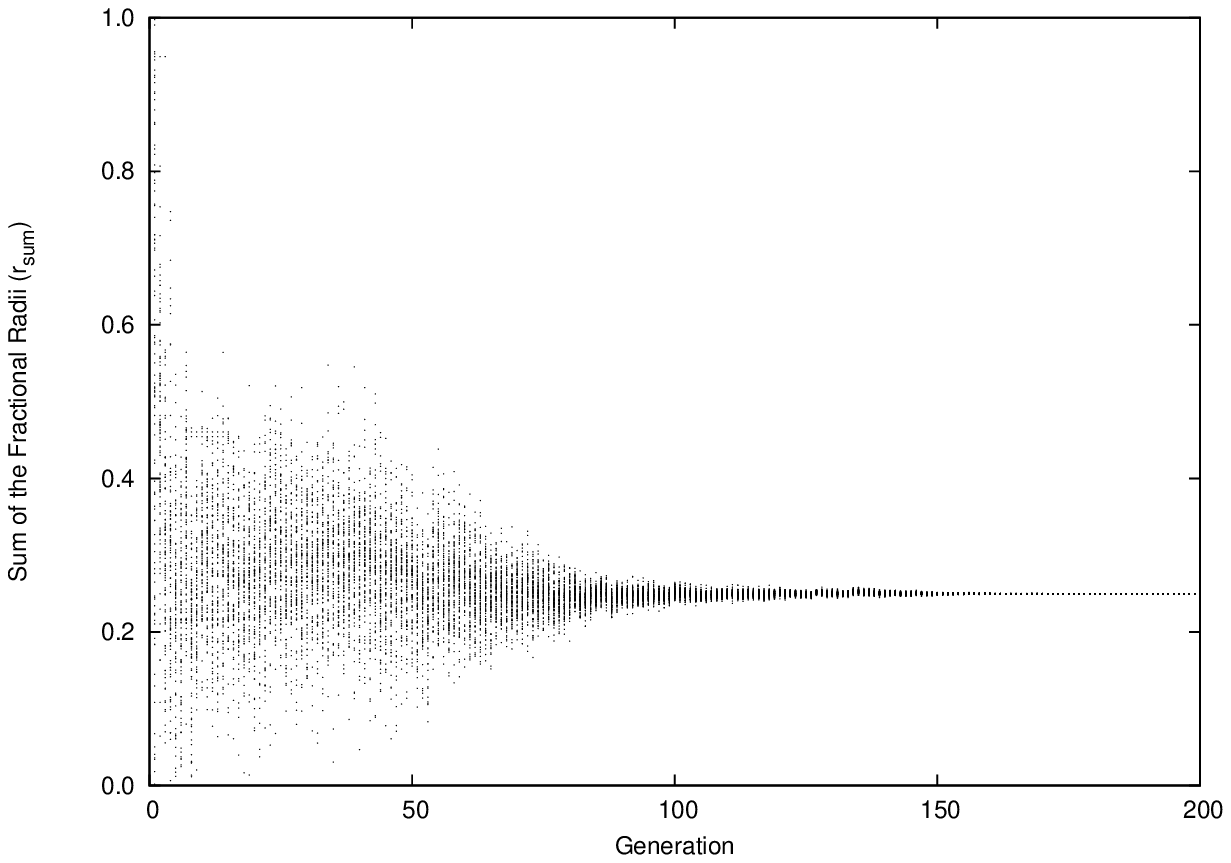} &
\epsfig{width=0.315\linewidth,file=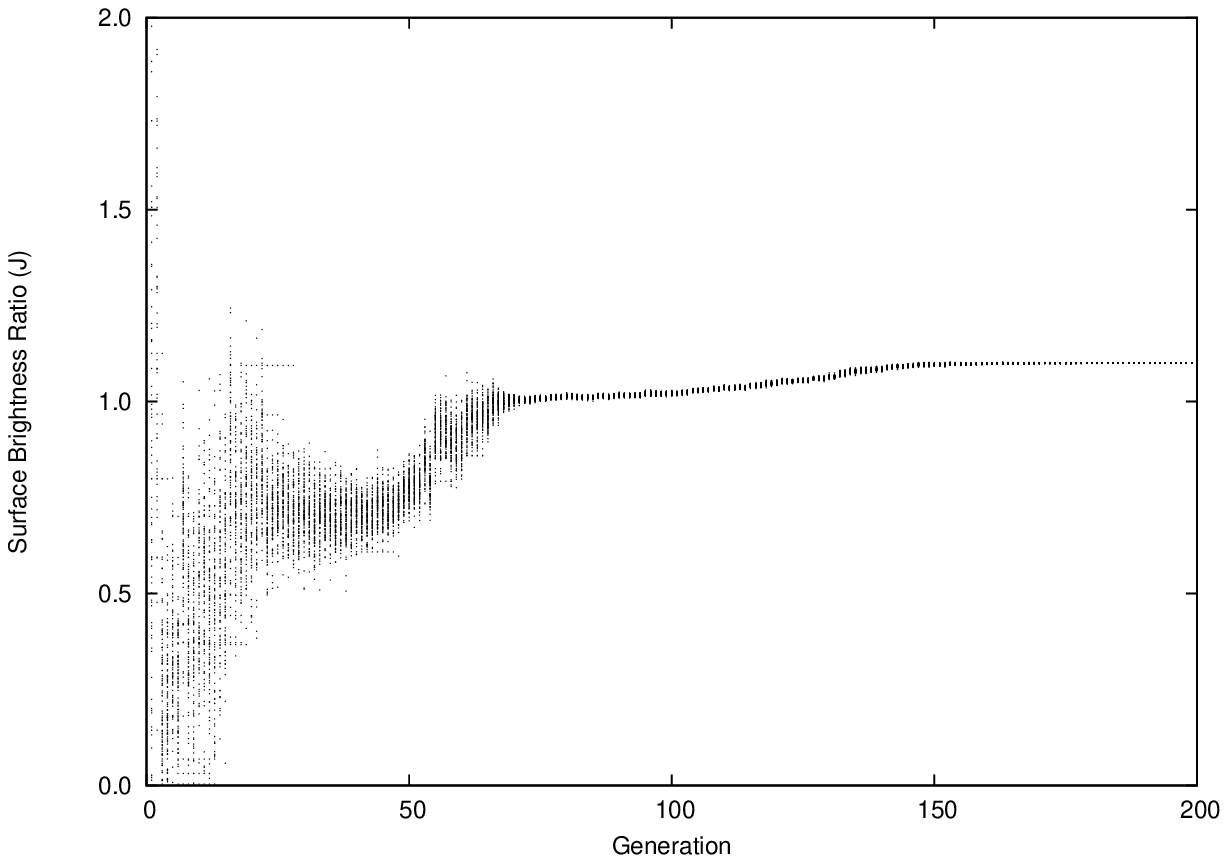} \\
\epsfig{width=0.315\linewidth,file=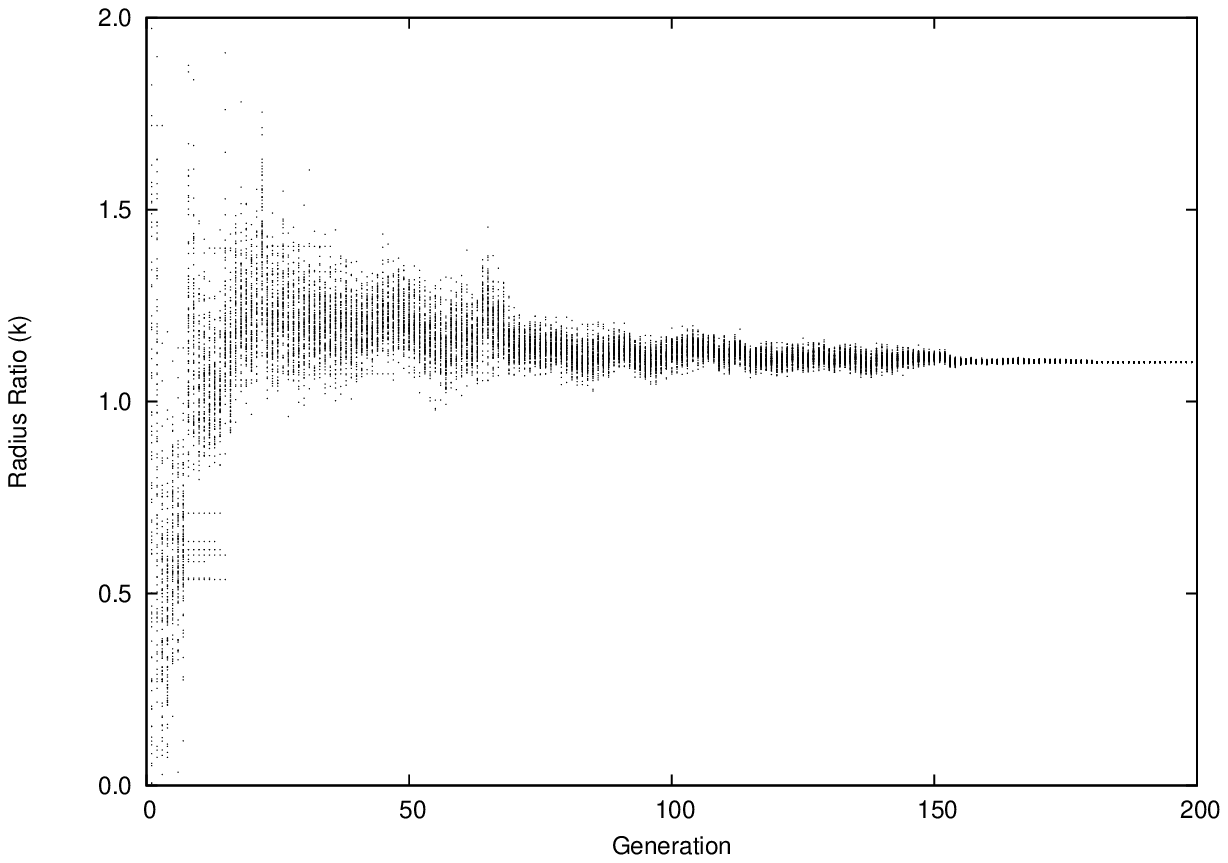} &
\epsfig{width=0.315\linewidth,file=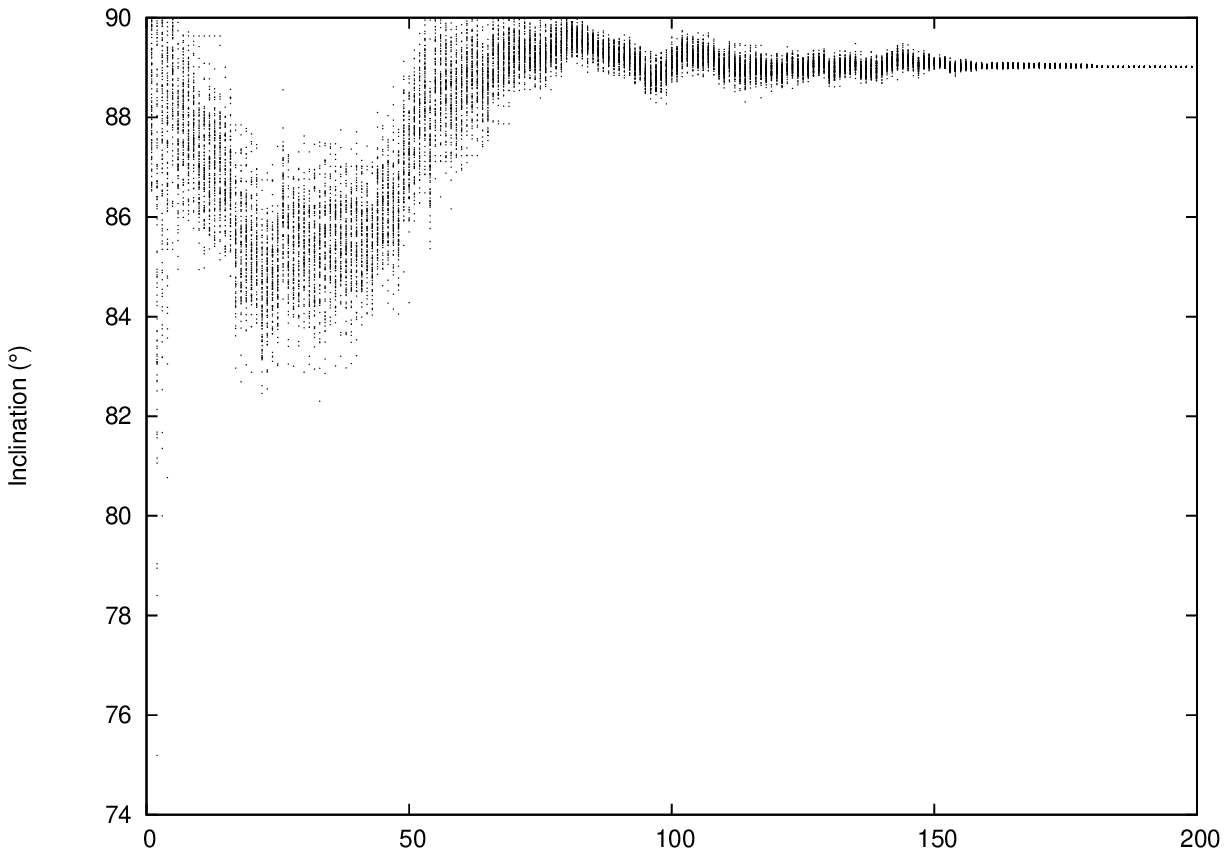} &
\epsfig{width=0.315\linewidth,file=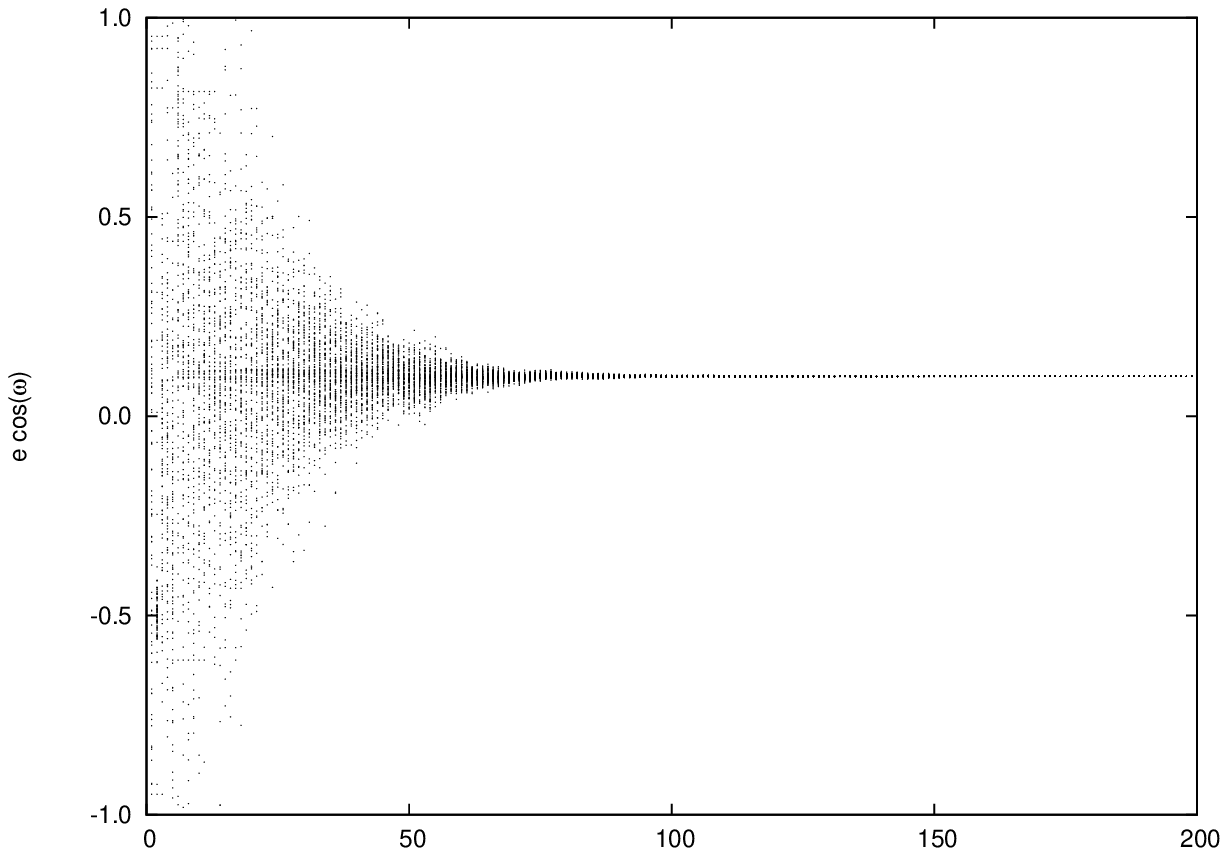} \\
\epsfig{width=0.315\linewidth,file=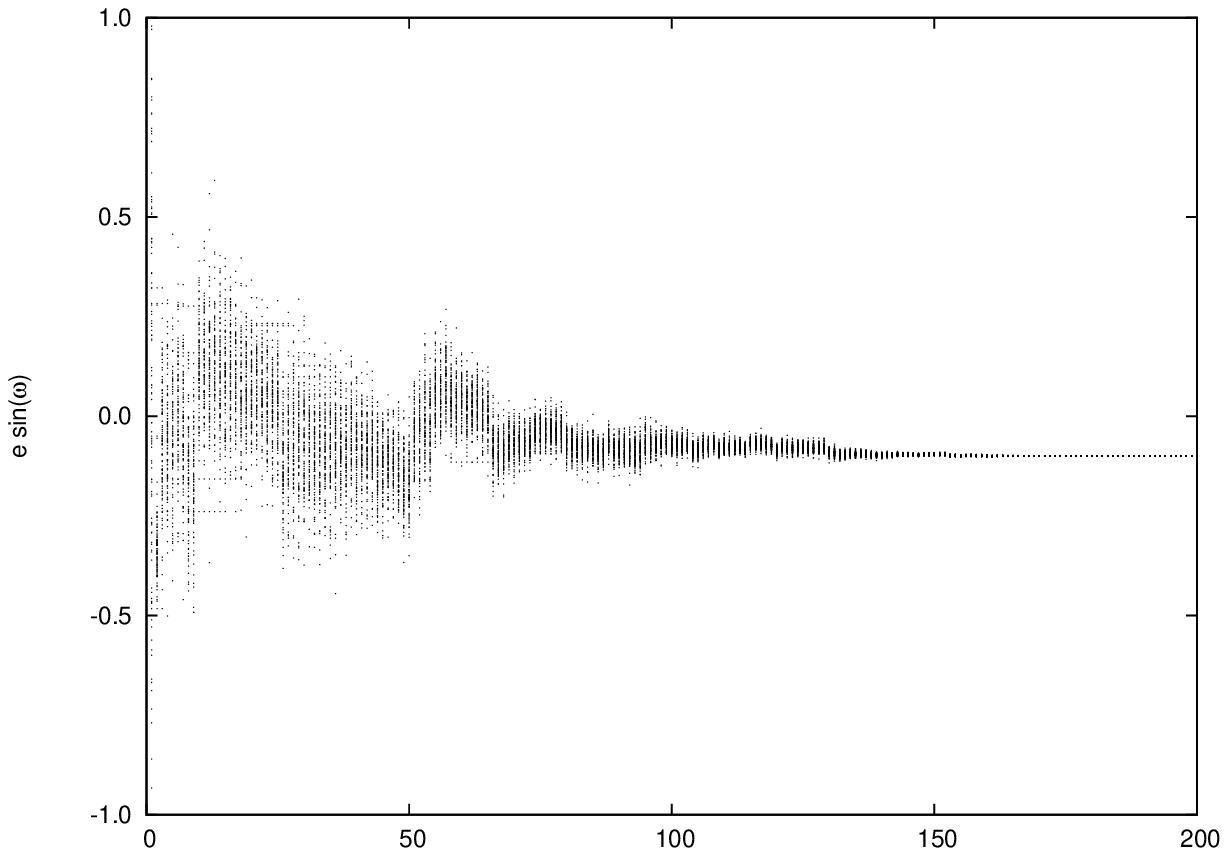} &
\epsfig{width=0.315\linewidth,file=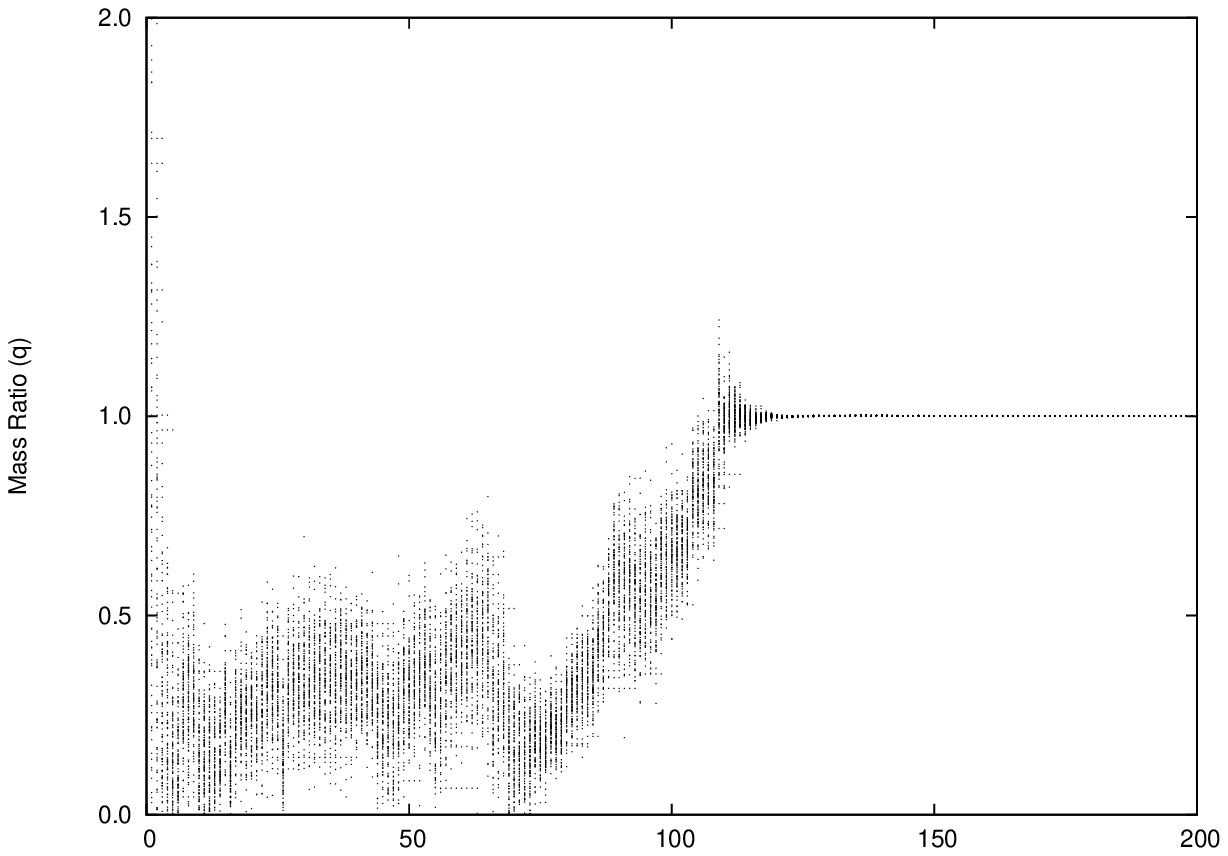} &
\epsfig{width=0.315\linewidth,file=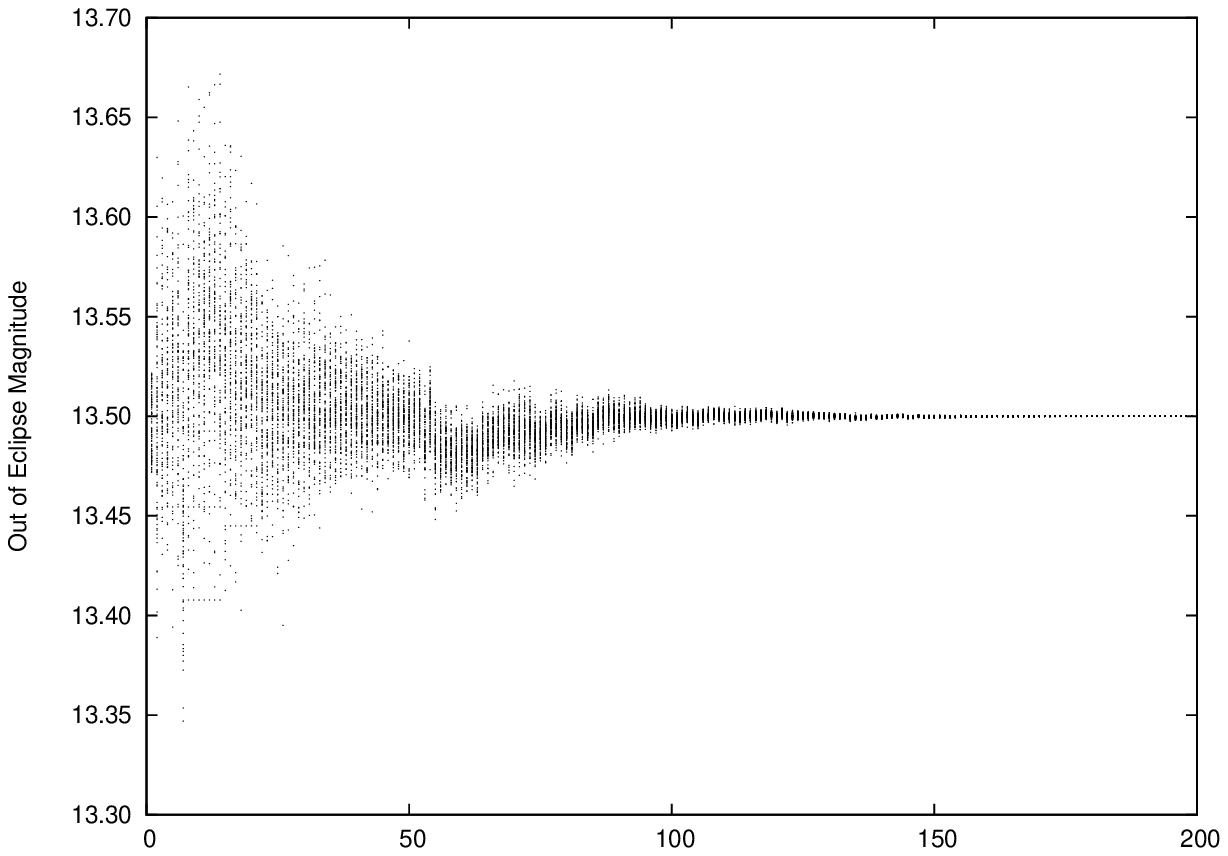} \\
\epsfig{width=0.315\linewidth,file=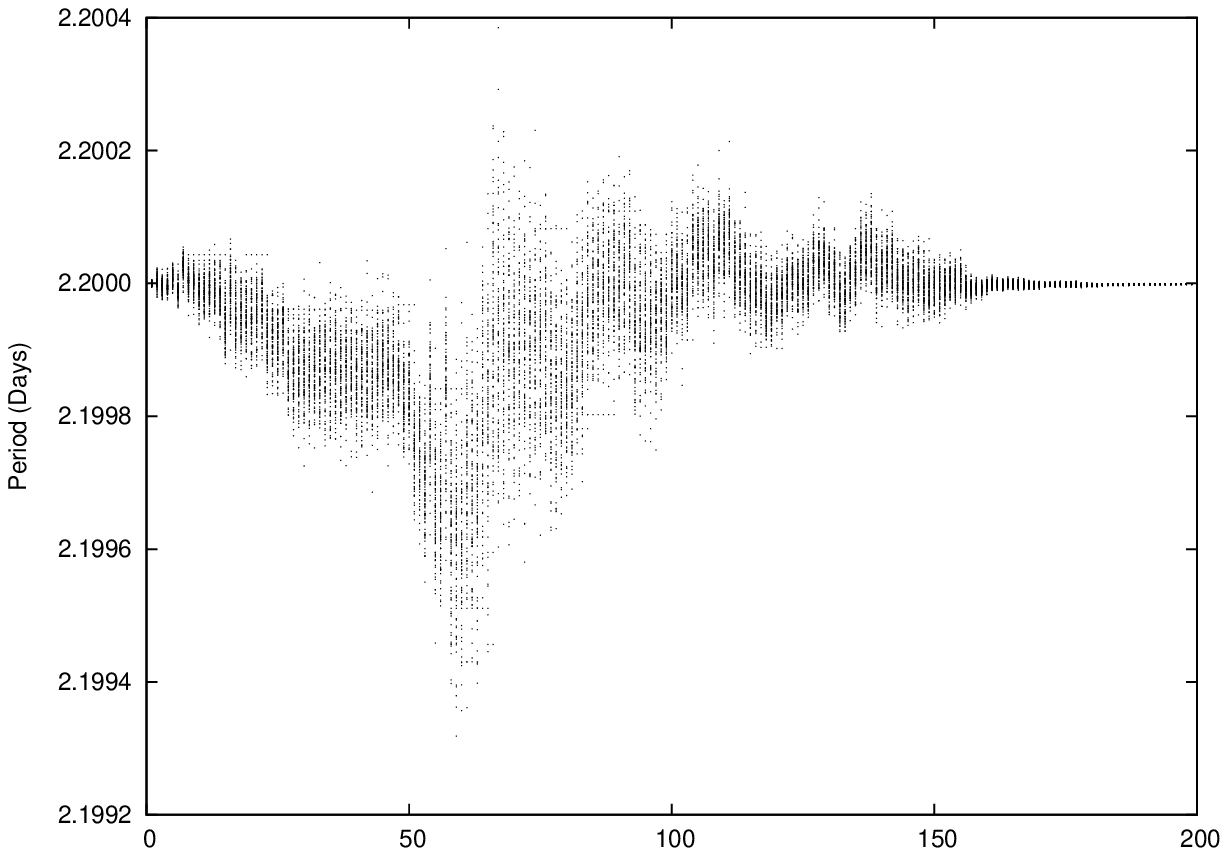} &
\epsfig{width=0.315\linewidth,file=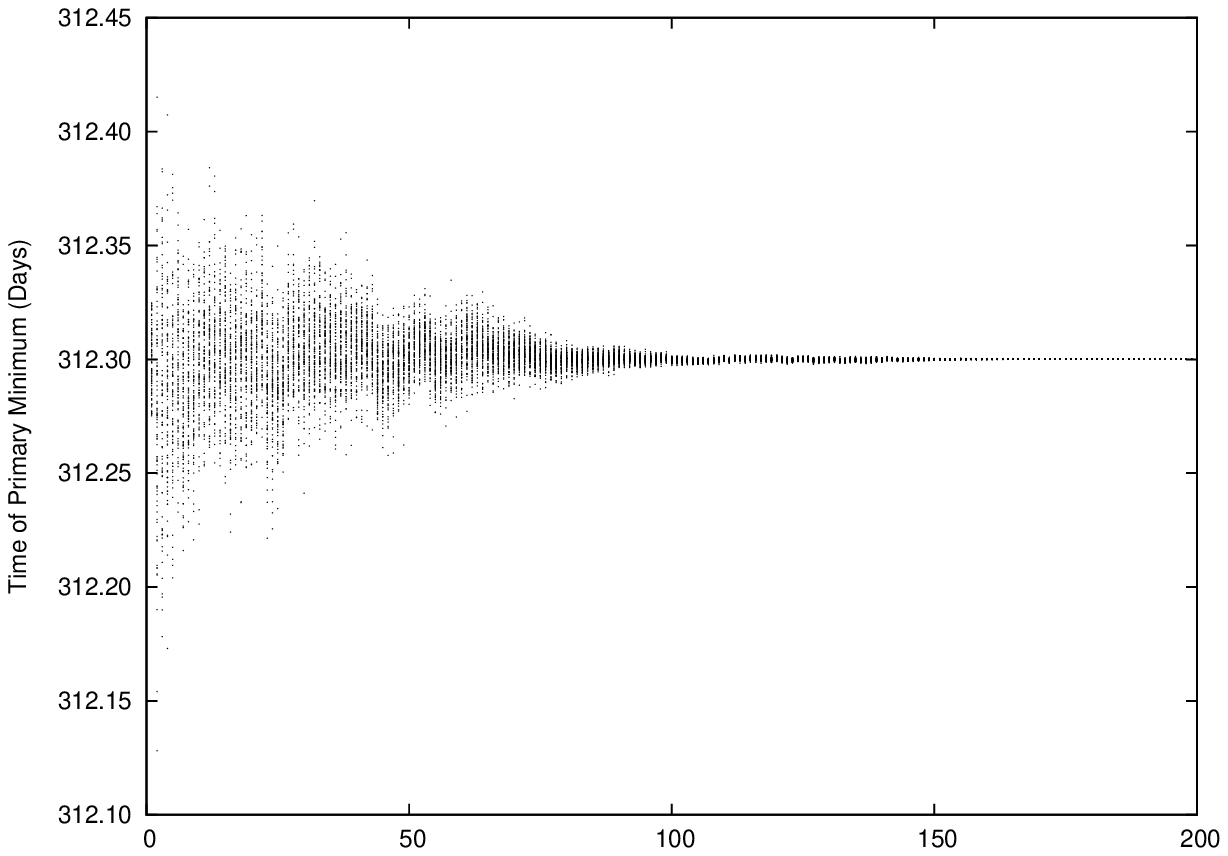} &
\epsfig{width=0.315\linewidth,file=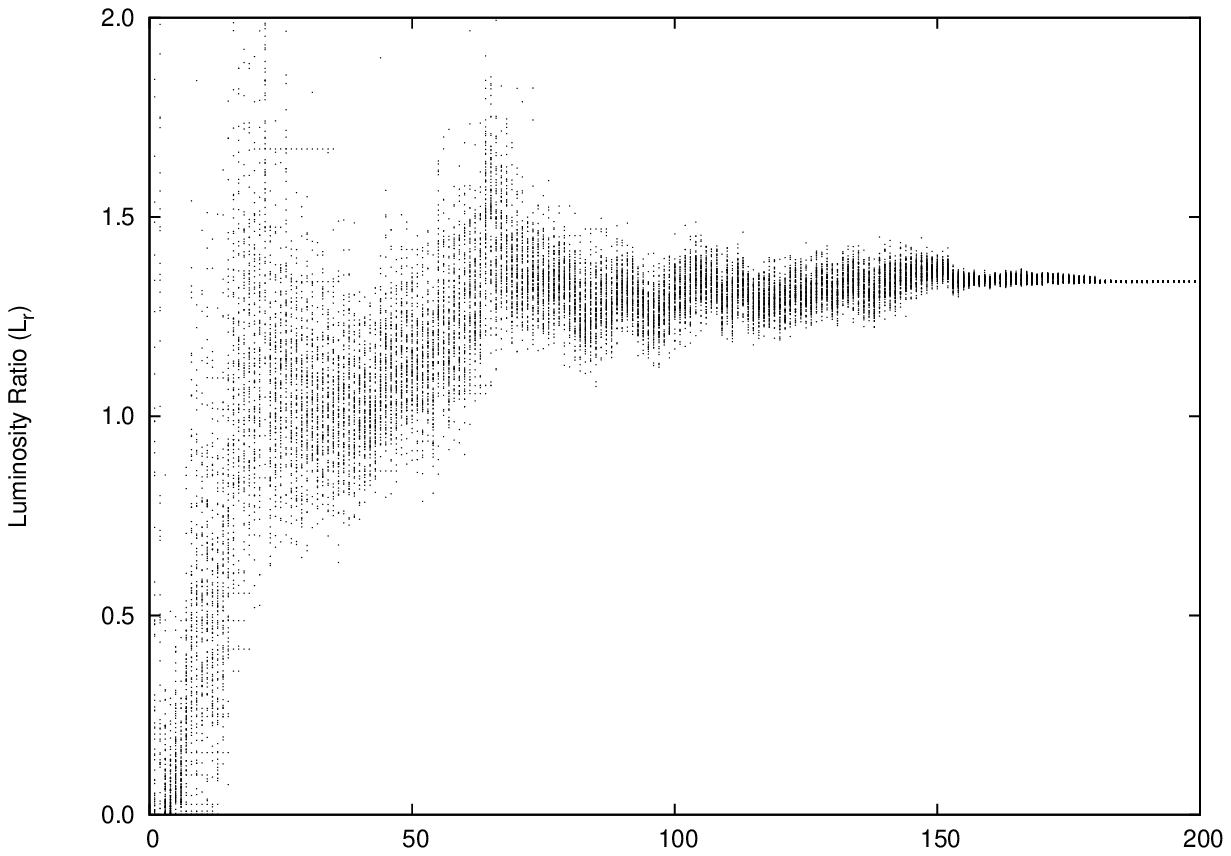} \\
\end{tabular}
\newpage
\caption{Illustration of how the AGA converges over subsequent generations by solving an artificially generated light curve, re-binned to the number of data points and error typical for a \emph{Kepler} light curve. The parameters of the system are r$_{sum}$ = 0.25, k = 1.1, i = 89.0$\degr$, q = 1.2, e$\cdot$cos($\omega$) = 0.1, e$\cdot$sin($\omega$) = -0.1, J = 1.1, P = 2.20 days, T$_{0}$ = 312.3 days, and out of eclipse magnitude = 13.5. The derived reduced $\chi^{2}$ and luminosity ratio are also plotted. The AGA converges rapidly, decreasing the lowest $\chi^{2}$ value found by an order of magnitude every $\sim$20 generations. It can be seen that the parameters that are most significant to the light curve converge the fastest.}
\label{agafig}
\end{figure}

We nominally found, for the eclipsing binaries in our sample, that a population of 100 individuals, bred for 200 generations, does an excellent job of solving the light curves. This only requires the generation of 20,000 light curves, which with the JKTEBOP code only required $\sim$3 minutes per light curve to solve on a single 2.0 GHz CPU. Of course, some systems may require a smaller or greater number of individuals and/or generations, but it should not be more than a factor of $\sim$2. One may substantially reduce the number of individuals or generations required, and thus the run time, if one can limit the range of parameter space. For example, if one knows, or wants to assume, the orbit is circular or nearly circular, one could constrain $|$e$\cdot$cos($\omega$)$|$ $<$ 0.1 and $|$e$\cdot$sin($\omega$)$|$ $<$ 0.1. Furthermore, the AGA code is extremely parallelizable, and thus with a multi-core computing cluster one could easily use this technique to model thousands of eclipsing binary lightcurves, as is to be expected from Pan-STARRS and other large photometric surveys, in a very reasonable time frame.

To visually demonstrate how the AGA works, we have generated a light curve with the following parameters: r$_{sum}$ = 0.25, k = 1.1, i = 89.0$\degr$, q = 1.2, e$\cdot$cos($\omega$) = 0.1, e$\cdot$sin($\omega$) = -0.1, J = 1.1, P = 2.20 days, T$_{0}$ = 312.3 days, and out of eclipse magnitude = 13.5. We then re-bin this data to match the number of data points in the Kepler Q1 data sets, and add typical Gaussian noise for a bright Kepler star of 0.1 mmag per data point. We then re-solve this light curve with the AGA, varying all the aforementioned parameters, and show in Figure~\ref{agafig} the value of each parameter for every individual in each generation, as well as the values for the derived reduced $\chi^{2}$ and luminosity ratio. One can see how even while searching over the entire global solution space, the AGA rapidly converges to the solution that was used to generate the light curve, with the $\chi^{2}$ decreasing by a factor of $\sim$10 every $\sim$20 generations. Even though the best solution of the 200$^{th}$ generation has $\chi^{2}$ $\sim$ 1.5, if allowed to continue for more generations, this run would eventually converge to $\chi^{2}$ = 1.0, and performing a simple Levenberg-Marquardt minimization from the best solution quickly produces a $\chi^{2}$ = 1.0 fit.

\clearpage

\bibliography{ref.bib}

\clearpage

\LongTables

\begin{deluxetable}{ccccccccrcccccc}
\tablenum{2}
\tablewidth{0pt}
\tabletypesize{\scriptsize}
\tablecaption{Model System Parameters via JKTEBOP for the 231 DDEBs with T$_{\rm eff}$ $<$ 5500 K}
\tablecolumns{15}
\tablehead{\emph{Kepler} ID & T$_{\rm eff}$ & m$_{\rm kep}$ & $\Delta$m$_{\rm kep}$ & Period & T$_{\rm 0}$ & i & e\tablenotemark{1} & $\omega$\tablenotemark{1} & r$_{\rm sum}$ & J & k & L$_{r}$ & L$_{1}$ Sine & L$_{3}$\\ & (K) & & & (Days) & (BJD-2450000) & ($\degr$) & & ($\degr$) & & & & & Amplitude & }
\startdata
\input{tab2.tex}
\enddata
\tablenotetext{1}{Although the values for e and $\omega$ are presented in this table for ease of reading, the values of e$\cdot$cos($\omega$) and e$\cdot$sin($\omega$) were actually solved for in the analysis.}
\label{ddebcandstab}
\end{deluxetable}

\clearpage

\begin{deluxetable}{ccccccccc}
\tablenum{3}
\tablewidth{0pt} 
\tabletypesize{\scriptsize}
\tablecaption{Temperature, Mass, and Radius Estimates for the 95 New LMMS DDEB Candidates with Amplitudes $\ge$ 0.1 Magnitudes and Both Masses $<$ 1.0 M$_{\sun}$}
\tablecolumns{9}
\tablehead{\emph{Kepler} ID & Period (Days) & T$_{\rm eff}$(K) & T$_{1}$(K) & T$_{2}$(K) & M$_{1}$(M$_{\sun}$) & M$_{2}$(M$_{\sun}$) & R$_{1}$(R$_{\sun}$) & R$_{2}$(R$_{\sun}$)}
\startdata
\input{tab3.tex}
\enddata
\label{lmbmrtab}
\end{deluxetable}

\end{document}

%% file: tab1.tex
003098197 & 38.3840\tablenotemark{3} & 5675 & 4.814 & 4.9 & 4.60\\
004178389\tablenotemark{1} & 45.2600\tablenotemark{3} & 5645 & 4.670 & 3.4 & 2.80\\
009016295\tablenotemark{2} & 19.9858 & 5819 & 4.582 & 4.1 & 0.17\\
009071386\tablenotemark{1} & 4.68513 & 6324 & 4.267 & 1.4 & 0.05\\
009838975\tablenotemark{1} & 18.7000 & 5018 & 4.802 & 5.7 & 0.21\\
012017140\tablenotemark{2} & 22.8624 & 6026 & 4.500 & 4.7 & 0.11\\
012504988\tablenotemark{1} & 5.09473 & 5985 & 4.464 & 2.9 & 0.06\\

%% file: tab4.tex
001571511 & 13.42 & 68.529019 & 14.02065 & 5804 & 89.28 & 1.08 & 0.14 & 1.43\\
003342592 & 14.92 & 69.190452 & 17.17864 & 5717 & 89.20 & 0.93 & 0.14 & 1.37\\
005372966 & 15.37 & 67.675070 & 9.286422 & 5464 & 88.91 & 0.92 & 0.19 & 1.87\\
006756669 & 15.33 & 65.860125 & 5.851827 & 5353 & 88.34 & 0.90 & 0.16 & 1.59\\
006805146 & 13.21 & 56.568771 & 13.77974 & 6214 & 89.14 & 1.41 & 0.21 & 2.11\\
008544996 & 15.20 & 65.898818 & 4.081488 & 5463 & 87.61 & 1.00 & 0.13 & 1.27\\
011974540$^{1}$ & 13.22 & 65.862352 & 24.67058 & 6507 & 89.53 & 0.69 & 0.06 & 0.56\\
012251650 & 14.76 & 71.657743 & 17.76233 & 4952 & 88.97 & 1.00 & 0.16 & 1.64\\

%% file: tab2.tex
002162994 & 5410 & 14.162 & 0.535 & 4.101544 & 5002.545861 & 89.87 &  0.01 & 270 & 0.199 &    0.991 &  0.702 & 0.4888 & 0.008 & 0.00\\
002437452 & 5398 & 16.981 & 0.256 & 14.47184 & 5003.759350 & 87.46 &  0.08 &  90 & 0.084 &    0.641 &   2.47 & 3.905 & 0.011 & 0.00\\
002580872 & 5293 & 14.880 & 0.374 & 15.92672 & 4978.550988 & 87.95 &  0.26 & 102 & 0.084 &     1.14 &   1.25 & 1.774 & 0.014 & 0.00\\
002719873 & 5086 & 15.160 & 0.235 & 17.27953 & 4968.273250 & 87.76 &  0.31 &  90 & 0.059 &    0.633 &   2.64 & 4.425 & 0.007 & 0.00\\
002852560 & 5381 & 15.308 & 0.460 & 11.96119 & 4964.912794 & 88.06 &  0.44 &  41 & 0.079 &     1.04 &  0.986 & 1.008 & 0.000 & 0.00\\
002860788 & 5319 & 14.043 & 0.137 & 5.259798 & 4965.066945 & 82.29 &  0.00 & 268 & 0.212 &    0.876 &  0.561 & 0.2755 & 0.009 & 0.00\\
003003991 & 5366 & 13.926 & 0.115 & 7.244790 & 4964.859062 & 86.88 &  0.28 & 270 & 0.083 &   0.0438 &   11.2 & 5.447 & 0.000 & 0.40\\
003102024 & 5117 & 12.809 & 0.351 & 13.78248 & 4958.697309 & 89.50 &  0.54 & 302 & 0.054 &    0.605 &   1.37 & 1.138 & 0.000 & 0.43\\
003113266 & 5077 & 15.577 & 0.011 & 0.9958567 & 5002.193202 & 72.96 &  0.01 & 266 & 0.325 &     1.24 &   9.95 & 123.3 & 0.026 & 0.00\\
003241344 & 5422 & 14.756 & 0.401 & 3.912656 & 4966.427889 & 90.00 &  0.02 & 256 & 0.121 &   0.0854 &  0.509 & 0.02209 & 0.012 & 0.00\\
003241619 & 5165 & 12.524 & 0.802 & 1.703368 & 4965.468231 & 85.35 &  0.03 &  88 & 0.267 &    0.303 &   1.04 & 0.3286 & 0.016 & 0.00\\
003344419 & 5348 & 14.997 & 0.005 & 0.6517609 & 4977.843744 & 50.18 &  0.01 & 269 & 0.694 &     1.08 &   23.4 & 592.4 & 0.068 & 0.00\\
003458919 & 5063 & 13.815 & 0.121 & 0.8920383 & 5002.281060 & 73.06 &  0.14 & 270 & 0.418 &    0.362 &   3.95 & 5.653 & 0.056 & 0.00\\
003543270 & 5288 & 15.220 & 0.130 & 4.177213 & 5003.789822 & 82.27 &  0.05 & 269 & 0.207 &    0.254 &  0.394 & 0.03937 & 0.024 & 0.00\\
003556742 & 4921 & 14.221 & 0.004 & 0.8229667 & 5003.017211 & 37.02 &  0.00 & 247 & 0.848 &     2.41 &   15.5 & 576.3 & 0.108 & 0.00\\
003656322 & 5075 & 13.061 & 0.150 & 3.660009 & 4989.330479 & 67.54 &  0.02 & 125 & 0.457 &    0.930 &   1.66 & 2.576 & 0.125 & 0.00\\
003730067 & 4099 & 14.610 & 0.594 & 0.2940818 & 4964.591764 & 75.71 &  0.03 &  88 & 0.590 &    0.420 &   1.25 & 0.6522 & 0.035 & 0.00\\
003830820 & 3902 & 15.368 & 0.044 & 15.58263 & 4999.277480 & 87.92 &  0.47 &  79 & 0.057 &     2.39 &   2.10 & 10.57 & 0.000 & 0.60\\
003834364 & 5449 & 14.661 & 0.089 & 2.908455 & 4965.315292 & 82.21 &  0.10 & 271 & 0.182 &   0.0407 &  0.874 & 0.03106 & 0.006 & 0.00\\
003848919 & 5226 & 13.901 & 0.636 & 1.047253 & 4964.766251 & 85.07 &  0.00 &  84 & 0.418 &    0.903 &   1.04 & 0.9684 & 0.017 & 0.00\\
003957477 & 5395 & 12.477 & 0.073 & 0.9789470 & 4964.726279 & 66.89 &  0.03 &  91 & 0.525 &     1.42 &   2.97 & 12.52 & 0.055 & 0.00\\
004049124 & 5349 & 14.654 & 0.175 & 4.804341 & 4969.004205 & 84.04 &  0.41 &  89 & 0.160 &     1.32 &   2.24 & 6.592 & 0.002 & 0.00\\
004077442 & 4523 & 13.512 & 0.153 & 0.6928736 & 5002.273033 & 69.03 &  0.01 & 277 & 0.499 &     1.94 &   1.34 & 3.499 & 0.183 & 0.00\\
004078693 & 5288 & 13.485 & 0.005 & 2.756407 & 5001.858784 & 85.40 &  0.32 & 270 & 0.119 &   0.0308 &  0.519 & 0.008294 & 0.001 & 0.79\\
004247791 & 4063 & 11.260 & 0.152 & 4.100862 & 5001.145258 & 77.90 &  0.00 &  87 & 0.326 &    0.928 &   1.48 & 2.035 & 0.001 & 0.00\\
004281895 & 5309 & 12.256 & 0.078 & 9.543591 & 5002.358654 & 87.52 &  0.30 &   4 & 0.065 &     1.18 &   3.41 & 13.80 & 0.000 & 0.00\\
004346875 & 5339 & 15.584 & 0.284 & 4.694341 & 5004.332965 & 87.08 &  0.02 & 267 & 0.135 &   0.0866 &  0.429 & 0.01592 & 0.014 & 0.00\\
004352168 & 5115 & 14.343 & 0.663 & 10.64334 & 4967.942159 & 89.58 &  0.18 & 213 & 0.072 &    0.268 &   1.29 & 0.4451 & 0.012 & 0.35\\
004484356 & 5080 & 14.235 & 0.177 & 1.144126 & 5002.774583 & 78.62 &  0.02 & 271 & 0.320 &    0.899 &  0.652 & 0.3817 & 0.025 & 0.00\\
004540632 & 4818 & 14.991 & 1.045 & 31.00996 & 4983.860841 & 89.93 &  0.66 &  98 & 0.030 &    0.315 &  0.826 & 0.2150 & 0.000 & 0.00\\
004579313 & 5363 & 14.811 & 0.008 & 2.112635 & 5002.947518 & 68.73 &  0.00 & 309 & 0.390 &    0.553 &   13.3 & 98.34 & 0.029 & 0.00\\
004633434 & 4902 & 15.362 & 0.233 & 22.27067 & 4967.759577 & 89.52 &  0.09 & 107 & 0.031 &   0.0806 &   1.58 & 0.2024 & 0.000 & 0.77\\
004672010 & 4655 & 14.602 & 0.049 & 0.9628780 & 5002.694396 & 41.63 &  0.02 & 271 & 0.904 &     1.97 &   7.26 & 103.9 & 0.038 & 0.00\\
004678171 & 4240 & 15.993 & 0.951 & 15.28859 & 4965.805465 & 89.68 &  0.01 & 111 & 0.045 &    0.474 &  0.996 & 0.4702 & 0.002 & 0.00\\
004737267 & 5156 & 15.145 & 0.471 & 9.523936 & 5001.185337 & 89.05 &  0.01 & 220 & 0.154 &    0.521 &   1.98 & 2.032 & 0.072 & 0.00\\
004757331 & 5092 & 15.725 & 0.087 & 2.362127 & 5001.736119 & 81.59 &  0.11 &  90 & 0.210 &     1.77 &   2.01 & 7.157 & 0.043 & 0.00\\
004758368 & 4594 & 10.805 & 0.044 & 3.750218 & 5003.200871 & 67.51 &  0.02 & 260 & 0.489 &     1.01 &   3.73 & 14.11 & 0.002 & 0.00\\
004773155 & 5447 & 13.592 & 0.733 & 25.70599 & 4989.643369 & 89.86 &  0.43 & 309 & 0.042 &    0.767 &   1.14 & 0.9988 & 0.000 & 0.03\\
004908495 & 4731 & 13.871 & 0.359 & 3.120583 & 4965.367280 & 86.08 &  0.01 & 265 & 0.153 &    0.732 &   1.01 & 0.7479 & 0.076 & 0.00\\
004940201 & 5284 & 14.984 & 0.050 & 8.816203 & 5002.550394 & 90.00 &  0.03 &  87 & 0.067 &    0.455 &   1.24 & 0.7023 & 0.000 & 0.93\\
004948863 & 5490 & 15.414 & 0.090 & 8.643652 & 4972.829522 & 87.28 &  0.26 &  89 & 0.070 &     1.33 &   3.05 & 12.37 & 0.001 & 0.00\\
005015913 & 5487 & 12.989 & 0.002 & 2.359939 & 4954.580900 & 72.85 &  0.00 & 277 & 0.312 &     1.06 &   22.9 & 553.2 & 0.004 & 0.00\\
005018787 & 5215 & 15.428 & 0.023 & 0.6071971 & 5002.664852 & 83.52 &  0.12 &  89 & 0.338 &    0.410 &   1.98 & 1.601 & 0.000 & 0.96\\
005036538 & 4199 & 13.349 & 0.758 & 2.122015 & 5001.594725 & 88.69 &  0.00 & 285 & 0.181 &    0.773 &   1.02 & 0.8082 & 0.033 & 0.00\\
005041975 & 5149 & 13.981 & 0.160 & 2.958502 & 5003.379626 & 58.09 &  0.00 & 285 & 0.680 &    0.427 &   2.93 & 3.661 & 0.054 & 0.00\\
005080652 & 5344 & 15.080 & 0.524 & 4.144388 & 5001.321781 & 86.69 &  0.01 &  91 & 0.165 &    0.556 &  0.839 & 0.3915 & 0.018 & 0.00\\
005193386 & 4797 & 13.998 & 0.397 & 21.37192 & 4980.205957 & 88.85 &  0.01 & 121 & 0.134 &    0.268 &   3.43 & 3.148 & 0.098 & 0.00\\
005218014 & 4752 & 12.923 & 0.010 & 10.84612 & 4971.331816 & 88.91 &  0.24 & 157 & 0.068 &    0.944 &   1.16 & 1.266 & 0.002 & 0.98\\
005266937 & 5483 & 14.352 & 0.987 & 5.916942 & 5001.400391 & 88.40 &  0.05 & 268 & 0.429 &    0.113 &  0.721 & 0.05881 & 0.020 & 0.00\\
005286786 & 4946 & 15.456 & 0.006 & 9.949612 & 4976.748845 & 88.88 &  0.05 & 230 & 0.049 &    0.817 &   2.81 & 6.433 & 0.002 & 0.98\\
005294739 & 5068 & 13.930 & 0.994 & 3.736174 & 5001.678732 & 76.49 &  0.02 & 278 & 0.555 &    0.138 &   1.85 & 0.4705 & 0.068 & 0.00\\
005300878 & 4631 & 14.767 & 0.823 & 1.279424 & 5002.597321 & 89.49 &  0.01 &  93 & 0.294 &    0.817 &   1.02 & 0.8576 & 0.047 & 0.00\\
005347784 & 5392 & 13.094 & 0.155 & 9.584026 & 5000.621695 & 85.60 &  0.01 & 148 & 0.121 &     1.05 &   1.34 & 1.885 & 0.007 & 0.00\\
005467126 & 4683 & 12.367 & 0.014 & 2.845694 & 5001.431451 & 77.08 &  0.15 &  85 & 0.575 &    0.243 &   1.69 & 0.6920 & 0.000 & 0.98\\
005597970 & 5179 & 12.778 & 0.218 & 6.717435 & 4970.209216 & 86.01 &  0.28 & 270 & 0.106 &   0.0127 &   2.74 & 0.09511 & 0.002 & 0.00\\
005598639 & 4847 & 10.201 & 0.135 & 1.297549 & 5003.022903 & 83.12 &  0.00 & 280 & 0.441 &    0.995 &  0.990 & 0.9752 & 0.003 & 0.69\\
005696909 & 5451 & 14.984 & 0.006 & 0.6430210 & 4964.688376 & 63.09 &  0.00 & 261 & 0.490 &    0.965 &   15.0 & 218.3 & 0.035 & 0.00\\
005731312 & 4658 & 13.811 & 0.388 & 7.946392 & 4968.092030 & 88.99 &  0.43 &  15 & 0.058 &    0.113 &  0.592 & 0.03952 & 0.000 & 0.20\\
005781192 & 5372 & 12.989 & 0.301 & 9.459957 & 4999.722660 & 88.15 &  0.07 & 295 & 0.077 &    0.518 &  0.645 & 0.2154 & 0.006 & 0.00\\
005802285 & 4791 & 15.349 & 0.017 & 2.417017 & 5003.656318 & 77.87 &  0.00 &  89 & 0.232 &    0.740 &   3.74 & 10.36 & 0.002 & 0.00\\
005802470 & 5418 & 13.764 & 0.337 & 3.791871 & 5001.260474 & 85.11 &  0.03 &  90 & 0.149 &    0.344 &  0.985 & 0.3337 & 0.013 & 0.00\\
005871918 & 4021 & 15.701 & 0.319 & 12.64175 & 4972.761250 & 90.00 &  0.16 & 180 & 0.058 &    0.246 &   1.29 & 0.4091 & 0.056 & 0.64\\
006029130 & 5160 & 14.832 & 0.421 & 12.59140 & 5005.516830 & 88.72 &  0.02 &  49 & 0.063 &    0.851 &   1.02 & 0.8859 & 0.002 & 0.00\\
006042116 & 4771 & 11.300 & 0.089 & 5.407156 & 5002.038929 & 80.98 &  0.11 &  54 & 0.211 &     1.60 &   1.32 & 2.789 & 0.004 & 0.00\\
006044064 & 5095 & 15.001 & 1.653 & 5.063280 & 5002.149463 & 83.72 &  0.03 &  75 & 0.389 &    0.145 &   1.78 & 0.4606 & 0.078 & 0.00\\
006060580 & 5308 & 13.460 & 0.019 & 2.313334 & 5003.212901 & 75.37 &  0.00 &  27 & 0.289 &    0.304 &  0.298 & 0.02701 & 0.001 & 0.00\\
006131659 & 4870 & 12.534 & 0.475 & 17.52783 & 4960.041441 & 89.37 &  0.02 & 270 & 0.044 &    0.316 &  0.593 & 0.1111 & 0.000 & 0.00\\
006187893 & 5103 & 11.702 & 0.077 & 0.7891775 & 5006.959004 & 64.02 &  0.01 & 274 & 0.634 &    0.135 &  0.349 & 0.01646 & 0.008 & 0.00\\
006191574 & 4208 & 14.353 & 0.233 & 0.000000 & -50000.000000 &  0.00 &  0.00 &   0 & 0.000 &     0.00 &   0.00 & 0.000 & 0.000 & 0.00\\
006197038 & 4937 & 13.531 & 0.798 & 9.752156 & 5000.794386 & 80.21 &  0.19 &  90 & 0.277 &    0.261 &   2.62 & 1.790 & 0.308 & 0.00\\
006205460 & 5242 & 12.746 & 0.796 & 3.722771 & 5001.134908 & 85.88 &  0.01 &  36 & 0.419 &    0.159 &   2.79 & 1.238 & 0.069 & 0.00\\
006307537 & 4253 & 11.753 & 0.193 & 29.74440 & 4960.659149 & 87.35 &  0.04 & 277 & 0.108 &    0.246 &   4.69 & 5.396 & 0.003 & 0.00\\
006312534 & 4897 & 15.583 & 0.024 & 3.015501 & 5002.128021 & 81.73 &  0.36 &  90 & 0.213 &     16.0 &   2.01 & 64.38 & 0.158 & 0.00\\
006359798 & 5452 & 12.932 & 0.071 & 14.15394 & 4959.543146 & 89.54 &  0.41 & 183 & 0.048 &    0.377 &   1.35 & 0.6877 & 0.000 & 0.90\\
006367628 & 5185 & 13.035 & 0.548 & 3.780139 & 5002.708087 & 76.36 &  0.02 & 267 & 0.550 &    0.213 &  0.840 & 0.1500 & 0.016 & 0.00\\
006449552 & 5357 & 14.904 & 0.946 & 20.14888 & 4968.810574 & 89.40 &  0.27 & 247 & 0.045 &    0.188 &   1.08 & 0.2176 & 0.001 & 0.00\\
006464285 & 5061 & 13.826 & 0.444 & 0.8436324 & 5003.755443 & 73.04 &  0.01 & 286 & 0.469 &    0.188 &   2.74 & 1.411 & 0.019 & 0.00\\
006466939 & 4920 & 14.454 & 0.733 & 2.285920 & 5003.760706 & 88.72 &  0.00 &  95 & 0.199 &    0.762 &   1.14 & 0.9858 & 0.021 & 0.00\\
006548447 & 5031 & 12.880 & 0.165 & 10.76541 & 5009.086049 & 89.90 &  0.11 & 184 & 0.158 &    0.617 &   3.29 & 6.693 & 0.004 & 0.00\\
006591789 & 5410 & 15.353 & 0.614 & 5.088435 & 5002.974423 & 88.48 &  0.01 & 171 & 0.128 &    0.318 &  0.664 & 0.1399 & 0.005 & 0.00\\
006620003 & 3955 & 15.686 & 0.037 & 3.428469 & 4997.172065 & 82.97 &  0.01 & 269 & 0.146 &    0.775 &   1.04 & 0.8436 & 0.000 & 0.00\\
006629332 & 5452 & 13.997 & 0.073 & 4.310363 & 5007.525591 & 84.14 &  0.05 &  90 & 0.122 &     2.95 &   1.46 & 6.279 & 0.130 & 0.00\\
006694186 & 5247 & 12.376 & 0.189 & 5.554204 & 5001.487264 & 80.66 &  0.29 & 271 & 0.223 &  0.00832 &   31.3 & 8.173 & 0.000 & 0.00\\
006697716 & 4898 & 14.424 & 0.279 & 1.443221 & 5008.877209 & 82.57 &  0.00 &  30 & 0.261 &    0.505 &  0.634 & 0.2030 & 0.018 & 0.00\\
006706287 & 5182 & 13.620 & 0.607 & 2.535431 & 5004.418517 & 87.19 &  0.00 & 260 & 0.196 &    0.697 &  0.905 & 0.5714 & 0.025 & 0.00\\
006778050 & 5091 & 14.514 & 0.420 & 0.9458108 & 4964.620719 & 81.63 &  0.01 & 276 & 0.389 &    0.822 &  0.855 & 0.6013 & 0.025 & 0.00\\
006841577 & 5478 & 14.875 & 0.270 & 15.53753 & 4973.272586 & 89.35 &  0.19 & 128 & 0.059 &   0.0828 &   1.72 & 0.2455 & 0.000 & 0.74\\
006863840 & 5024 & 15.138 & 0.668 & 3.852650 & 4964.746207 & 88.78 &  0.00 &  81 & 0.142 &    0.830 &   1.05 & 0.9234 & 0.060 & 0.00\\
006939670 & 5436 & 14.858 & 0.152 & 4.238755 & 4968.201178 & 79.83 &  0.05 &  91 & 0.235 &    0.889 &   2.80 & 6.980 & 0.145 & 0.00\\
007049486 & 5498 & 13.144 & 0.088 & 26.71855 & 4971.051326 & 88.56 &  0.29 &  21 & 0.052 &    0.684 &  0.498 & 0.1696 & 0.004 & 0.64\\
007097571 & 5266 & 11.267 & 0.153 & 2.213962 & 5005.416674 & 80.18 &  0.03 &  91 & 0.385 &    0.217 &  0.314 & 0.02135 & 0.001 & 0.00\\
007119757 & 5072 & 15.608 & 0.249 & 0.7429393 & 4980.909123 & 71.04 &  0.02 &  86 & 0.541 &     1.10 &   1.58 & 2.750 & 0.058 & 0.00\\
007125636 & 4358 & 15.507 & 0.266 & 6.490765 & 4978.048116 & 87.67 &  0.02 & 214 & 0.081 &    0.835 &  0.929 & 0.7198 & 0.005 & 0.00\\
007128918 & 5386 & 15.758 & 0.142 & 7.118892 & 4984.394210 & 88.47 &  0.01 &  54 & 0.085 &    0.573 &  0.885 & 0.4486 & 0.004 & 0.70\\
007129465 & 5182 & 15.316 & 0.427 & 5.491840 & 4966.171031 & 87.83 &  0.00 & 271 & 0.107 &    0.856 &  0.941 & 0.7591 & 0.004 & 0.00\\
007200102 & 5207 & 15.213 & 0.538 & 14.66695 & 4972.573070 & 88.65 &  0.42 & 124 & 0.054 &    0.564 &  0.751 & 0.3181 & 0.000 & 0.04\\
007220322 & 4887 & 11.884 & 0.009 & 0.7521433 & 5002.397213 & 53.82 &  0.01 &  85 & 0.657 &     1.55 &   11.1 & 190.1 & 0.130 & 0.00\\
007257373 & 5311 & 13.424 & 0.745 & 10.46686 & 4955.658505 & 89.72 &  0.00 &  89 & 0.113 &     1.02 &  0.922 & 0.8700 & 0.001 & 0.00\\
007284688 & 4191 & 11.234 & 0.088 & 0.6461003 & 5002.783432 & 64.97 &  0.00 & 124 & 0.552 &     1.22 &   2.29 & 6.374 & 0.055 & 0.00\\
007624297 & 5135 & 14.928 & 0.222 & 18.01846 & 4981.666593 & 88.97 &  0.05 &  76 & 0.042 &   0.0661 &   1.73 & 0.1989 & 0.000 & 0.74\\
007670617 & 4876 & 15.517 & 0.450 & 24.70190 & 4969.146845 & 89.85 &  0.28 & 308 & 0.033 &    0.201 &  0.711 & 0.1018 & 0.000 & 0.35\\
007671594 & 3717 & 15.815 & 0.165 & 1.410329 & 4965.398972 & 84.54 &  0.00 & 302 & 0.138 &    0.612 &   1.19 & 0.8632 & 0.039 & 0.00\\
007691527 & 5354 & 15.431 & 0.463 & 4.800056 & 5002.382912 & 87.47 &  0.10 & 130 & 0.115 &     1.07 &   1.21 & 1.573 & 0.073 & 0.00\\
007749318 & 5211 & 14.528 & 0.341 & 2.371784 & 5003.689256 & 80.47 &  0.18 & 270 & 0.246 &   0.0569 &   3.32 & 0.6277 & 0.078 & 0.00\\
007769072 & 4858 & 13.886 & 0.003 & 0.6088726 & 5002.799849 & 57.25 &  0.00 &  72 & 0.583 &     1.18 &   21.8 & 559.9 & 0.016 & 0.00\\
007798259 & 4619 & 15.726 & 0.411 & 1.734306 & 5005.720952 & 84.29 &  0.03 & 270 & 0.200 &    0.310 &   1.22 & 0.4600 & 0.070 & 0.00\\
007830321 & 5347 & 15.476 & 0.008 & 2.027248 & 5003.200747 & 73.45 &  0.00 & 242 & 0.301 &    0.672 &   6.41 & 27.63 & 0.010 & 0.00\\
007842610 & 5375 & 15.289 & 0.021 & 1.943760 & 5001.841852 & 76.41 &  0.00 & 270 & 0.271 &     1.32 &   4.25 & 23.80 & 0.027 & 0.00\\
007846730 & 5476 & 12.956 & 0.381 & 11.02825 & 4969.966521 & 88.40 &  0.02 & 151 & 0.101 &    0.599 &   1.88 & 2.126 & 0.000 & 0.12\\
007885570 & 5398 & 11.679 & 0.223 & 1.729021 & 5001.851398 & 74.44 &  0.05 &  85 & 0.454 &    0.747 &   3.41 & 8.661 & 0.237 & 0.00\\
007947631 & 4823 & 15.179 & 0.022 & 2.516590 & 4987.092316 & 79.51 &  0.01 &  89 & 0.209 &     1.13 &   3.03 & 10.33 & 0.008 & 0.00\\
007955301 & 4821 & 12.672 & 0.007 & 15.30817 & 4960.464666 & 86.04 &  0.01 & 174 & 0.078 &     1.42 &   8.19 & 95.41 & 0.001 & 0.00\\
007987749 & 5349 & 14.461 & 0.095 & 17.03109 & 4978.541107 & 88.17 &  0.14 & 187 & 0.046 &    0.398 &  0.495 & 0.09742 & 0.000 & 0.00\\
008075618 & 5288 & 15.674 & 0.119 & 17.56154 & 4970.923092 & 88.76 &  0.02 &  90 & 0.031 &     1.11 &  0.961 & 1.027 & 0.000 & 0.00\\
008075755 & 4075 & 13.407 & 0.009 & 0.4962355 & 4964.752415 & 66.36 &  0.01 &  94 & 0.471 &    0.516 & 0.0939 & 0.004556 & 0.000 & 0.00\\
008076905 & 4214 & 15.613 & 0.011 & 0.4180906 & 5003.192377 & 51.07 &  0.01 & 279 & 0.715 &     1.67 &   11.8 & 231.1 & 0.083 & 0.00\\
008094140 & 4200 & 15.159 & 0.606 & 0.7064196 & 4973.624150 & 86.37 &  0.01 & 265 & 0.326 &    0.274 &  0.628 & 0.1079 & 0.038 & 0.00\\
008095110 & 5370 & 13.171 & 0.036 & 4.206510 & 4974.280246 & 76.85 &  0.02 &  91 & 0.300 &     1.56 &   4.03 & 25.26 & 0.035 & 0.00\\
008097825 & 5329 & 13.283 & 0.261 & 2.937050 & 4966.633044 & 78.40 &  0.00 & 286 & 0.343 &    0.645 &   1.50 & 1.449 & 0.028 & 0.00\\
008143170 & 4957 & 12.850 & 0.058 & 28.78627 & 4970.110463 & 85.83 &  0.20 & 255 & 0.103 &    0.269 &   6.55 & 11.55 & 0.001 & 0.00\\
008145789 & 4829 & 15.314 & 0.027 & 1.670636 & 5000.039740 & 75.30 &  0.01 & 272 & 0.311 &    0.917 &   5.66 & 29.34 & 0.016 & 0.00\\
008181016 & 5179 & 14.677 & 0.008 & 0.7090293 & 4965.187959 & 75.06 &  0.20 & 270 & 0.434 &    0.197 &   1.37 & 0.3690 & 0.000 & 0.98\\
008210721 & 5412 & 14.274 & 0.080 & 22.67256 & 4971.153407 & 87.76 &  0.29 &  64 & 0.057 &    0.234 &   3.46 & 2.809 & 0.000 & 0.75\\
008231877 & 4956 & 14.932 & 0.024 & 2.615519 & 4975.237630 & 83.45 &  0.15 &  90 & 0.162 &     1.87 &  0.511 & 0.4880 & 0.000 & 0.79\\
008279765 & 5464 & 15.235 & 0.051 & 2.757746 & 4965.474866 & 83.85 &  0.01 & 282 & 0.169 &   0.0539 &  0.208 & 0.002322 & 0.003 & 0.00\\
008288719 & 5090 & 13.276 & 0.043 & 1.510074 & 4972.744897 & 75.00 &  0.01 &  95 & 0.312 &     1.04 &   2.58 & 6.916 & 0.007 & 0.00\\
008296467 & 5316 & 15.177 & 0.987 & 10.30327 & 4970.167785 & 89.99 &  0.26 & 320 & 0.069 &    0.623 &   1.06 & 0.7061 & 0.006 & 0.00\\
008358008 & 5020 & 14.674 & 0.012 & 10.06506 & 4968.250048 & 89.42 &  0.06 &  79 & 0.054 &    0.533 &   1.51 & 1.216 & 0.000 & 0.98\\
008364119 & 5443 & 12.408 & 0.462 & 7.735857 & 4970.986699 & 88.29 &  0.03 &  44 & 0.093 &    0.897 &  0.849 & 0.6471 & 0.003 & 0.00\\
008379547 & 4861 & 13.373 & 0.174 & 6.041994 & 4959.163251 & 81.83 &  0.35 & 270 & 0.222 &   0.0668 &   7.25 & 3.510 & 0.083 & 0.00\\
008397675 & 5462 & 13.501 & 0.002 & 0.5532564 & 5001.856348 & 83.35 &  0.16 &  91 & 0.230 &    0.933 &   2.95 & 8.128 & 0.092 & 0.99\\
008411947 & 5086 & 15.300 & 0.860 & 1.797734 & 5003.785574 & 88.27 &  0.02 &  96 & 0.265 &    0.607 &   1.12 & 0.7612 & 0.050 & 0.00\\
008444552 & 5388 & 13.643 & 0.083 & 1.178041 & 4964.597354 & 77.49 &  0.11 &  90 & 0.323 &     2.28 &   1.96 & 8.734 & 0.021 & 0.00\\
008453324 & 4733 & 11.516 & 0.010 & 2.524694 & 5001.646619 & 72.45 &  0.00 &  82 & 0.341 &     1.42 &   5.66 & 45.59 & 0.016 & 0.00\\
008543278 & 4950 & 14.608 & 0.073 & 7.549631 & 4998.208506 & 88.40 &  0.12 & 276 & 0.052 &    0.187 &  0.451 & 0.03811 & 0.000 & 0.57\\
008559863 & 5154 & 12.723 & 0.055 & 22.46892 & 4953.814854 & 88.21 &  0.04 & 217 & 0.054 &    0.720 &  0.437 & 0.1377 & 0.002 & 0.56\\
008574270 & 5061 & 15.166 & 0.031 & 15.11963 & 4972.699012 & 87.27 &  0.29 & 321 & 0.059 &   0.0360 &  0.955 & 0.03285 & 0.000 & 0.00\\
008580438 & 5307 & 14.502 & 0.152 & 6.495852 & 5000.947823 & 90.00 &  0.01 &  80 & 0.108 &   0.0314 &  0.315 & 0.003107 & 0.004 & 0.00\\
008581232 & 4314 & 15.381 & 0.037 & 4.012679 & 5003.764787 & 87.32 &  0.33 & 133 & 0.086 &    0.279 &  0.138 & 0.005326 & 0.000 & 0.00\\
008616873 & 5486 & 15.237 & 0.015 & 0.5760785 & 5002.245893 & 81.58 &  0.14 &  86 & 0.437 &    0.140 &   2.17 & 0.6544 & 0.089 & 0.98\\
008655458 & 5210 & 14.585 & 0.008 & 1.594193 & 5002.299400 & 88.86 &  0.03 &  78 & 0.397 &    0.183 &  0.472 & 0.04073 & 0.000 & 0.98\\
008718273 & 4577 & 10.565 & 0.006 & 6.958070 & 4997.699036 & 89.55 &  0.03 & 269 & 0.050 &    0.740 & 0.0474 & 0.001664 & 0.000 & 0.00\\
008719897 & 4905 & 12.392 & 0.262 & 3.151596 & 4955.232895 & 80.22 &  0.02 &  90 & 0.315 &     1.02 &   1.13 & 1.291 & 0.015 & 0.00\\
008841616 & 4550 & 12.833 & 0.133 & 1.679564 & 4966.238497 & 61.97 &  0.02 &  71 & 0.650 &   0.0436 &  0.639 & 0.01781 & 0.022 & 0.00\\
008846978 & 5191 & 13.371 & 0.225 & 1.379281 & 4970.036969 & 64.68 &  0.06 & 312 & 0.556 &   0.0263 &   3.72 & 0.3635 & 0.196 & 0.58\\
008848104 & 5447 & 12.372 & 0.041 & 0.8248496 & 4972.049484 & 61.47 &  0.01 & 100 & 0.538 &    0.307 &   6.94 & 14.79 & 0.029 & 0.00\\
008906676 & 5249 & 12.121 & 0.167 & 8.209521 & 4967.062429 & 88.28 &  0.03 &  89 & 0.075 &    0.118 &   1.71 & 0.3441 & 0.000 & 0.78\\
008971432 & 5057 & 15.487 & 0.063 & 0.6243809 & 5001.634891 & 65.63 &  0.03 &  89 & 0.533 &   0.0555 &  0.299 & 0.004965 & 0.001 & 0.00\\
009001468 & 4949 & 15.200 & 0.339 & 17.32833 & 4975.727756 & 89.41 &  0.52 & 239 & 0.043 &    0.216 &   1.54 & 0.5119 & 0.000 & 0.57\\
009029486 & 5368 & 13.630 & 0.342 & 6.277180 & 4965.329729 & 89.54 &  0.00 & 279 & 0.094 &    0.903 &  0.981 & 0.8686 & 0.007 & 0.45\\
009098810 & 5126 & 13.448 & 0.443 & 8.258238 & 4972.758295 & 88.41 &  0.16 &  87 & 0.079 &    0.957 &  0.832 & 0.6620 & 0.006 & 0.00\\
009210828 & 4893 & 13.221 & 0.205 & 1.656351 & 4977.002807 & 80.24 &  0.02 &  90 & 0.269 &    0.756 &   1.16 & 1.016 & 0.004 & 0.00\\
009266285 & 4184 & 14.072 & 0.072 & 5.613843 & 4965.571978 & 82.83 &  0.27 &  91 & 0.182 &     3.49 &   1.65 & 9.496 & 0.014 & 0.00\\
009284741 & 5085 & 14.807 & 0.516 & 20.72910 & 4974.226975 & 89.42 &  0.37 &  42 & 0.041 &     1.15 &   1.05 & 1.265 & 0.006 & 0.00\\
009291629 & 4629 & 13.957 & 0.168 & 20.69085 & 4966.893246 & 84.62 &  0.13 & 271 & 0.214 &    0.346 &   4.57 & 7.243 & 0.151 & 0.00\\
009328852 & 4338 & 15.330 & 0.550 & 0.6458239 & 5008.159243 & 84.39 &  0.04 &  82 & 0.410 &   0.0705 &  0.519 & 0.01901 & 0.047 & 0.00\\
009334490 & 5105 & 15.695 & 0.017 & 18.84520 & 4982.944981 & 89.46 &  0.02 & 129 & 0.038 &    0.797 &  0.547 & 0.2389 & 0.000 & 0.96\\
009346655 & 4183 & 14.299 & 0.144 & 0.8716196 & 4965.119502 & 81.56 &  0.50 &  90 & 0.262 &     6.12 &   1.50 & 13.73 & 0.380 & 0.00\\
009412462 & 5350 & 14.846 & 0.518 & 10.18653 & 4965.527836 & 87.39 &  0.03 & 230 & 0.143 &    0.747 &   1.19 & 1.065 & 0.016 & 0.00\\
009418994 & 5053 & 13.396 & 0.021 & 32.00590 & 4969.494447 & 89.65 &  0.23 &  54 & 0.024 &    0.136 &  0.668 & 0.06079 & 0.000 & 0.96\\
009474485 & 4469 & 14.884 & 0.668 & 1.025164 & 4965.292428 & 87.18 &  0.00 &   5 & 0.329 &    0.864 &   1.02 & 0.9024 & 0.033 & 0.00\\
009574614 & 5276 & 15.933 & 0.011 & 1.964342 & 5002.018849 & 78.41 &  0.01 & 238 & 0.220 &     1.01 &   4.37 & 19.23 & 0.001 & 0.00\\
009597095 & 5331 & 15.945 & 0.073 & 2.745608 & 5003.145923 & 81.81 &  0.02 & 269 & 0.203 &    0.121 &  0.377 & 0.01723 & 0.001 & 0.00\\
009632895 & 5425 & 13.552 & 0.097 & 27.32202 & 4965.424356 & 87.89 &  0.03 & 257 & 0.046 &   0.0428 &   11.9 & 6.042 & 0.000 & 0.46\\
009639265 & 5004 & 15.575 & 0.370 & 0.5063492 & 4964.814722 & 75.19 &  0.02 & 275 & 0.520 &    0.888 &  0.765 & 0.5194 & 0.068 & 0.00\\
009658832 & 4545 & 13.638 & 0.029 & 0.4568510 & 5002.649683 & 56.54 &  0.01 &  80 & 0.683 &    0.335 &  0.163 & 0.008909 & 0.002 & 0.00\\
009665503 & 5141 & 15.217 & 0.656 & 11.56806 & 4970.339984 & 89.46 &  0.28 & 330 & 0.063 &    0.383 &  0.694 & 0.1844 & 0.000 & 0.00\\
009714358 & 4825 & 14.998 & 0.283 & 6.479757 & 4999.785837 & 86.63 &  0.26 & 272 & 0.089 &   0.0295 &   3.94 & 0.4572 & 0.000 & 0.13\\
009761199 & 4060 & 15.692 & 0.014 & 1.383998 & 4964.727100 & 74.47 &  0.00 & 128 & 0.289 &     1.14 &   3.00 & 10.29 & 0.009 & 0.00\\
009762519 & 5435 & 13.711 & 0.152 & 7.515083 & 4971.079973 & 86.05 &  0.18 & 282 & 0.095 &   0.0521 &   2.31 & 0.2778 & 0.007 & 0.00\\
009837578 & 5359 & 15.726 & 0.698 & 20.73369 & 4965.845828 & 89.44 &  0.16 &  87 & 0.048 &    0.681 &   1.26 & 1.083 & 0.000 & 0.00\\
009851126 & 4164 & 13.183 & 0.097 & 8.480306 & 4968.853813 & 89.89 &  0.21 &  18 & 0.129 &    0.137 &  0.257 & 0.009052 & 0.000 & 0.03\\
009912977 & 5158 & 13.726 & 0.442 & 1.887885 & 5002.578442 & 79.82 &  0.01 &  91 & 0.473 &     1.01 &   1.17 & 1.380 & 0.002 & 0.00\\
009913798 & 4659 & 14.945 & 0.218 & 2.143443 & 5002.935126 & 83.44 &  0.01 & 271 & 0.269 &    0.173 &  0.409 & 0.02886 & 0.001 & 0.00\\
009934208 & 4258 & 15.507 & 0.166 & 9.058852 & 4970.337139 & 85.96 &  0.14 &  51 & 0.091 &    0.193 &   5.49 & 5.819 & 0.002 & 0.00\\
009944201 & 4737 & 15.069 & 0.032 & 0.7215318 & 5002.227862 & 86.33 &  0.07 &  90 & 0.307 &    0.215 &   1.71 & 0.6296 & 0.091 & 0.96\\
009944421 & 5304 & 15.137 & 0.349 & 7.095304 & 4968.370748 & 86.24 &  0.26 &  69 & 0.100 &    0.637 &   1.18 & 0.8853 & 0.024 & 0.00\\
010014830 & 4324 & 14.827 & 0.897 & 3.030715 & 5003.492462 & 85.95 &  0.00 & 185 & 0.549 &    0.279 &   1.80 & 0.9018 & 0.058 & 0.00\\
010026457 & 5222 & 15.390 & 0.089 & 9.934463 & 5005.612121 & 89.71 &  0.13 & 320 & 0.109 &    0.990 &  0.970 & 0.9321 & 0.000 & 0.84\\
010090246 & 5442 & 13.567 & 0.171 & 2.285607 & 5003.110556 & 56.54 &  0.03 &  89 & 0.695 &    0.560 &   2.32 & 3.022 & 0.038 & 0.00\\
010095484 & 5486 & 14.382 & 0.008 & 0.6777383 & 5002.579926 & 29.31 &  0.01 &  97 & 0.906 &    0.536 &   16.2 & 140.0 & 0.060 & 0.00\\
010129482 & 4558 & 15.994 & 0.268 & 0.8462873 & 5002.429877 & 80.44 &  0.00 &  76 & 0.326 &    0.190 &  0.615 & 0.07199 & 0.011 & 0.00\\
010189523 & 5002 & 15.856 & 0.117 & 1.013960 & 5002.929802 & 74.89 &  0.05 &  89 & 0.325 &     1.29 &  0.530 & 0.3627 & 0.039 & 0.00\\
010215422 & 5427 & 14.608 & 0.444 & 24.39590 & 4987.127475 & 89.06 &  0.29 &   9 & 0.045 &    0.207 &   1.43 & 0.4202 & 0.000 & 0.44\\
010264202 & 5207 & 15.777 & 0.144 & 1.035161 & 5002.815429 & 75.46 &  0.01 & 272 & 0.373 &    0.672 &  0.955 & 0.6130 & 0.004 & 0.00\\
010292465 & 5258 & 14.956 & 0.152 & 1.353325 & 5002.832143 & 73.64 &  0.00 &   0 & 0.348 &   0.0940 &   4.64 & 2.023 & 0.068 & 0.34\\
010330495 & 5132 & 14.724 & 0.075 & 18.06030 & 4971.608312 & 85.34 &  0.14 & 254 & 0.117 &    0.150 &   10.5 & 16.46 & 0.003 & 0.00\\
010346522 & 5286 & 14.404 & 1.204 & 3.988565 & 5001.472319 & 85.90 &  0.01 &  63 & 0.585 &    0.188 &  0.814 & 0.1246 & 0.008 & 0.00\\
010491544 & 4835 & 13.436 & 0.031 & 22.77214 & 4973.487861 & 86.42 &  0.54 &  56 & 0.089 &     15.6 &   1.34 & 27.91 & 0.017 & 0.00\\
010592163 & 5482 & 15.095 & 0.098 & 14.76289 & 4966.772333 & 88.98 &  0.32 & 329 & 0.057 &    0.486 &  0.628 & 0.1916 & 0.000 & 0.73\\
010613718 & 5080 & 12.735 & 0.010 & 1.175802 & 4966.821353 & 74.39 &  0.01 & 269 & 0.307 &    0.966 &   8.52 & 70.13 & 0.006 & 0.00\\
010711646 & 4339 & 15.787 & 0.204 & 0.7376206 & 4997.149560 & 78.07 &  0.08 & 270 & 0.343 &    0.134 &   1.16 & 0.1808 & 0.046 & 0.42\\
010753734 & 5446 & 13.564 & 0.725 & 19.40624 & 4982.807297 & 89.74 &  0.52 &  18 & 0.051 &    0.823 &  0.849 & 0.5942 & 0.005 & 0.00\\
010794242 & 5459 & 14.170 & 0.269 & 7.143779 & 4970.803174 & 89.19 &  0.08 & 247 & 0.102 &   0.0887 &  0.425 & 0.01604 & 0.010 & 0.00\\
010794405 & 5479 & 14.713 & 0.005 & 0.9522659 & 4979.620590 & 40.08 &  0.00 & 237 & 0.821 &     3.37 &   12.9 & 557.6 & 0.192 & 0.00\\
010809677 & 4995 & 13.942 & 0.008 & 7.042849 & 4970.731749 & 80.78 &  0.00 & 271 & 0.172 &    0.968 &   3.55 & 12.20 & 0.000 & 0.00\\
010936427 & 5082 & 14.419 & 0.756 & 14.35935 & 4971.843223 & 88.64 &  0.02 &  84 & 0.116 &    0.401 &   1.72 & 1.192 & 0.042 & 0.00\\
010979716 & 3932 & 15.774 & 0.125 & 10.68394 & 4967.091349 & 88.05 &  0.15 & 278 & 0.054 &    0.307 &  0.718 & 0.1586 & 0.006 & 0.00\\
010991989 & 5021 & 10.282 & 0.012 & 0.9744771 & 4965.368901 & 85.97 &  0.09 & 268 & 0.358 &    0.568 &  0.597 & 0.2028 & 0.000 & 0.97\\
010992733 & 5274 & 15.124 & 0.728 & 18.52628 & 4977.193722 & 89.99 &  0.38 &  26 & 0.055 &    0.704 &  0.777 & 0.4255 & 0.006 & 0.00\\
011124509 & 5417 & 14.735 & 0.018 & 8.893240 & 4968.729831 & 85.98 &  0.01 & 269 & 0.080 &    0.393 &   1.34 & 0.7016 & 0.001 & 0.74\\
011134079 & 5201 & 14.864 & 0.232 & 1.260506 & 4965.061893 & 73.28 &  0.01 & 265 & 0.368 &    0.134 &   4.48 & 2.688 & 0.043 & 0.00\\
011147460 & 4855 & 13.912 & 0.009 & 4.107429 & 4965.943730 & 73.69 &  0.00 &  71 & 0.316 &    0.548 &  0.139 & 0.01052 & 0.000 & 0.00\\
011232745 & 5204 & 15.973 & 0.056 & 9.633799 & 4970.918869 & 89.91 &  0.03 &  90 & 0.037 &    0.574 &   1.67 & 1.609 & 0.058 & 0.91\\
011233911 & 5193 & 14.742 & 0.285 & 4.959761 & 4969.120822 & 85.16 &  0.01 &  90 & 0.171 &    0.789 &   2.19 & 3.776 & 0.103 & 0.00\\
011235323 & 5071 & 13.486 & 0.496 & 19.67035 & 4965.522562 & 89.28 &  0.06 & 270 & 0.137 &    0.458 &  0.598 & 0.1636 & 0.006 & 0.00\\
011287726 & 5167 & 14.176 & 0.159 & 4.736985 & 4970.064587 & 78.94 &  0.07 & 271 & 0.258 &    0.175 &   1.53 & 0.4081 & 0.005 & 0.00\\
011350389 & 5124 & 15.724 & 0.037 & 1.512708 & 4969.562269 & 83.00 &  0.05 & 270 & 0.172 &    0.213 &   8.74 & 16.24 & 0.019 & 0.59\\
011391181 & 5218 & 15.257 & 0.276 & 8.617414 & 4972.068740 & 87.23 &  0.20 &  30 & 0.081 &    0.923 &   1.15 & 1.228 & 0.024 & 0.00\\
011391667 & 5394 & 12.923 & 0.011 & 1.083646 & 4954.131479 & 74.63 &  0.00 & 292 & 0.320 &    0.497 &  0.116 & 0.006724 & 0.000 & 0.00\\
011455795 & 4477 & 15.414 & 0.072 & 1.057351 & 4964.791193 & 81.87 &  0.00 &   0 & 0.221 &   0.0229 &   7.44 & 1.271 & 0.108 & 0.90\\
011546211 & 3682 & 15.155 & 0.083 & 2.194447 & 4966.712688 & 85.75 &  0.67 &  90 & 0.133 &     88.2 &  0.454 & 18.17 & 0.756 & 0.00\\
011671660 & 4867 & 13.350 & 0.089 & 8.702917 & 4956.587112 & 72.47 &  0.03 & 271 & 0.368 &    0.718 &   5.29 & 20.08 & 0.233 & 0.00\\
011768970 & 5038 & 12.658 & 0.012 & 15.54223 & 4959.412961 & 87.15 &  0.87 &  82 & 0.089 &     28.7 &   1.42 & 57.58 & 0.000 & 0.42\\
011858541 & 5375 & 14.215 & 0.045 & 5.674410 & 4968.755298 & 81.16 &  0.06 & 355 & 0.180 &    0.421 &   4.60 & 8.922 & 0.006 & 0.00\\
011968514 & 4940 & 11.449 & 0.005 & 2.073289 & 5002.408007 & 73.94 &  0.01 & 259 & 0.304 &     1.32 &   16.0 & 338.4 & 0.037 & 0.00\\
011975363 & 5482 & 15.409 & 0.578 & 3.518364 & 4967.411791 & 88.02 &  0.01 &  89 & 0.180 &    0.916 &   1.08 & 1.061 & 0.018 & 0.00\\
012004679 & 5432 & 13.231 & 0.833 & 5.042429 & 4955.770424 & 89.85 &  0.01 &  77 & 0.115 &    0.803 &  0.997 & 0.7988 & 0.008 & 0.00\\
012004834 & 3576 & 14.718 & 0.333 & 0.2623169 & 4964.398367 & 72.47 &  0.06 & 269 & 0.517 &     1.04 &   1.56 & 2.532 & 0.018 & 0.00\\
012105785 & 5349 & 13.032 & 0.021 & 31.95107 & 4975.706638 & 87.34 &  0.18 & 114 & 0.056 &     2.81 &  0.507 & 0.7219 & 0.000 & 0.66\\
012351927 & 4641 & 15.520 & 0.086 & 10.11594 & 4972.982326 & 85.41 &  0.01 & 119 & 0.087 &   0.0699 &   7.19 & 3.613 & 0.000 & 0.00\\
012356914 & 5368 & 15.529 & 0.621 & 27.30710 & 4976.502419 & 89.94 &  0.44 &  74 & 0.035 &    0.157 &  0.636 & 0.06344 & 0.001 & 0.00\\
012365000 & 5080 & 13.573 & 0.028 & 1.262660 & 4986.658328 & 77.85 &  0.11 & 269 & 0.283 &    0.148 &  0.912 & 0.1230 & 0.025 & 0.70\\
012367017 & 5004 & 14.730 & 0.008 & 1.222133 & 4970.925867 & 59.96 &  0.00 & 290 & 0.540 &    0.806 &   18.6 & 277.5 & 0.026 & 0.00\\
012367310 & 4965 & 13.835 & 0.044 & 8.627137 & 4972.995164 & 81.33 &  0.02 & 314 & 0.167 &    0.147 &   5.54 & 4.518 & 0.017 & 0.00\\
012400729 & 4949 & 15.227 & 0.149 & 0.9317268 & 4965.479857 & 72.86 &  0.13 & 196 & 0.366 & 0.000549 &   8.92 & 0.04368 & 0.219 & 0.88\\
012418816 & 4583 & 12.402 & 0.581 & 1.521896 & 4965.395396 & 87.12 &  0.01 &  88 & 0.248 &     1.00 &   1.04 & 1.082 & 0.038 & 0.00\\
012470530 & 4725 & 15.300 & 0.658 & 8.207057 & 4968.824442 & 88.44 &  0.38 & 347 & 0.072 &    0.302 &  0.979 & 0.2889 & 0.000 & 0.03\\
012557713 & 4594 & 14.853 & 0.068 & 7.214603 & 4965.498124 & 87.06 &  0.43 &  92 & 0.077 &     12.6 &  0.412 & 2.134 & 0.000 & 0.17\\
012599700 & 3887 & 15.784 & 0.120 & 1.017821 & 4968.317001 & 87.78 &  0.03 & 268 & 0.136 &    0.433 &   1.24 & 0.6686 & 0.363 & 0.87\\
012645761 & 4844 & 13.368 & 0.018 & 5.419663 & 4958.954807 & 81.82 &  0.34 &  90 & 0.212 &     23.8 &   1.91 & 86.59 & 0.185 & 0.00\\

%% file: tab3.tex
002162994 & 4.102 & 5410 & 5593 & 5038 & 0.96 & 0.86 & 1.39 & 1.24 \\
002437452 & 14.47 & 5398 & 5591 & 4647 & 0.96 & 0.79 & 1.40 & 1.13 \\
002719873 & 17.28 & 5086 & 5246 & 4382 & 0.90 & 0.73 & 1.08 & 0.86 \\
002852560 & 11.96 & 5381 & 5385 & 5378 & 0.93 & 0.92 & 1.06 & 1.06 \\
003003991 & 7.245 & 5366 & 5554 & 4598 & 0.96 & 0.78 & 0.83 & 0.67 \\
003102024 & 13.78 & 5117 & 5160 & 5069 & 0.89 & 0.87 & 0.79 & 0.78 \\
003241344 & 3.913 & 5422 & 5461 & 3688 & 0.94 & 0.52 & 0.94 & 0.49 \\
003241619 & 1.703 & 5165 & 5344 & 4622 & 0.92 & 0.79 & 1.04 & 0.88 \\
003458919 & 0.8920 & 5063 & 5206 & 4254 & 0.89 & 0.70 & 1.08 & 0.83 \\
003730067 & 0.2941 & 4099 & 4158 & 4010 & 0.68 & 0.64 & 0.62 & 0.58 \\
003848919 & 1.047 & 5226 & 5238 & 5214 & 0.90 & 0.90 & 1.10 & 1.10 \\
004049124 & 4.804 & 5349 & 5501 & 4347 & 0.95 & 0.73 & 1.30 & 0.97 \\
004077442 & 0.6929 & 4523 & 4643 & 4094 & 0.79 & 0.66 & 1.03 & 0.84 \\
004346875 & 4.694 & 5339 & 5367 & 3599 & 0.92 & 0.46 & 1.21 & 0.56 \\
004352168 & 10.64 & 5115 & 5281 & 4744 & 0.91 & 0.81 & 0.93 & 0.82 \\
004484356 & 1.144 & 5080 & 5250 & 4636 & 0.90 & 0.79 & 0.94 & 0.81 \\
004540632 & 31.01 & 4818 & 4953 & 4190 & 0.85 & 0.69 & 0.80 & 0.63 \\
004633434 & 22.27 & 4902 & 5041 & 4219 & 0.86 & 0.69 & 0.67 & 0.52 \\
004678171 & 15.29 & 4240 & 4331 & 4048 & 0.72 & 0.65 & 0.68 & 0.60 \\
004773155 & 25.71 & 5447 & 5448 & 5447 & 0.94 & 0.94 & 0.96 & 0.96 \\
004908495 & 3.121 & 4731 & 4791 & 4655 & 0.82 & 0.79 & 0.82 & 0.79 \\
005036538 & 2.122 & 4199 & 4236 & 4155 & 0.70 & 0.68 & 0.71 & 0.69 \\
005080652 & 4.144 & 5344 & 5536 & 4858 & 0.95 & 0.83 & 1.17 & 1.01 \\
005300878 & 1.279 & 4631 & 4667 & 4590 & 0.80 & 0.78 & 0.87 & 0.85 \\
005597970 & 6.717 & 5179 & 5284 & 4060 & 0.91 & 0.65 & 1.08 & 0.74 \\
005731312 & 7.946 & 4658 & 4701 & 3583 & 0.80 & 0.45 & 0.68 & 0.36 \\
005781192 & 9.460 & 5372 & 5546 & 4482 & 0.95 & 0.76 & 0.97 & 0.75 \\
005802470 & 3.792 & 5418 & 5620 & 4859 & 0.97 & 0.83 & 1.00 & 0.86 \\
005871918 & 12.64 & 4021 & 4052 & 3983 & 0.65 & 0.63 & 0.79 & 0.76 \\
006029130 & 12.59 & 5160 & 5201 & 5114 & 0.89 & 0.88 & 0.88 & 0.86 \\
006131659 & 17.53 & 4870 & 4970 & 3972 & 0.85 & 0.63 & 0.84 & 0.59 \\
006449552 & 20.15 & 5357 & 5537 & 4532 & 0.95 & 0.77 & 0.93 & 0.74 \\
006464285 & 0.8436 & 5061 & 5159 & 4923 & 0.89 & 0.84 & 1.09 & 1.03 \\
006466939 & 2.286 & 4920 & 4925 & 4916 & 0.84 & 0.84 & 0.87 & 0.86 \\
006591789 & 5.088 & 5410 & 5560 & 4342 & 0.96 & 0.73 & 1.09 & 0.81 \\
006697716 & 1.443 & 4898 & 5036 & 4215 & 0.86 & 0.69 & 0.91 & 0.71 \\
006706287 & 2.535 & 5182 & 5327 & 4931 & 0.91 & 0.85 & 0.96 & 0.89 \\
006778050 & 0.9458 & 5091 & 5223 & 4872 & 0.90 & 0.83 & 0.98 & 0.91 \\
006841577 & 15.54 & 5478 & 5676 & 4676 & 0.98 & 0.80 & 1.03 & 0.83 \\
006863840 & 3.853 & 5024 & 5050 & 4997 & 0.87 & 0.86 & 0.88 & 0.87 \\
007119757 & 0.7429 & 5072 & 5242 & 4607 & 0.90 & 0.78 & 1.20 & 1.03 \\
007125636 & 6.491 & 4358 & 4417 & 4277 & 0.74 & 0.71 & 0.69 & 0.66 \\
007128918 & 7.119 & 5386 & 5574 & 4968 & 0.96 & 0.85 & 0.86 & 0.76 \\
007129465 & 5.492 & 5182 & 5269 & 5069 & 0.90 & 0.87 & 0.87 & 0.83 \\
007200102 & 14.67 & 5207 & 5390 & 4643 & 0.93 & 0.79 & 0.88 & 0.74 \\
007624297 & 18.02 & 5135 & 5291 & 4352 & 0.91 & 0.73 & 0.81 & 0.63 \\
007670617 & 24.70 & 4876 & 4971 & 3945 & 0.85 & 0.62 & 0.80 & 0.56 \\
007671594 & 1.410 & 3717 & 3773 & 3597 & 0.56 & 0.46 & 0.40 & 0.32 \\
007691527 & 4.800 & 5354 & 5492 & 5138 & 0.94 & 0.88 & 0.87 & 0.81 \\
007749318 & 2.372 & 5211 & 5347 & 4991 & 0.92 & 0.86 & 1.16 & 1.07 \\
007798259 & 1.734 & 4619 & 4735 & 4386 & 0.81 & 0.74 & 0.74 & 0.67 \\
007846730 & 11.03 & 5476 & 5667 & 5079 & 0.98 & 0.87 & 1.37 & 1.22 \\
008075618 & 17.56 & 5288 & 5301 & 5275 & 0.91 & 0.91 & 0.55 & 0.54 \\
008094140 & 0.7064 & 4200 & 4266 & 3598 & 0.71 & 0.46 & 0.70 & 0.44 \\
008296467 & 10.30 & 5316 & 5427 & 5159 & 0.93 & 0.89 & 0.86 & 0.82 \\
008364119 & 7.736 & 5443 & 5581 & 5232 & 0.96 & 0.90 & 0.97 & 0.91 \\
008411947 & 1.798 & 5086 & 5168 & 4980 & 0.89 & 0.85 & 1.01 & 0.97 \\
008580438 & 6.496 & 5307 & 5314 & 3348 & 0.91 & 0.23 & 1.31 & 0.35 \\
008906676 & 8.210 & 5249 & 5436 & 4709 & 0.93 & 0.80 & 0.84 & 0.71 \\
009001468 & 17.33 & 4949 & 5089 & 4676 & 0.87 & 0.80 & 0.76 & 0.69 \\
009029486 & 6.277 & 5368 & 5421 & 5309 & 0.93 & 0.91 & 0.83 & 0.81 \\
009098810 & 8.258 & 5126 & 5240 & 4956 & 0.90 & 0.85 & 0.84 & 0.79 \\
009210828 & 1.656 & 4893 & 4898 & 4888 & 0.84 & 0.84 & 0.94 & 0.94 \\
009284741 & 20.73 & 5085 & 5156 & 4998 & 0.88 & 0.86 & 0.80 & 0.78 \\
009328852 & 0.6458 & 4338 & 4357 & 3375 & 0.73 & 0.25 & 0.94 & 0.34 \\
009346655 & 0.8716 & 4183 & 4232 & 3512 & 0.70 & 0.39 & 0.67 & 0.37 \\
009474485 & 1.025 & 4469 & 4492 & 4444 & 0.76 & 0.75 & 0.81 & 0.80 \\
009639265 & 0.5063 & 5004 & 5147 & 4730 & 0.88 & 0.81 & 0.87 & 0.79 \\
009665503 & 11.57 & 5141 & 5293 & 4321 & 0.91 & 0.72 & 0.90 & 0.69 \\
009714358 & 6.480 & 4825 & 4964 & 4522 & 0.85 & 0.77 & 0.81 & 0.72 \\
009762519 & 7.515 & 5435 & 5528 & 4050 & 0.95 & 0.65 & 0.95 & 0.62 \\
009837578 & 20.73 & 5359 & 5390 & 5327 & 0.93 & 0.91 & 0.95 & 0.94 \\
009934208 & 9.059 & 4258 & 4347 & 3743 & 0.73 & 0.55 & 1.04 & 0.76 \\
009944421 & 7.095 & 5304 & 5348 & 5255 & 0.92 & 0.90 & 0.96 & 0.94 \\
010129482 & 0.8463 & 4558 & 4622 & 3669 & 0.79 & 0.51 & 0.83 & 0.51 \\
010189523 & 1.014 & 5002 & 5143 & 4239 & 0.88 & 0.70 & 0.91 & 0.70 \\
010215422 & 24.40 & 5427 & 5625 & 4944 & 0.97 & 0.85 & 1.04 & 0.91 \\
010264202 & 1.035 & 5207 & 5347 & 4971 & 0.92 & 0.85 & 1.01 & 0.93 \\
010292465 & 1.353 & 5258 & 5417 & 4965 & 0.93 & 0.85 & 1.15 & 1.05 \\
010711646 & 0.7376 & 4339 & 4440 & 3877 & 0.75 & 0.59 & 0.74 & 0.57 \\
010753734 & 19.41 & 5446 & 5603 & 5183 & 0.97 & 0.89 & 0.99 & 0.91 \\
010794242 & 7.144 & 5459 & 5490 & 3633 & 0.94 & 0.49 & 1.20 & 0.58 \\
010979716 & 10.68 & 3932 & 3996 & 3530 & 0.63 & 0.41 & 0.68 & 0.43 \\
010992733 & 18.53 & 5274 & 5457 & 4848 & 0.94 & 0.83 & 1.04 & 0.91 \\
011134079 & 1.261 & 5201 & 5381 & 4732 & 0.92 & 0.81 & 1.17 & 1.01 \\
011233911 & 4.960 & 5193 & 5370 & 4531 & 0.92 & 0.77 & 1.38 & 1.12 \\
011391181 & 8.617 & 5218 & 5288 & 5133 & 0.91 & 0.88 & 0.88 & 0.85 \\
011975363 & 3.518 & 5482 & 5507 & 5457 & 0.95 & 0.94 & 1.09 & 1.08 \\
012004679 & 5.042 & 5432 & 5514 & 5330 & 0.95 & 0.92 & 0.89 & 0.86 \\
012004834 & 0.2623 & 3576 & 3620 & 3468 & 0.48 & 0.34 & 0.48 & 0.35 \\
012356914 & 27.31 & 5368 & 5455 & 4003 & 0.94 & 0.63 & 0.93 & 0.60 \\
012400729 & 0.9317 & 4949 & 5005 & 3715 & 0.86 & 0.54 & 1.03 & 0.61 \\
012418816 & 1.522 & 4583 & 4603 & 4563 & 0.78 & 0.77 & 0.81 & 0.80 \\
012470530 & 8.207 & 4725 & 4863 & 4245 & 0.83 & 0.70 & 0.78 & 0.64 \\
012599700 & 1.018 & 3887 & 3936 & 3816 & 0.61 & 0.57 & 0.32 & 0.30 \\